\pdfoutput=1

\documentclass[a4paper,nopdfoutputerror,accepted=2024-03-12]{quantumarticle}

\usepackage{tikz}
\usetikzlibrary{decorations,decorations.pathmorphing,decorations.markings,calc,snakes,positioning,fixedpointarithmetic,shapes,shapes.geometric,patterns}

\usepackage{etoolbox}
\AtBeginEnvironment{tikzpicture}{\catcode`$=3 } 

\tikzset{alignmid/.style={baseline={([yshift=-.5ex]current bounding box.center)}}} 
\tikzset{every picture/.append style=alignmid}

\tikzset{
bottomzigzag/.style={postaction={draw,decorate, decoration={zigzag,amplitude=1pt,segment length=3pt,raise=1pt}}},
zigzag/.style={draw,decorate, decoration={zigzag,amplitude=1pt,segment length=3pt}},
rc/.style=rounded corners,
}

\tikzset{
    -|/.style={to path={-| (\tikztotarget)}},
    |-/.style={to path={|- (\tikztotarget)}},
}

\usepackage{ifthen}
\usepackage{xparse,pgffor}

\tikzset{
mark/.code={
\tikzset{postaction={/network/mark/.cd,#1,/tikz/.cd,decorate,decoration={name=markings,mark=at position \netmarkpos with{
\begin{scope}[netmarktrafo]
\netmarkcode
\end{scope}
}}}}
\def\netmarkpos{0.5}
},
}

\def\netmarkpos{0.5}
\def\netmarkposoff{0cm}
\def\netmarkcode{}

\tikzset{
netmarktrafo/.style={},
netmarkstyle/.style={solid,semithick,sharp corners},
}

\pgfkeys{
/network/mark/.cd,
sty/.code={
\tikzset{netmarkstyle/.style={#1}}
},
asty/.code={
\tikzset{netmarkstyle/.append style={#1}}
},
f/.style={asty=fill},
p/.code={ 
\def\netmarkpos{#1}
},
poff/.code={ 
\def\netmarkposoff{#1cm}
},
e/.code={ 
\def\netmarkpos{\pgfdecoratedpathlength-0.005cm-\netmarkposoff}

\tikzset{netmarktrafo/.append style={shift={(-\netmarkwidth,0)}}}
},
s/.code={ 
\def\netmarkpos{0.005cm+\netmarkposoff}

\tikzset{netmarktrafo/.append style={shift={(\netmarkwidth,0)},xscale=-1,yscale=-1}}
},
a/.code={ 
\def\netmarkpos{\pgfdecoratedpathlength-0.005cm}

\tikzset{netmarktrafo/.append style={xscale=-1,shift={(-\netmarkwidth,0)}}}
},
b/.code={ 
\def\netmarkpos{0.005cm}

\tikzset{netmarktrafo/.append style={xscale=-1,shift={(\netmarkwidth,0),yscale=-1}}}
},
-/.code={ 
\tikzset{netmarktrafo/.append style={xscale=-1}}
},
r/.code={ 
\tikzset{netmarktrafo/.append style={yscale=-1}}
},
sideoff/.code={ 
\tikzset{netmarktrafo/.append style={shift={(0,#1)}}}
},
slab/.code={ 
\def\netmarkwidth{0}
\def\netmarkcode{
\node[inner sep=0.04cm,netmarkstyle,draw=none] (mylabelwidthtest) at (0,0){\phantom{#1}};
\path let \p1=(mylabelwidthtest.north east), \p2=(mylabelwidthtest.south east), \n1 = {max(abs(\y1),abs(\y2))} in node[inner sep=0.04cm,netmarkstyle] at (0,\n1) {#1};
}
},
lab/.code={ 
\def\netmarkwidth{0}
\def\netmarkcode{
\node[inner sep=0.04cm,anchor=\netmarkanchor] (mylabelwidthtest) at (0,0) {\phantom{#1}};
\draw[white] (mylabelwidthtest.\pgfdecoratedangle)--(mylabelwidthtest.\pgfdecoratedangle+180);
\node[inner sep=0.04cm,anchor=\netmarkanchor,netmarkstyle] at (0,0) {#1};
}
},
rotlab/.code={ 
\def\netmarkwidth{0}
\def\netmarkcode{
\node[inner sep=0.04cm,fill=white,transform shape,rotate=90,anchor=\netmarkrotanchor,netmarkstyle] (mydecorationnodename) at (0,0) {#1};
}
},
mrotlab/.style={ 
rotlab={$\scriptstyle{#1}$}
},
arr/.code={ 
\def\netmarkwidth{0.04}
\def\netmarkcode{\draw[netmarkstyle] (-0.04,0.08)--(0.04,0)--(-0.04,-0.08);}
},
arrbar/.code={ 
\def\netmarkwidth{0.08}
\def\netmarkcode{\draw[netmarkstyle] (-0.08,0.08)--(0,0)--(-0.08,-0.08) (0.04,0.08)--(0.04,-0.08);}
},
rarr/.code={ 
\def\netmarkwidth{0.04}
\def\netmarkcode{\draw[netmarkstyle] (-0.04,-0.08)arc(90-180:90:0.08);}
},
dot/.code={ 
\def\netmarkwidth{0.08}
\def\netmarkcode{\draw[netmarkstyle] (0,0)circle(0.08);}
},
flag/.code={ 
\def\netmarkwidth{0.06}
\def\netmarkcode{\draw[netmarkstyle] (-0.06,0)--(0,0.09)--(0.06,0)--cycle;}
},
darr/.code={ 
\def\netmarkwidth{0.08}
\def\netmarkcode{\draw[netmarkstyle] (-0.04,0)--(0.04,0)--(-0.04,0.08)--cycle;}
},
circ/.code={
\def\netmarkwidth{0.1}
\def\netmarkcode{\draw[netmarkstyle] (0,0) circle (0.1);}
},
hcirc/.code={ 
\def\netmarkwidth{0.1}
\def\netmarkcode{\draw[netmarkstyle] (-0.1,0) arc (180:0:0.1);}
},
diamond/.code={
\def\netmarkwidth{0.1}
\def\netmarkcode{\draw[netmarkstyle] (-0.1,0)--(0,-0.1)--(0.1,0)--(0,0.1)--cycle;}
},
bar/.code={ 
\def\netmarkwidth{0.05}
\def\netmarkcode{
\draw[netmarkstyle] (0,-0.08cm-0.5*\pgflinewidth)--(0,0.08cm+0.5*\pgflinewidth);
}
},
dbar/.code={ 
\def\netmarkwidth{0.13}
\def\netmarkcode{
\draw[netmarkstyle] (-0.04cm,-0.08cm-0.5*\pgflinewidth)--(-0.04cm,0.08cm+0.5*\pgflinewidth) (0.04cm,-0.08cm-0.5*\pgflinewidth)--(0.04cm,0.08cm+0.5*\pgflinewidth);
}
},
hbar/.code={ 
\def\netmarkwidth{0.05}
\def\netmarkcode{
\draw[netmarkstyle] (0, 0.5*\pgflinewidth)--++(0,0.12);
}
},
mill/.code={
\def\netmarkwidth{0.16}
\def\netmarkcode{
\draw[netmarkstyle] (0,-0.5*\pgflinewidth)--++(-0.08,-0.08)--++(0,0.08);
\draw[netmarkstyle] (0,0.5*\pgflinewidth)--++(0.08,0.08)--++(0,-0.08);
}
},
three dots/.code={
\def\netmarkwidth{0.2}
\def\netmarkcode{
\fill (-0.12,0) circle (0.5*0.05) (0,0) circle (0.5*0.05) (0.12,0) circle (0.5*0.05);
}
},
}

\tikzset{wid/.style={minimum width=#1cm}}
\tikzset{hei/.style={minimum height=#1cm}}

\tikzset{sx/.style={xshift=#1cm}}
\tikzset{sy/.style={yshift=#1cm}}

\tikzset{box/.style={draw,rectangle}}
\tikzset{fbox/.style={draw,rectangle, line width=1.1}}
\tikzset{roundbox/.style={draw,rectangle,rounded corners}}
\tikzset{froundbox/.style={draw,rectangle, rounded corners, line width=1.1}}
\tikzset{rounddiamond/.style={draw,diamond,rounded corners}}
\tikzset{dot/.style={draw, shape=circle, fill=black, scale=0.5}}

\tikzset{
netbox/.code={
\node[draw,netbdstyle] (\atomname) at (0,0) {#1};
\coordinate (\atomname-r) at (\atomname.east);
\coordinate (\atomname-l) at (\atomname.west);
\coordinate (\atomname-t) at (\atomname.north);
\coordinate (\atomname-b) at (\atomname.south);
\coordinate (\atomname-tr) at (\atomname.north east);
\coordinate (\atomname-br) at (\atomname.south east);
\coordinate (\atomname-tl) at (\atomname.north west);
\coordinate (\atomname-bl) at (\atomname.south west);
},
}

\tikzset{bdlw/.code={\tikzset{mybdstyle/.style={draw, line width=#1}}}}
\tikzset{bdcol/.code={\tikzset{mybdstyle/.append style={#1}}}}

\newcommand\setelements[1]{
\pgfkeys{/network/atom/.cd,#1}
}

\newcommand\setmarks[1]{
\pgfkeys{/network/mark/.cd,#1}
}

\newcommand\atoms[2]{
\foreach \name/\keys in {#2}{
\expandafter\atom\expandafter{\keys,#1}{\name}
}
}

\newcommand\atom[2]{
\def\atomname{#2}
\tikzset{
nettrafo/.style={},
netatompos/.style={},
netdeco/.style={},
netpostdeco/.style={},
}

\pgfkeys{/network/atom/.cd,#1}

\begin{scope}[netatompos] 
\begin{scope}[nettrafo] 
\netshapecoords 
\fill[netbackstyle] \netshapepath;
\clip \netshapepath;
\tikzset{netdeco}
\draw[netbdstyle] \netshapepath;
\end{scope}
\tikzset{netpostdeco} 
\end{scope}

}

\pgfkeys{
/network/atom/.cd,
square/.code={

\def\netshapepath{(-\tempsize,-\tempsize)rectangle (\tempsize,\tempsize)}
\def\netshapecoords{
\node[rectangle,wid=2*\tempsize,hei=2*\tempsize,inner sep=0,transform shape](\atomname)at(0,0){};
\coordinate(\atomname-c) at (0,0);
\coordinate(\atomname-r) at (\tempsize,0);
\coordinate(\atomname-l) at (-\tempsize,0);
\coordinate(\atomname-t) at (0,\tempsize);
\coordinate(\atomname-b) at (0,-\tempsize);
\coordinate(\atomname-br) at (\tempsize,-\tempsize);
\coordinate(\atomname-tr) at (\tempsize,\tempsize);
\coordinate(\atomname-bl) at (-\tempsize,-\tempsize);
\coordinate(\atomname-tl) at (-\tempsize,\tempsize);
}},
circ/.code={

\def\netshapepath{(0,0)circle(\tempsize)}
\def\netshapecoords{
\node[circle,wid=2*\tempsize,hei=2*\tempsize,inner sep=0,transform shape](\atomname)at(0,0){};
\coordinate(\atomname-c) at (0,0);
\coordinate(\atomname-r) at (\tempsize,0);
\coordinate(\atomname-l) at (-\tempsize,0);
\coordinate(\atomname-t) at (0,\tempsize);
\coordinate(\atomname-b) at (0,-\tempsize);
}},
triang/.code={

\def\netshapepath{(-30:\tempsize)--(90:\tempsize)--(-150:\tempsize)--cycle}
\def\netshapecoords{
\node[regular polygon,regular polygon sides=3,wid=2*\tempsize,inner sep=0,transform shape](\atomname)at(0,0){};
\coordinate(\atomname-c) at (0,0);
\coordinate(\atomname-cr) at (-30:\tempsize);
\coordinate(\atomname-cl) at (-150:\tempsize);
\coordinate(\atomname-ct) at (90:\tempsize);
\coordinate(\atomname-mb) at (-90:0.5*\tempsize);
\coordinate(\atomname-mr) at (30:0.5*\tempsize);
\coordinate(\atomname-ml) at (150:0.5*\tempsize);
}},
diamond/.code={

\def\netshapepath{(0,-\tempsize)--(\tempsize,0)--(0,\tempsize)--(-\tempsize,0)--cycle}
\def\netshapecoords{
\node[rotate=45,rectangle,wid=sqrt(2)*\tempsize,hei=sqrt(2)*\tempsize,inner sep=0,transform shape](\atomname)at(0,0){};
\coordinate(\atomname-c) at (0,0);
\coordinate(\atomname-r) at (\tempsize,0);
\coordinate(\atomname-l) at (-\tempsize,0);
\coordinate(\atomname-t) at (0,\tempsize);
\coordinate(\atomname-b) at (0,-\tempsize);
}},
pentagon/.code={

\def\netshapepath{(-126:\tempsize)--(-54:\tempsize)--(18:\tempsize)--(90:\tempsize)--(162:\tempsize)--cycle}
\def\netshapecoords{
\node[regular polygon,regular polygon sides=5,wid=2*\tempsize,inner sep=0,transform shape](\atomname)at(0,0){};
\coordinate(\atomname-c) at (0,0);
\coordinate (\atomname-mb)at(-90:{\tempsize*cos(36)});
\coordinate (\atomname-mbr)at(-18:{\tempsize*cos(36)});
\coordinate (\atomname-mtr)at(54:{\tempsize*cos(36)});
\coordinate (\atomname-mtl)at(126:{\tempsize*cos(36)});
\coordinate (\atomname-mbl)at(-162:{\tempsize*cos(36)});
\coordinate (\atomname-cbr)at(-54:\tempsize);
\coordinate (\atomname-cr)at(18:\tempsize);
\coordinate (\atomname-ct)at(90:\tempsize);
\coordinate (\atomname-cl)at(162:\tempsize);
\coordinate (\atomname-cbl)at(-126:\tempsize);
}},
halfcirc/.code={

\def\netshapepath{(\tempsize,0)arc(0:180:\tempsize)--++(0,-0.04)-|cycle}
\def\netshapecoords{
\node[circle,wid=2*\tempsize,hei=2*\tempsize,inner sep=0,transform shape](\atomname)at(0,0){};
\coordinate(\atomname-c) at (0,0);
\coordinate(\atomname-r) at (\tempsize,0);
\coordinate(\atomname-l) at (-\tempsize,0);
\coordinate(\atomname-t) at (0,\tempsize);
\coordinate(\atomname-b) at (0,0);
}},
void/.code={
\def\netshapepath{}
\def\netshapecoords{
\coordinate(\atomname) at (0,0);
\coordinate(\atomname-c) at (0,0);
}},
labbox/.code={
\def\netshapepath{(0,0)}
\def\netshapecoords{}
\tikzset{netpostdeco/.append style={netbox=#1}}
},
}

\pgfkeys{
/network/atom/.cd,
p/.code={\tikzset{netatompos/.append style={shift={(#1)}}}},
xscale/.code={\tikzset{nettrafo/.append style={xscale=#1}}},
yscale/.code={\tikzset{nettrafo/.append style={yscale=#1}}},
scale/.code={\tikzset{nettrafo/.append style={scale=#1}}},
rot/.code={\tikzset{nettrafo/.append style={rotate=#1}}},
hflip/.style={xscale=-1},
vflip/.style={yscale=-1},
big/.style={scale=3/2},
small/.style={scale=2/3},
huge/.style={scale=9/4},
tiny/.style={scale=4/9},
flat/.style={yscale=2/3},
fflat/.style={yscale=4/9},
}

\tikzset{
netbdstyle/.style={line width=0.15em}, 
netdecstyle/.style={},
netpostdecstyle/.style={},
netbackstyle/.style={white},
}
\pgfkeys{
/network/atom/.cd,
bdstyle/.code={\tikzset{netbdstyle/.style={#1}}},
bdastyle/.code={\tikzset{netbdstyle/.append style={#1}}},
decstyle/.code={\tikzset{netdecstyle/.style={#1}}},
decastyle/.code={\tikzset{netdecstyle/.append style={#1}}},
postdecstyle/.code={\tikzset{netpostdecstyle/.style={#1}}},
postdecastyle/.code={\tikzset{netpostdecstyle/.append style={#1}}},
backstyle/.code={\tikzset{netbackstyle/.style={#1}}},
backastyle/.code={\tikzset{netbackstyle/.append style={#1}}},
style/.style={bdstyle=#1, decstyle=#1, postdecstyle=#1},
astyle/.style={bdastyle=#1, decastyle=#1, postdecastyle=#1},
whiteblack/.style={decstyle=white,backstyle=black},
nobd/.style={bdstyle={line width=0}},
}

\tikzset{
netbscope/.code={\begin{scope}[#1]},
netescope/.code={\end{scope}},
}

\pgfkeys{
/network/atom/.cd,
bscope/.code={\tikzset{netdeco/.append style={netbscope={#1}}}},
escope/.code={\tikzset{netdeco/.append style={netescope}}},
dec/.style 2 args={bscope={#1},#2,escope}, 
vbar/.style={dec={rotate=90}{hbar}},
dcross/.style={dec={rotate=45}{hbar},dec={rotate=-45}{hbar}},
cross/.style={hbar,vbar},
sect6/.style={sect=60},
sect8/.style={sect=45},
sect10/.style={sect=36},
}

\def\regdec#1{\pgfkeys{/network/atom/.cd,#1/.code={\tikzset{netdeco/.append style={net#1}}}}}
\regdec{all}\regdec{rhalf}\regdec{rquart}\regdec{brquart}\regdec{dot}\regdec{spiral}\regdec{swirl}\regdec{hstripe}\regdec{hbar}\regdec{rrey}
\pgfkeys{/network/atom/.cd,sect/.code={\tikzset{netdeco/.append style={netsect=#1}}}}

\tikzset{
netall/.code={\fill[netdecstyle] (-0.3,-0.3)rectangle (0.3,0.3);}, 
netrhalf/.code={\fill[netdecstyle] (0,-0.3)rectangle (0.3,0.3);}, 
netrquart/.code={\fill[netdecstyle] (0.075,-0.3)rectangle (0.3,0.3);}, 
netbrquart/.code={\fill[netdecstyle] (0,0)rectangle (0.3,-0.3);}, 
netsect/.code={\fill[netdecstyle] (0,0)--(0,-0.3)arc(-90:-90+#1:0.3)--cycle;}, 
netdot/.code={\fill[netdecstyle] (0,0)circle(0.07);}, 
netspiral/.code={\draw[netdecstyle] plot [variable=\t,domain=0:4] ({0.075*\t*cos(pi*(\t-0.5) r)},{0.075*\t*sin(pi*(\t-0.5) r)});}, 
netswirl/.code={\fill[netdecstyle] plot [variable=\t,domain=0:2] ({0.15*\t*cos(pi*(\t-0.5) r)},{0.15*\t*sin(pi*(\t-0.5) r)}) arc(-90:-450:0.3)--cycle;}, 
nethstripe/.code={\fill[netdecstyle] (-0.3,-0.05)rectangle(0.3,0.05);}, 
nethbar/.code={\draw[netdecstyle] (-0.3,0)--(0.3,0);}, 
netrrey/.code={\draw[netdecstyle] (0,0)--(0.3,0);} 
}

\pgfkeys{
/network/atom/.cd,
postbscope/.code={\tikzset{netpostdeco/.append style={netbscope={#1}}}},
postescope/.code={\tikzset{netpostdeco/.append style={netescope}}},
postdec/.style 2 args={postbscope={#1},#2,postescope}, 
arc/.code={\tikzset{netpostdeco/.append style={netarc=#1}}},
lab/.code={\tikzset{netpostdeco/.append style={netlab={#1}}}},
shadecirc/.code={\tikzset{netpostdeco/.append style=netshadecirc}},
shaderect/.code={\tikzset{netpostdeco/.append style={netshaderect={#1}}}},
debug/.code={\tikzset{netpostdeco/.append style=netdebug}},
markline/.code 2 args={\tikzset{netpostdeco/.append style={netmarkline={#1}{#2}}}},
postcirc/.code={\tikzset{netpostdeco/.append style=netpostcirc}},
}

\tikzset{
netlab/.code={
\pgfkeys{/network/atom/lab/.cd,#1}
\node[netpostdecstyle] at (\ifdefined\netlabpos\netlabpos\else\netlabang:\netlabdist\fi) {\netlabwrap{\netlabtext}};
},
netarc/.code args={#1:#2:#3}{
\draw[netpostdecstyle] (#1:#3) arc (#1:#2:#3);
},
netshadecirc/.code= {
\fill[opacity=0.4,netpostdecstyle] (0,0)circle(0.4);
},
netpostcirc/.code= {
\draw[netpostdecstyle] (0,0)circle(0.15);
},
netshaderect/.code= {
\fill[rc,opacity=0.4,netpostdecstyle] ($-1*(#1)$) rectangle (#1);
},
netdebug/.code= {
\node[red] at (0,0){\atomname};
},
netmarkline/.code 2 args= {
\draw (\atomname)edge[mark={#2}]++(#1);
},
}

\def\netlabwrap#1{#1}
\pgfkeys{
/network/atom/lab/.cd,
t/.code={\def\netlabtext{#1}}, 
p/.code={\def\netlabpos{#1}}, 
ang/.code={\def\netlabang{#1}},
dist/.code={\def\netlabdist{#1}},
wrap/.code={\def\netlabwrap##1{#1}}, 
}

\usetikzlibrary{patterns.meta}

\usepackage{xcolor}
\usepackage{amsmath}
\usepackage{amssymb}
\usepackage{amsthm}
\theoremstyle{definition}
\newtheorem{mydef}{Informal definition}
\newtheorem{myprop}{Proposition}

\usepackage{bbold}
\usepackage{mathtools}
\usepackage{hyperref}
\usepackage{makecell}
\usepackage{braket}

\def\zz{\mathbb{Z}}
\def\cc{\mathbb{C}}
\def\pcoc{P_{\text{cocycle}}}

\definecolor{darkgreen}{rgb}{0,0.6,0.2}
\definecolor{darkyellow}{rgb}{0.7,0.5,0}
\newcommand{\vca}[1]{\textcolor{red}{#1}}
\newcommand{\vcb}[1]{\textcolor{darkgreen}{#1}}
\newcommand{\vcc}[1]{\textcolor{blue}{#1}}
\newcommand{\vcd}[1]{\textcolor{darkyellow}{#1}}

\tikzset{
ind/.style={mark={lab=#1,a}}, 
startind/.style={mark={lab=#1,b}}, 
classical/.style={line width=1.5},
quantum/.style={double,double distance=0.5mm},
actualedge/.style={line width=2},
anyon1gon/.style={fill=red,fill opacity=0.2,text opacity=1},
e or not/.style={preaction={draw,very thick,red!50}},
m or not/.style={pattern={Hatch[distance=1.2mm,line width=0.3mm,angle=45]},pattern color=red,opacity=0.3},
front/.style={preaction={draw,white,line width=2}},
worldline/.style={red,line width=5,opacity=0.3,line join=round},
}

\setmarks{
ar/.style={arr,asty=fill},
}

\setelements{
z2/.style={small,circ},
delta/.style={small,circ,all}, 
deltatens/.style={tiny,circ,all}, 
cc/.style={small,circ,dcross},
hadamard/.style={small,square,dcross},
smat/.style={small,square,all},
umat/.style={small,square},
vertex/.style={tiny,circ,all},
vca/.style={astyle=red},
vcb/.style={style=darkgreen},
vcc/.style={style=blue},
vcd/.style={style=darkyellow},
}

\begin{document}

\title{Topological error correcting processes from fixed-point path integrals}
\author{Andreas Bauer}
\email{andibauer@zedat.fu-berlin.de}
\affiliation{Freie Universit{\"a}t Berlin, Arnimallee 14, 14195 Berlin, Germany}

\begin{abstract}
We propose a unifying paradigm for analyzing and constructing topological quantum error correcting codes as dynamical circuits of geometrically local channels and measurements.
To this end, we relate such circuits to discrete fixed-point path integrals in Euclidean spacetime, which describe the underlying topological order:
If we fix a history of measurement outcomes, we obtain a fixed-point path integral carrying a pattern of topological defects.
As an example, we show that the stabilizer toric code, subsystem toric code, and CSS honeycomb Floquet code can be viewed as one and the same code on different spacetime lattices, and the honeycomb Floquet code is equivalent to the CSS honeycomb Floquet code under a change of basis.
We also use our formalism to derive two new error-correcting codes, namely a Floquet version of the $3+1$-dimensional toric code using only 2-body measurements, as well as a dynamic code based on the double-semion string-net path integral.
\end{abstract}

\maketitle
\tableofcontents

\section{Introduction}
One of the most promising routes towards scalable fault-tolerant quantum computation is topological quantum computation.
Thereby, logical quantum information is stored in the ground state space of a topological phase on a topologically non-trivial spatial configuration, which may contain \emph{computational defects} such as anyons, twist defects, or boundaries \cite{Kitaev1997,Dennis2001,Nayak2007}.
Topological order has been shown to be robust under arbitrary local perturbations \cite{Bravyi2010}.
In a similar vein, topological quantum error correction (QEC) is believed to provide a threshold for arbitrary local noise.
Despite the similarity, these two notions of robustness are technically very different:
Whereas topological order concerns ground-state properties, captured by the imaginary-time evolution, topological QEC executes a dissipative real-time evolution including syndrome measurements and corrections.

Both topological phases and topological QEC can be described by path integrals that are discrete in space and time.
For QEC, these are mixed-state circuits of quantum channels and measurements.
For topological phases, they are fixed-point models in the form of state-sum TQFTs \cite{Fukuma1992,Dijkgraaf1990,Turaev1992,Barrett1993,Crane1993} or tensor-network path integrals \cite{liquid_intro}.
In this paper, we present a picture for topological QEC, at whose core is the relation between the two types of discrete path integrals.
Concretely, there exists a history of ``trivial'' measurement outcomes (often $+1$ for Pauli based codes) such that the QEC circuit becomes a fixed-point path integral.
The other histories of measurement outcomes then correspond to the same path integral including different patterns of topological defects such as anyons, which will be called \emph{syndrome defects}.
The path integral is locally invariant under certain changes of the positions of the syndrome defects, giving rise to equivalences between different defect patterns.
The corrections correspond to the insertion of additional segments of defects, which are chosen by the classical decoder to ensure that the total pattern of defects is equivalent to the trivial one.
The correspondence between QEC circuits and path integrals with defects provides a single simple criterion for topological fault tolerance.

Our formalism has two major practical applications.
The first application is to systematically analyze existing codes.
In particular, the correspondence to path integrals can be used to assign a topological phase to any topological code.
This phase determines the logical dimension on different topologies, the possible boundary conditions, anyons, or other sorts of computational defects that can be introduced, as well as the logical operations that can be performed.
Codes within the same phase can be seen as distinct microscopic representations of one another.
To illustrate this, we focus on recently developed \emph{Floquet codes} \cite{Hastings2021,Haah2021,Kesselring2022,Davydova2022,Aasen2022,Aasen2023,Sullivan2023,Zhang2022}.
These are specified by a sequence of Pauli measurements, usually 2-body, measured in a fixed schedule.
Since the checks are non-commuting, they can be analyzed using the formalism of \emph{subsystem codes} \cite{Kribs2004,Bombin2009,Bravyi2012}.
However, due to the fixed schedule, they manage to protect a certain number of logical qubits even though the subsystem formalism predicts less or none at all.
Lately, there has been a quest to better understand the relation between stabilizer, subsystem, and Floquet codes, and between different Floquet codes among each other.
Our formalism helps to establish direct relations between different codes, often by finding that they belong to a common phase.
Concretely, we find that the stabilizer toric code \cite{Kitaev1997,Dennis2001}, the subsystem toric code \cite{Bravyi2012}, and the CSS honeycomb Floquet code \cite{Kesselring2022,Davydova2022,Aasen2022} correspond to the same path integral on different spacetime lattices, and thus belong to a single code family in our spacetime perspective.
This can be seen as a spacetime analogue of viewing stabilizer toric codes on different spatial lattices as part of a single code family.
We also find that the underlying path integrals of the CSS honeycomb Floquet code and the original honeycomb Floquet code \cite{Hastings2021} are equal up to a local change of basis.
All four codes belong to the toric code phase.
They only differ by the microscopic representation of the underlying path integral, as well as by the locations of the syndrome defect segments corresponding to the non-trivial measurement outcomes.

The second application is to systematically construct new codes.
Roughly speaking, we start with a fixed-point path integral and interpret it as a non-unitary circuit.
Then we turn each non-unitary operator into an instrument that measures the absence or presence of a syndrome defect.
The circuit of instruments then defines a fault tolerant dynamic code.
By making use of the rich and developed mathematical theory of fixed-point models, this yields a great variety of new dynamic topological codes.
First, we can start from different models in different families of fixed-point path integrals, corresponding to different phases.
Further, topological fixed-point path integrals have a notion of exact combinatorial topological invariance, which is at the heart of their success in classifying topological phases.
So we can put the fixed-point path integrals on arbitrary spacetime lattices.
Finally, even if the path integral and lattice are fixed, they can be turned into a non-unitary circuit in various ways by choosing different causal orderings.
An interesting feature of our approach that goes beyond much of the quantum error-correction literature is that there is no necessity for the resulting codes to be based on Pauli/Clifford measurements or operations.
Concretely, we illustrate the capability of finding new codes through two examples.
First, we construct a Floquet version of the $3+1$-dimensional toric code that uses only 2-body $XX$ and $ZZ$ measurements.
The code lives on a triangulation with 4-colorable vertices, with a qubit on every left-handed tetrahedron.
In each of 8 rounds we perform measurements on the qubits adjacent to each edge of a certain type.
Second, we present a non-Pauli dynamic code based on the double-semion Turaev-Viro path integral \cite{Levin2004,Hu2012,Dijkgraaf1990,Turaev1992,Barrett1993}.
We sketch a presentation of this code as a circuit of common 2 and 3-qubit gates and measurements.
Due to its relatively large spacetime overhead it merely serves as an illustrative example rather than as a practical QEC code.

The structure of the remainder of this work is as follows.
In Section~\ref{sec:general_formalism}, we review fixed-point path integrals and their defects, and introduce the main definition of a fixed-point path integral code.
In Section~\ref{sec:examples}, we use our formalism to analyze four examples of existing codes as mentioned above.
In Section~\ref{sec:3d_toric_code}, construct the two new dynamic codes mentioned above.

\section{From fixed-point path integrals to error-correcting codes}
\label{sec:general_formalism}
\subsection{Fixed-point path integrals}
\label{sec:fixed_points}
In this section, we review fixed-point path integrals for topological phases, which are the key to understanding and constructing topological QEC codes in this work.
Fixed-point path integrals are defined on lattices representing a discrete Euclidean-signature spacetime.
The most common formulation of such path integrals is as \emph{state sums} \cite{Fukuma1992,Dijkgraaf1990,Turaev1992,Barrett1993,Crane1993}.
To this end we associate discrete variables to certain types of places (for example, all edges) in the lattice, and weights to other places (for example, all volumes).
Each weight depends on the configuration of the nearby variables.
We then perform a sum over all configurations of the variables, where the summand is given by the product of all the weights,
\begin{equation}
Z=\sum_{\mathbf c} \prod_i \omega_i(c_{i_0}, c_{i_1}, \ldots)\;.
\end{equation}
Here, $\mathbf c$ is a configuration of variables, and $i_0,i_1,\ldots$ label the variables on which the weight $\omega_i$ at the location $i$ depends.
Such path integrals are commonly used as partition functions in classical statistical physics on a space-only lattice, for example in the classical Ising model.
Here we will use them on a spacetime lattice to represent the imaginary-time evolution of a geometrically local Hamiltonian.
To this end, note that the imaginary-time evolution can be approximately discretized to a non-unitary circuit through Trotterization.
This non-unitary circuit yields a discrete path integral where the configurations of variables are the histories of qubit configurations, and the weights are the amplitudes of the non-unitary gates.
Even though the path integrals we consider do not directly correspond to discretized imaginary-time evolutions, they share the same qualitative properties and have the same physical interpretation.
Note that in this paper, all models will be translation invariant or uniformly defined on some lattice, and fully specified by a single unit cell.
If we want to evaluate the number $Z$, we need to choose a finite spacetime lattice with periodic boundary conditions.
As such, the number $Z$ alone has no immediate physical relevance.
However, consider evaluating the path integral on some patch with boundary, and only summing over variables in the interior while keeping these near the boundary fixed.
The number $Z$ can then be interpreted as the amplitude of a state for the fixed configuration of boundary variables.
We will thus refer to such a boundary as a \emph{spatial boundary}.
Physically, this state is a ground state of the model on the spatial boundary.

An equivalent formulation that is better suited for our purpose are \emph{tensor-network path integrals} \cite{liquid_intro}.
These are tensor networks whose tensors are located at some places (for example, all volumes) of the spacetime lattice, and nearby tensors share bonds (for example, at every face, connecting the two adjacent volumes).
Note that this is different from MPS or PEPS which live in space only and describe states, not path integrals.
In particular, tensor-network path integrals have no open indices except for when we cut the tensor network at a spatial boundary.
Contracting the resulting tensor network with open indices yields a tensor, which analogously to the previous paragraph can be interpreted as a ground state living on the spatial boundary.
Also in contrast to MPS or PEPS, path integrals have no distinction between \emph{virtual indices} that are contracted and \emph{physical indices} that remain open.

Topological fixed-point path integrals have one single powerful property that makes them exactly solvable, namely discrete topological invariance.
To this end, we first define the path integral not only on regular lattices, but on arbitrary triangulations or cellulations.
Let us consider here the case of $2+1$ dimensions where most of topological error correction takes place.
One possibility is to put the same 4-index tensor (in black) onto every tetrahedron of a 3-dimensional triangulation (in orange),
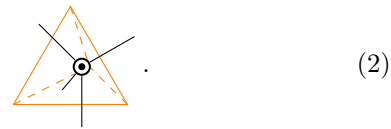
\begin{equation}
\begin{tikzpicture}
\atoms{void}{0/, 1/p={1.5,0}, 2/p={1,0.5}, 3/p={0.75,1.3}}
\draw[orange] (0)--(1)--(3)--(0);
\draw[orange,dashed] (0)--(2) (1)--(2) (3)--(2);
\atoms{circ,small,dot}{t/p={0.9,0.5}}
\draw (t)--++(135:0.8) (t)--++(-90:0.8) (t)--++(30:0.8) (t)--++(-130:0.4);
\end{tikzpicture}\;.
\end{equation}
Then, we choose a set of local deformations, which can arbitrarily change the lattice while keeping its overall topology fixed.
The local deformations of the lattice correspond to changing the tensor network by cutting out a small patch and gluing in another patch.
Topological invariance of the path integral is imposed by equating the cut-out and glued-in patches.
In our above example, we can demand invariance under \emph{Pachner moves} \cite{Pachner1991}, such as
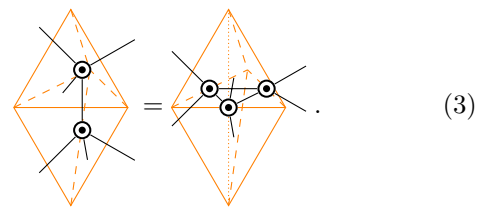
\begin{equation}
\label{eq:pachner_move_equation}
\begin{tikzpicture}
\atoms{void}{0/, 1/p={1.5,0}, 2/p={1,0.5}, 3/p={0.75,1.3}, 4/p={0.75,-1.3}}
\draw[orange] (0)--(1)--(3)--(0) (0)--(4)--(1);
\draw[orange,dashed] (0)--(2) (1)--(2) (3)--(2) (4)--(2);
\atoms{circ,small,dot}{t1/p={0.9,0.5}, t2/p={0.9,-0.3}}
\draw (t1)--++(135:0.8) (t1)--++(30:0.8) (t1)--++(-130:0.4) (t2)--++(-135:0.8) (t2)--++(-30:0.8) (t2)--++(-80:0.4) (t1)--(t2);
\end{tikzpicture}
=
\begin{tikzpicture}
\atoms{void}{0/, 1/p={1.5,0}, 2/p={1,0.5}, 3/p={0.75,1.3}, 4/p={0.75,-1.3}}
\draw[orange] (0)--(1)--(3)--(0) (0)--(4)--(1);
\draw[orange,dashed] (0)--(2) (1)--(2) (3)--(2) (4)--(2);
\draw[orange,densely dotted] (3)--(4);
\atoms{circ,small,dot}{t1/p={0.75,0}, t2/p={0.5,0.25}, t3/p={1.25,0.25}}
\draw (t2)--++(135:0.7) (t2)--++(-135:0.7) (t1)--++(-80:0.4) (t1)--++(80:0.4) (t3)--++(-30:0.6) (t3)--++(30:0.6) (t1)--(t2)--(t3)--(t1);
\end{tikzpicture}\;.
\end{equation}
On the left hand side, we have two tetrahedra stacked on top of each other.
On the right, there are three tetrahedra surrounding the vertical dotted line in the center, such that each tetrahedron is spanned by one of the three edges ``around the equator'' with the edge ``connecting the north and south pole''.
Note that in addition to the ``standard'' way of implementing discrete topological invariance above, there are other ways of doing this.
A general implementation of topological invariance is a way to assign tensor networks to cellulations such that tensors and the geometry of the network at one place depends only on the combinatorial structure of the cellulation within a constant-size neighborhood.
Another implementation would be to put tensors on the edges and faces instead of the volumes, which is how topological invariance is formulated for the toric code below.
Different implementations are equivalent in the sense that they describe the same phases.
However, these different microscopic representations of the same phase are interesting in practice, since yield different QEC codes via the formalism we propose.
For a more detailed discussion of discrete topological invariance in fixed-point path integrals, we refer the reader to Ref.~\cite{liquid_intro}.

As an example, let us consider the toric code path integral in 2+1 spacetime dimensions, which will be used in all examples in Section~\ref{sec:examples}.
As a stabilizer code, the toric code is defined with qubits on the edges of a 2-dimensional square lattice, a $Z_0Z_1Z_2Z_3$ stabilizer at each face acting on the four adjacent qubits, and a $X_0X_1X_2X_3$ stabilizer at each vertex acting on the four adjacent qubits.
The ground states (that is, the stabilized states) are equal-weight superpositions of all configurations of one $\zz_2$-variable at every edge such that at every plaquette the surrounding variables sum to $0$ (mod $2$ with $\zz_2$ written additively).
On the Poincar\'e dual lattice, these configurations can be pictured as equal-weight superpositions of all closed-loop patterns.
Mathematically, the closed-loop patterns are $\zz_2$-valued \emph{1-cocycles}.
The path integral in the 2+1-dimensional spacetime cellulation is just the same, namely a sum over all 1-cocycles.
In 2+1 dimensions, 1-cocycles can be pictured as closed-membrane configurations on the Poincar\'e dual lattice.
Indeed, consider the state obtained by evaluating the path integral on some cellulation with spatial boundary as discussed above:
Each bulk (closed membrane) 1-cocycle restricts to a (closed loop) 1-cocycle on the boundary, and so the resulting ground state is the expected toric code ground state.
So the toric code state sum has one $\zz_2$-variable on each edge, and one weight at each face which is $1$ if the surrounding edge variables sum to $0$, and $0$ otherwise.
As a tensor network, the summation over each $\zz_2$-variable at an edge is implemented by a \emph{$\delta$-tensor},
\begin{equation}
\label{eq:delta_definition}
\begin{tikzpicture}
\atoms{delta}{0/}
\draw (0)edge[ind=$a$]++(0:0.5) (0)edge[ind=$b$]++(90:0.5) (0)edge[ind=$c$]++(180:0.5);
\node at (-90:0.5){$\ldots$};
\end{tikzpicture}
=
\begin{cases}
1 & \text{if } a=b=c=\ldots\\
0 & \text{otherwise}
\end{cases}\;.
\end{equation}
Each weight at a face is implemented by a \emph{$\zz_2$-tensor},
\begin{equation}
\begin{tikzpicture}
\atoms{z2}{0/}
\draw (0)edge[ind=$a$]++(0:0.5) (0)edge[ind=$b$]++(90:0.5) (0)edge[ind=$c$]++(180:0.5);
\node at (-90:0.5){$\ldots$};
\end{tikzpicture}
=
\begin{cases}
1 & \text{if } a+b+c+\ldots=0\mod 2\\
0 & \text{otherwise}
\end{cases}\;.
\end{equation}
So in total, there is one $\zz_2$-tensor at every face and one $\delta$-tensor at every edge,
\begin{equation}
\begin{tikzpicture}
\atoms{void}{1/p={-30:0.6}, 2/p={90:0.6}, 3/p={-150:0.6}}
\fill[orange,opacity=0.35] (3)--(2)--(1)--cycle;
\draw[dashed,orange] (3)--(2) (2)--(1) (3)--(1);
\atoms{z2}{a/}
\draw (a)--++(30:0.5) (a)--++(150:0.5) (a)--++(-90:0.5);
\end{tikzpicture}
,\qquad
\begin{tikzpicture}
\fill[orange,opacity=0.35] (0,0)--(1,0)--++(150:0.6)--++(-1,0)--cycle;
\fill[orange,opacity=0.35] (0,0)--(1,0)--(1,-0.4)--(0,-0.4)--cycle;
\draw[dashed,orange,actualedge] (0,0)--(1,0);
\atoms{delta}{a/p={0.5,0}}
\draw (a)--++(150:0.7) (a)--++(-90:0.5);
\fill[orange,opacity=0.35] (0,0)--(1,0)--++(30:0.4)--++(-1,0)--cycle;
\draw (a)--++(30:0.5);
\end{tikzpicture}
\;,
\end{equation}
and for each pair of adjacent face and edge the tensors are connected by a bond.
For example, on a patch of a cubic lattice (in orange), the tensor network (in black/gray) looks like:
\begin{equation}
\begin{tikzpicture}
\atoms{void}{x/p={2,0}, y/p={0,2}, z/p={1.6,0.8}}
\foreach \x in {0,1,2}{
\foreach \y in {0,1,2}{
\foreach \z in {0,1}{
\atoms{void}{\x\y\z/p={$\x*(x)+\y*(y)+\z*(z)$}}
}}}
\draw[orange] (000)--(200) (010)--(210) (020)--(220) (000)--(020) (100)--(120) (200)--(220) (201)--(221) (021)--(221);
\draw[orange,dashed] (001)--(201) (011)--(211) (001)--(021) (101)--(121);
\draw[orange, dashed] (000)--(001) (100)--(101) (010)--(011) (110)--(111);
\draw[orange] (200)--(201) (210)--(211) (220)--(221) (020)--(021) (120)--(121);
\foreach \x/\y/\z in {1/0/0, 3/0/0, 1/2/0, 3/2/0, 1/4/0, 3/4/0, 0/1/0, 0/3/0, 2/1/0, 2/3/0, 4/1/0, 4/3/0, 4/0/1, 4/2/1, 0/4/1, 2/4/1, 4/4/1, 4/1/2, 4/3/2, 1/4/2, 3/4/2}{
\atoms{delta}{d\x\y\z/p={$0.5*\x*(x)+0.5*\y*(y)+0.5*\z*(z)$}}
}
\foreach \x/\y/\z in {1/0/2, 3/0/2, 1/2/2, 3/2/2, 0/1/2, 0/3/2, 2/1/2, 2/3/2, 2/2/1, 0/2/1, 0/0/1, 2/0/1}{
\atoms{delta,astyle=gray}{d\x\y\z/p={$0.5*\x*(x)+0.5*\y*(y)+0.5*\z*(z)$}}
}
\foreach \x/\y/\z in {1/1/0, 3/1/0, 1/3/0, 3/3/0, 1/4/1, 3/4/1, 4/1/1, 4/3/1}{
\atoms{z2}{z\x\y\z/p={$0.5*\x*(x)+0.5*\y*(y)+0.5*\z*(z)$}}
}
\foreach \x/\y/\z in {1/1/2, 3/1/2, 1/3/2, 3/3/2, 1/2/1, 3/2/1, 2/1/1, 2/3/1, 0/1/1, 0/3/1, 1/0/1, 3/0/1}{
\atoms{z2,astyle=gray}{z\x\y\z/p={$0.5*\x*(x)+0.5*\y*(y)+0.5*\z*(z)$}}
}
\foreach \x/\xx/\xxx in {0/1/2,2/3/4}{
\foreach \y/\yy/\yyy in {0/1/2,2/3/4}{
\foreach \z in {0}{
\draw (z\xx\yy\z)--(d\x\yy\z) (z\xx\yy\z)--(d\xxx\yy\z) (z\xx\yy\z)--(d\xx\y\z) (z\xx\yy\z)--(d\xx\yyy\z);
}
\foreach \z in {2}{
\draw[gray] (z\xx\yy\z)--(d\x\yy\z) (z\xx\yy\z)--(d\xxx\yy\z) (z\xx\yy\z)--(d\xx\y\z) (z\xx\yy\z)--(d\xx\yyy\z);
}}}
\foreach \x/\xx/\xxx in {0/1/2,2/3/4}{
\foreach \y in {0,2}{
\draw[gray] (z\xx\y1)--(d\x\y1) (z\xx\y1)--(d\xxx\y1) (z\xx\y1)--(d\xx\y0) (z\xx\y1)--(d\xx\y2);
}
\foreach \y in {4}{
\draw (z\xx\y1)--(d\x\y1) (z\xx\y1)--(d\xxx\y1) (z\xx\y1)--(d\xx\y0) (z\xx\y1)--(d\xx\y2);
}}
\foreach \y/\yy/\yyy in {0/1/2,2/3/4}{
\foreach \x in {0,2}{
\draw[gray] (z\x\yy1)--(d\x\y1) (z\x\yy1)--(d\x\yyy1) (z\x\yy1)--(d\x\yy0) (z\x\yy1)--(d\x\yy2);
}
\foreach \x in {4}{
\draw (z\x\yy1)--(d\x\y1) (z\x\yy1)--(d\x\yyy1) (z\x\yy1)--(d\x\yy0) (z\x\yy1)--(d\x\yy2);
}}
\end{tikzpicture}\;.
\end{equation}
As shown later in Eq.~\eqref{eq:toric_code_marked_stabilizer}, this tensor network is also directly related to the toric code stabilizers by taking a periodic product of the $+1$ postselectsd stabilizer measurements.

2-index $\zz_2$-tensors or $\delta$-tensors are equal to the identity matrix, so at $2$-gons and $2$-valent edges we can just put a bond instead of a tensor, for example,
\begin{equation}
\label{eq:2gon_trivial}
\begin{tikzpicture}
\fill[orange,opacity=0.35] (0,0)to[bend left=45](1.5,0)--++(135:0.7)to[bend right=45]++(-1.5,0)--cycle;
\fill[orange,opacity=0.35] (0,0)to[bend right=45](1.5,0)--++(-135:0.7)to[bend left=45]++(-1.5,0)--cycle;
\fill[orange,opacity=0.35] (0,0)to[bend left=45](1.5,0)to[bend left=45](0,0);
\draw[dashed,orange,actualedge] (0,0)to[bend left=45](1.5,0);
\draw[dashed,orange,actualedge] (0,0)to[bend right=45](1.5,0);
\atoms{delta}{0/p={0.75,0.3}, 1/p={0.75,-0.3}}
\atoms{z2}{z/p={0.75,0}}
\draw (0)--(z) (z)--(1) (1)--++(-135:0.5) (1)--++(-45:0.5);
\fill[orange,opacity=0.35] (0,0)to[bend left=45](1.5,0)--++(45:0.6)to[bend right=45]++(-1.5,0)--cycle;
\fill[orange,opacity=0.35] (0,0)to[bend right=45](1.5,0)--++(-45:0.6)to[bend left=45]++(-1.5,0)--cycle;
\draw (0)--++(135:0.5)  (0)--++(45:0.5);
\end{tikzpicture}
=
\begin{tikzpicture}
\fill[orange,opacity=0.35] (0,0)to[bend left](1.5,0)--++(135:0.6)to[bend right]++(-1.5,0)--cycle;
\fill[orange,opacity=0.35] (0,0)to[bend right](1.5,0)--++(-135:0.6)to[bend left]++(-1.5,0)--cycle;
\fill[orange,opacity=0.35] (0,0)to[bend left](1.5,0)to[bend left](0,0);
\draw[dashed,orange,actualedge] (0,0)to[bend left](1.5,0);
\draw[dashed,orange,actualedge] (0,0)to[bend right](1.5,0);
\atoms{delta}{0/p={0.75,0.2}, 1/p={0.75,-0.2}}
\draw (0)--(1) (1)--++(-135:0.5) (1)--++(-45:0.5);
\fill[orange,opacity=0.35] (0,0)to[bend left](1.5,0)--++(45:0.5)to[bend right]++(-1.5,0)--cycle;
\fill[orange,opacity=0.35] (0,0)to[bend right](1.5,0)--++(-45:0.5)to[bend left]++(-1.5,0)--cycle;
\draw (0)--++(135:0.5)  (0)--++(45:0.5);
\end{tikzpicture}\;.
\end{equation}
Note here that the cellulations we consider are combinatorial and not geometrical, so it is possible that we have to bend some edges or faces in order to embed the lattice into Euclidean space.
Let us now define combinatorial moves that impose the topological invariance of the path integral.
While we could again use Pachner moves, there is a set of moves that is more elegant for the present path integral.
First, we impose equality of different ways of splitting faces into triangles along 2-valent edges,
\begin{equation}
\label{eq:22pachner}
\begin{tikzpicture}
\atoms{void}{0/, 1/p={0.5,-0.5}, 2/p={1,0}, 3/p={0.5,0.5}}
\fill[orange,opacity=0.35] (0)--(3)--(2)--(1)--cycle;
\draw[dashed,orange] (0)--(3) (3)--(2) (2)--(1) (0)--(1) (0)--(2);
\atoms{z2}{a/p={0.5,-0.2}, b/p={0.5,0.2}}
\draw (a)--(b) (b)--++(45:0.4) (b)--++(135:0.4) (a)--++(-45:0.4) (a)--++(-135:0.4);
\end{tikzpicture}
=
\begin{tikzpicture}
\atoms{void}{0/, 1/p={0.5,-0.5}, 2/p={1,0}, 3/p={0.5,0.5}}
\fill[orange,opacity=0.35] (0)--(3)--(2)--(1)--cycle;
\draw[dashed,orange] (0)--(3) (3)--(2) (2)--(1) (0)--(1) (3)--(1);
\atoms{z2}{a/p={0.3,0}, b/p={0.7,0}}
\draw (a)--(b) (b)--++(45:0.4) (a)--++(135:0.4) (b)--++(-45:0.4) (a)--++(-135:0.4);
\end{tikzpicture}\;.
\end{equation}
Poincar\'e dually, we also equate different ways of splitting $n$-valent edges into $3$-valent ones separated by $2$-gon faces such as the splitting shown in Eq.~\eqref{eq:2gon_trivial}.
Finally, we add a move involving both face and edge tensors,
\begin{equation}
\label{eq:bialgebra_move}
\begin{tikzpicture}
\atoms{void}{2/p={-30:1}, 1/p={90:1}, 0/p={-150:1}}
\fill[orange,opacity=0.35] (0)to[bend right=20](2)--(1)--cycle;
\fill[orange,opacity=0.35] (1)--(2)--++(30:0.4)--($(1)+(30:0.4)$)--cycle;
\fill[orange,opacity=0.35] (0)--(1)--++(150:0.4)--($(0)+(150:0.4)$)--cycle;
\atoms{z2}{a/p={0.1,-0.1}}
\draw[dashed,orange,actualedge] (0)--(1) (1)--(2);
\atoms{delta}{c/p={30:0.5}, d/p={150:0.5}}
\draw[dashed,orange] (0)to[bend left=20](2) (0)to[bend right=20](2);
\draw (a)to[bend left=20](c) (a)to[bend right=20](d) (a)--++(-90:0.7);
\fill[orange,opacity=0.35] (0)to[bend left=20](2)--(1)--cycle;
\atoms{z2}{b/p={-0.2,-0.1}}
\draw (b)to[bend left=20](c) (b)to[bend right=20](d) (b)--++(-90:0.4);
\draw (c)--++(30:0.4) (d)--++(150:0.4);
\end{tikzpicture}
=
\begin{tikzpicture}
\atoms{void}{2/p={-30:1}, 1/p={90:1}, 0/p={-150:1}}
\fill[orange,opacity=0.35] (0)--(2)--(1)--cycle;
\fill[orange,opacity=0.35] (0)--(2)--++(-150:0.5)--($(0)+(-150:0.6)$)--cycle;
\draw[dashed,orange] (0)--(1) (1)--(2);
\draw[dashed,orange,actualedge] (0)--(2);
\atoms{delta}{b/p={-90:0.5}}
\draw (b)--++(-150:0.7);
\fill[orange,opacity=0.35] (0)--(2)--++(-30:0.4)--($(0)+(-30:0.4)$)--cycle;
\atoms{z2}{a/}
\draw (a)--(b) (a)--++(30:0.6) (a)--++(150:0.6) (b)--++(-30:0.6);
\end{tikzpicture}\;.
\end{equation}
On the left-hand side, there are two triangles with two shared $3$-valent edges.
On the right-hand side, there is one triangle with one adjacent $3$-valent edge.
Like for Pachner moves, applying the moves above allows us to arbitrarily change the cellulation while leaving the topology invariant.
Note that this type of state sum (for general choices of face and edge tensors) can describe not only the toric code but any non-chiral topological phase, and is closely related to weak Hopf algebras \cite{liquid_intro}.
Also note that for the toric code, the equations above are a subset of the $ZX$ calculus \cite{Coecke2017,Wetering2020}.

We could also extend the definition of the path integral to manifolds with boundary, for example by adding two additional tensors associated to boundary edges and faces and imposing a boundary version of topological invariance.
More generally, we could introduce any sort of defects, which are lower-dimensional manifolds along which the path integral is altered, including domain walls, twist defects, corners between boundaries, anyons, and so on.
In order to turn path integrals into fault tolerant circuits, we will use a subset of all the defects that we will refer to as \emph{syndrome defects}.
For most examples in this paper, the syndrome defects will be the anyons, which are implemented by altering the path integral on 1-dimensional worldlines inside a 3-dimensional spacetime.
For the toric code, there are two generating types of anyon worldlines, namely $e$ and $m$ anyons.
$e$ anyon worldlines are closed paths of edges in the lattice.
We can introduce an $e$ anyon by replacing all $\delta$-tensors on the worldline by a \emph{charged $\delta$-tensor},
\begin{equation}
\label{eq:e_anyon_tensor}
\begin{tikzpicture}
\atoms{delta,bdastyle=red}{0/}
\draw (0)edge[ind=$a$]++(0:0.5) (0)edge[ind=$b$]++(90:0.5) (0)edge[ind=$c$]++(180:0.5);
\node at (-90:0.5){$\ldots$};
\end{tikzpicture}
=
\begin{cases}
1 & \text{if } a=b=c=\ldots=0\\
-1 & \text{if } a=b=c=\ldots=1\\
0 & \text{otherwise}
\end{cases}\;.
\end{equation}
$m$ anyon worldlines are closed paths of edges in the Poincar\'e dual lattice.
For an $m$ anyon, we replace all $\zz_2$-tensors on the worldline by a \emph{charged $\zz_2$-tensor},
\begin{equation}
\label{eq:m_anyon_tensor}
\begin{tikzpicture}
\atoms{z2,bdastyle=red}{0/}
\draw (0)edge[ind=$a$]++(0:0.5) (0)edge[ind=$b$]++(90:0.5) (0)edge[ind=$c$]++(180:0.5);
\node at (-90:0.5){$\ldots$};
\end{tikzpicture}
=
\begin{cases}
0 & \text{if } a+b+c+\ldots=0\mod 2\\
1 & \text{otherwise}
\end{cases}\;.
\end{equation}
An example for a configuration of $e$ and $m$ worldlines inside the cubic lattice (with worldlines marked in semi-transparent red) is:
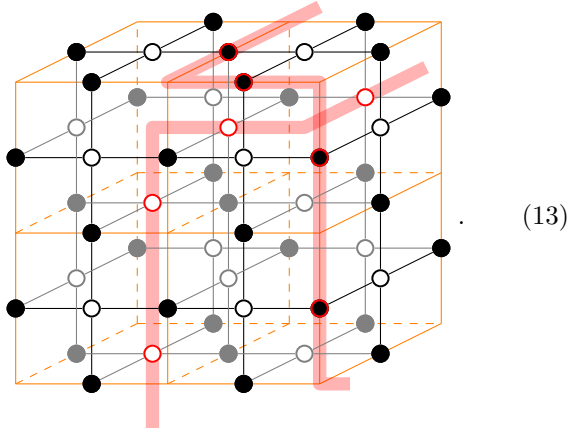
\begin{equation}
\label{eq:anyon_configuration}
\begin{tikzpicture}
\atoms{void}{x/p={2,0}, y/p={0,2}, z/p={1.6,0.8}}
\foreach \x in {0,1,2}{
\foreach \y in {0,1,2}{
\foreach \z in {0,1}{
\atoms{void}{\x\y\z/p={$\x*(x)+\y*(y)+\z*(z)$}}
}}}
\foreach \x in {-1,1,3,5}{
\foreach \y in {-1,1,3,5}{
\foreach \z in {-1,1,3}{
\atoms{void}{m\x\y\z/p={$0.5*\x*(x)+0.5*\y*(y)+0.5*\z*(z)$}}
}}}
\draw[orange] (000)--(200) (010)--(210) (020)--(220) (000)--(020) (100)--(120) (200)--(220) (201)--(221) (021)--(221);
\draw[orange,dashed] (001)--(201) (011)--(211) (001)--(021) (101)--(121);
\draw[orange, dashed] (000)--(001) (100)--(101) (010)--(011) (110)--(111);
\draw[orange] (200)--(201) (210)--(211) (220)--(221) (020)--(021) (120)--(121);
\draw[worldline] (m1-11)--(m111)--(m131)--(m331)--(m333);
\draw[worldline] ($(200)+(0:0.4)$)--(200)--(210)--(220)--(120)--(121)--++(0.4,0.2);
\foreach \x/\y/\z in {1/0/0, 3/0/0, 1/2/0, 3/2/0, 1/4/0, 3/4/0, 0/1/0, 0/3/0, 2/1/0, 2/3/0, 4/1/0, 4/3/0, 4/0/1, 4/2/1, 0/4/1, 2/4/1, 4/4/1, 4/1/2, 4/3/2, 1/4/2, 3/4/2}{
\atoms{delta}{d\x\y\z/p={$0.5*\x*(x)+0.5*\y*(y)+0.5*\z*(z)$}}
}
\foreach \x/\y/\z in {1/0/2, 3/0/2, 1/2/2, 3/2/2, 0/1/2, 0/3/2, 2/1/2, 2/3/2, 2/2/1, 0/2/1, 0/0/1, 2/0/1}{
\atoms{delta,astyle=gray}{d\x\y\z/p={$0.5*\x*(x)+0.5*\y*(y)+0.5*\z*(z)$}}
}
\foreach \x/\y/\z in {1/1/0, 3/1/0, 1/3/0, 3/3/0, 1/4/1, 3/4/1, 4/1/1, 4/3/1}{
\atoms{z2}{z\x\y\z/p={$0.5*\x*(x)+0.5*\y*(y)+0.5*\z*(z)$}}
}
\foreach \x/\y/\z in {1/1/2, 3/1/2, 1/3/2, 3/3/2, 1/2/1, 3/2/1, 2/1/1, 2/3/1, 0/1/1, 0/3/1, 1/0/1, 3/0/1}{
\atoms{z2,astyle=gray}{z\x\y\z/p={$0.5*\x*(x)+0.5*\y*(y)+0.5*\z*(z)$}}
}
\foreach \x/\xx/\xxx in {0/1/2,2/3/4}{
\foreach \y/\yy/\yyy in {0/1/2,2/3/4}{
\foreach \z in {0}{
\draw (z\xx\yy\z)--(d\x\yy\z) (z\xx\yy\z)--(d\xxx\yy\z) (z\xx\yy\z)--(d\xx\y\z) (z\xx\yy\z)--(d\xx\yyy\z);
}
\foreach \z in {2}{
\draw[gray] (z\xx\yy\z)--(d\x\yy\z) (z\xx\yy\z)--(d\xxx\yy\z) (z\xx\yy\z)--(d\xx\y\z) (z\xx\yy\z)--(d\xx\yyy\z);
}}}
\foreach \x/\xx/\xxx in {0/1/2,2/3/4}{
\foreach \y in {0,2}{
\draw[gray] (z\xx\y1)--(d\x\y1) (z\xx\y1)--(d\xxx\y1) (z\xx\y1)--(d\xx\y0) (z\xx\y1)--(d\xx\y2);
}
\foreach \y in {4}{
\draw (z\xx\y1)--(d\x\y1) (z\xx\y1)--(d\xxx\y1) (z\xx\y1)--(d\xx\y0) (z\xx\y1)--(d\xx\y2);
}}
\foreach \y/\yy/\yyy in {0/1/2,2/3/4}{
\foreach \x in {0,2}{
\draw[gray] (z\x\yy1)--(d\x\y1) (z\x\yy1)--(d\x\yyy1) (z\x\yy1)--(d\x\yy0) (z\x\yy1)--(d\x\yy2);
}
\foreach \x in {4}{
\draw (z\x\yy1)--(d\x\y1) (z\x\yy1)--(d\x\yyy1) (z\x\yy1)--(d\x\yy0) (z\x\yy1)--(d\x\yy2);
}}
\foreach \x/\y/\z in {1/0/1, 1/2/1, 2/3/1, 3/3/2}{
\atoms{z2,bdastyle=red}{d\x\y\z/p={$0.5*\x*(x)+0.5*\y*(y)+0.5*\z*(z)$}}
}
\foreach \x/\y/\z in {4/1/0, 4/3/0, 3/4/0, 2/4/1}{
\atoms{delta,bdastyle=red}{z\x\y\z/p={$0.5*\x*(x)+0.5*\y*(y)+0.5*\z*(z)$}}
}
\end{tikzpicture}\;.
\end{equation}
More generally, $e$ anyons can be supported on any \emph{1-cycle}.
A 1-cycle is a map that associates to every edge an element of $\zz_2=\{0,1\}$ depending on whether the edge carries an anyon or not, fulfilling a Gauss law:
For every vertex, the sum of $\zz_2$-values on the incident edges must be $0$ (mod 2).
In other words, the configuration of anyon worldlines must satisfy the anyon fusion rules.
Dually, $m$ anyons live on any \emph{2-cocycle}, a map that associates a $\zz_2$-element to every face such that the faces adjacent to every volume sum to $0$.
In principle, one can define $e$ ($m$) anyon worldline patterns on arbitrary \emph{1-chains} (\emph{2-cochains}), that is arbitrary maps from the edges (faces) to $\zz_2$.
However, if the cycle (cocycle) condition is violated, the path integral evaluates to $0$, for example:
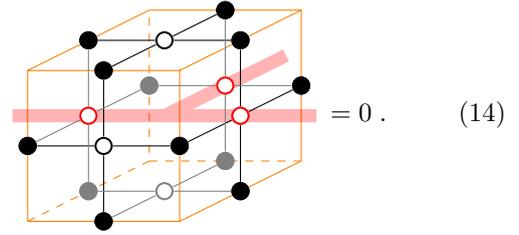
\begin{equation}
\label{eq:cocycle_constraint}
\begin{tikzpicture}
\atoms{void}{x/p={2,0}, y/p={0,2}, z/p={1.6,0.8}}
\foreach \x in {0,1}{
\foreach \y in {0,1}{
\foreach \z in {0,1}{
\atoms{void}{\x\y\z/p={$\x*(x)+\y*(y)+\z*(z)$}}
}}}
\foreach \x in {-1,1,3}{
\foreach \y in {-1,1,3}{
\foreach \z in {-1,1,3}{
\atoms{void}{m\x\y\z/p={$0.5*\x*(x)+0.5*\y*(y)+0.5*\z*(z)$}}
}}}
\draw[orange] (000)--(100)--(110)--(010)--(000) (100)--(101)--(111)--(110) (010)--(011)--(111);
\draw[orange,dashed] (000)--(001)--(011) (001)--(101);
\draw[worldline] (m-111)--(m111) (m311)--(m111) (m111)--(m113);
\foreach \x/\y/\z in {1/0/0, 1/2/0, 0/1/0, 2/1/0, 2/0/1, 0/2/1, 2/2/1, 1/2/2, 2/1/2}{
\atoms{delta}{d\x\y\z/p={$0.5*\x*(x)+0.5*\y*(y)+0.5*\z*(z)$}}
}
\foreach \x/\y/\z in {1/0/2, 0/1/2, 0/0/1}{
\atoms{delta,astyle=gray}{d\x\y\z/p={$0.5*\x*(x)+0.5*\y*(y)+0.5*\z*(z)$}}
}
\foreach \x/\y/\z in {1/1/0, 1/2/1, 2/1/1}{
\atoms{z2}{z\x\y\z/p={$0.5*\x*(x)+0.5*\y*(y)+0.5*\z*(z)$}}
}
\foreach \x/\y/\z in {1/1/2, 1/0/1, 0/1/1}{
\atoms{z2,astyle=gray}{z\x\y\z/p={$0.5*\x*(x)+0.5*\y*(y)+0.5*\z*(z)$}}
}
\draw (z110)--(d100) (z110)--(d210) (z110)--(d120) (z110)--(d010) (z211)--(d201) (z211)--(d210) (z211)--(d212) (z211)--(d221) (z121)--(d221) (z121)--(d120) (z121)--(d021) (z121)--(d122);
\draw[gray] (z112)--(d102) (z112)--(d212) (z112)--(d122) (z112)--(d012) (z011)--(d001) (z011)--(d010) (z011)--(d012) (z011)--(d021) (z101)--(d201) (z101)--(d100) (z101)--(d001) (z101)--(d102);
\foreach \x/\y/\z in {0/1/1, 2/1/1, 1/1/2}{
\atoms{z2,bdastyle=red}{d\x\y\z/p={$0.5*\x*(x)+0.5*\y*(y)+0.5*\z*(z)$}}
}
\end{tikzpicture}
=
0\;.
\end{equation}
To see this, we note that the path integral with a 2-cochain $b$ of $m$-anyons corresponds to a sum over all 1-cochains $a$ with $da=b$, where $da$ denotes the \emph{coboundary} of $a$ associating to each face the sum $a$ on the surrounding edges.
Since coboundaries are always cocycles, $b$ must be a 2-cocycle, otherwise we sum over the empty set of configurations and the path integral evaluates to $0$.
Dually, the path integral with a 1-chain $c$ of $e$-anyons corresponds to a sum over all 1-cocycles $a$, with a weight $(-1)^{ac}$ at every edge.
If $c$ violates the cycle condition at a vertex $v$, then adding the cocycle consisting of all edges adjacent to $v$ to $a$ will change the weight by a factor of $-1$.
Thus, the weights for all configurations cancel and the path integral again evaluates to $0$.
There are additional tensor-network equations for the tensors of the bulk together with their charged versions, such as
\begin{equation}
\label{eq:anyon_invariance}
\begin{gathered}
\begin{tikzpicture}
\atoms{z2,bdastyle=red}{0/}
\draw (0)--++(0:0.5) (0)--++(90:0.5) (0)--++(180:0.5) (0)--++(-90:0.5);
\end{tikzpicture}
=
\begin{tikzpicture}
\atoms{z2}{0/}
\atoms{z2,bdastyle=red}{1/p=180:0.4}
\draw (0)--++(0:0.5) (0)--++(90:0.5) (0)--(1) (1)--++(180:0.4) (0)--++(-90:0.5);
\end{tikzpicture}\;,
\qquad
\begin{tikzpicture}
\atoms{delta}{0/}
\atoms{z2,bdastyle=red}{1/p=180:0.4}
\draw (0)--++(0:0.5) (0)--++(90:0.5) (0)--(1) (1)--++(180:0.4) (0)--++(-90:0.5);
\end{tikzpicture}
=
\begin{tikzpicture}
\atoms{delta}{0/}
\atoms{z2,bdastyle=red}{1/p=90:0.4, 2/p=-90:0.4, 3/p=0:0.4}
\draw (0)--(3) (3)--++(0:0.4) (0)--(1) (1)--++(90:0.4) (0)--++(180:0.5) (0)--(2) (2)--++(-90:0.4);
\end{tikzpicture}\;,\\
\begin{tikzpicture}
\atoms{z2,bdastyle=red}{0/, 1/p={0.5,0}}
\draw (0)--(1) (1)--++(0:0.5) (0)--++(180:0.5);
\end{tikzpicture}
=
\begin{tikzpicture}
\draw (0,0)--++(0:0.8);
\end{tikzpicture}\;,\qquad
\begin{tikzpicture}
\atoms{delta,bdastyle=red}{0/}
\atoms{z2,bdastyle=red}{1/p=180:0.4}
\draw (0)--++(0:0.5) (0)--++(90:0.5) (0)--(1) (1)--++(180:0.4) (0)--++(-90:0.5);
\end{tikzpicture}
=
(-1)\cdot
\begin{tikzpicture}
\atoms{delta,bdastyle=red}{0/}
\atoms{z2,bdastyle=red}{1/p=90:0.4, 2/p=-90:0.4, 3/p=0:0.4}
\draw (0)--(3) (3)--++(0:0.4) (0)--(1) (1)--++(90:0.4) (0)--++(180:0.5) (0)--(2) (2)--++(-90:0.4);
\end{tikzpicture}\;,
\end{gathered}
\end{equation}
and the same equations with full and empty circles exchanged.
Note that the equations in Eq.~\eqref{eq:anyon_invariance} are also part of the $ZX$ calculus, where the charged tensors in Eqs.~\eqref{eq:e_anyon_tensor} and \eqref{eq:m_anyon_tensor} are the $Z$ and $X$-type tensor labeled by a phase of $\pi$.
These equations can be used to freely move the worldlines through the bulk, at least up to a $\pm1$ prefactor, for example,
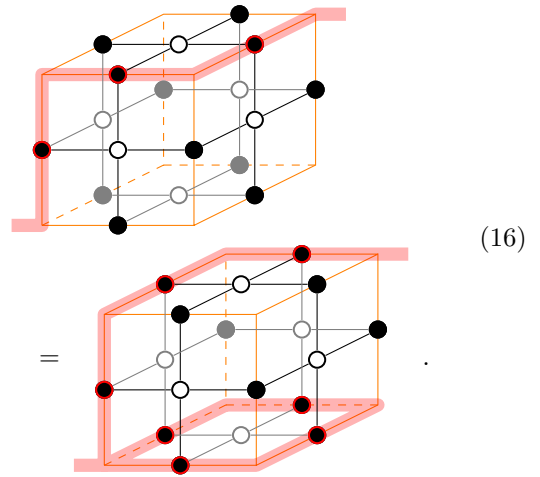
\begin{equation}
\begin{multlined}
\begin{tikzpicture}
\atoms{void}{x/p={2,0}, y/p={0,2}, z/p={1.6,0.8}}
\foreach \x in {0,1}{
\foreach \y in {0,1}{
\foreach \z in {0,1}{
\atoms{void}{\x\y\z/p={$\x*(x)+\y*(y)+\z*(z)$}}
}}}
\foreach \x in {-1,1,3}{
\foreach \y in {-1,1,3}{
\foreach \z in {-1,1,3}{
\atoms{void}{m\x\y\z/p={$0.5*\x*(x)+0.5*\y*(y)+0.5*\z*(z)$}}
}}}
\draw[orange] (000)--(100)--(110)--(010)--(000) (100)--(101)--(111)--(110) (010)--(011)--(111);
\draw[orange,dashed] (000)--(001)--(011) (001)--(101);
\draw[worldline] ($(000)+(180:0.4)$)--(000)--(010)--(110)--(111)--++(0:0.4);
\foreach \x/\y/\z in {1/0/0, 1/2/0, 0/1/0, 2/1/0, 2/0/1, 0/2/1, 2/2/1, 1/2/2, 2/1/2}{
\atoms{delta}{d\x\y\z/p={$0.5*\x*(x)+0.5*\y*(y)+0.5*\z*(z)$}}
}
\foreach \x/\y/\z in {1/0/2, 0/1/2, 0/0/1}{
\atoms{delta,astyle=gray}{d\x\y\z/p={$0.5*\x*(x)+0.5*\y*(y)+0.5*\z*(z)$}}
}
\foreach \x/\y/\z in {1/1/0, 1/2/1, 2/1/1}{
\atoms{z2}{z\x\y\z/p={$0.5*\x*(x)+0.5*\y*(y)+0.5*\z*(z)$}}
}
\foreach \x/\y/\z in {1/1/2, 1/0/1, 0/1/1}{
\atoms{z2,astyle=gray}{z\x\y\z/p={$0.5*\x*(x)+0.5*\y*(y)+0.5*\z*(z)$}}
}
\draw (z110)--(d100) (z110)--(d210) (z110)--(d120) (z110)--(d010) (z211)--(d201) (z211)--(d210) (z211)--(d212) (z211)--(d221) (z121)--(d221) (z121)--(d120) (z121)--(d021) (z121)--(d122);
\draw[gray] (z112)--(d102) (z112)--(d212) (z112)--(d122) (z112)--(d012) (z011)--(d001) (z011)--(d010) (z011)--(d012) (z011)--(d021) (z101)--(d201) (z101)--(d100) (z101)--(d001) (z101)--(d102);
\foreach \x/\y/\z in {0/1/0, 1/2/0, 2/2/1}{
\atoms{delta,bdastyle=red}{d\x\y\z/p={$0.5*\x*(x)+0.5*\y*(y)+0.5*\z*(z)$}}
}
\end{tikzpicture}\\
=
\begin{tikzpicture}
\atoms{void}{x/p={2,0}, y/p={0,2}, z/p={1.6,0.8}}
\foreach \x in {0,1}{
\foreach \y in {0,1}{
\foreach \z in {0,1}{
\atoms{void}{\x\y\z/p={$\x*(x)+\y*(y)+\z*(z)$}}
}}}
\foreach \x in {-1,1,3}{
\foreach \y in {-1,1,3}{
\foreach \z in {-1,1,3}{
\atoms{void}{m\x\y\z/p={$0.5*\x*(x)+0.5*\y*(y)+0.5*\z*(z)$}}
}}}
\draw[orange] (000)--(100)--(110)--(010)--(000) (100)--(101)--(111)--(110) (010)--(011)--(111);
\draw[orange,dashed] (000)--(001)--(011) (001)--(101);
\draw[worldline] ($(000)+(180:0.4)$)--(000)--(010)--(011)--(111)--++(0:0.4) (000)--(100)--(101)--(001)--cycle;
\foreach \x/\y/\z in {1/0/0, 1/2/0, 0/1/0, 2/1/0, 2/0/1, 0/2/1, 2/2/1, 1/2/2, 2/1/2}{
\atoms{delta}{d\x\y\z/p={$0.5*\x*(x)+0.5*\y*(y)+0.5*\z*(z)$}}
}
\foreach \x/\y/\z in {1/0/2, 0/1/2, 0/0/1}{
\atoms{delta,astyle=gray}{d\x\y\z/p={$0.5*\x*(x)+0.5*\y*(y)+0.5*\z*(z)$}}
}
\foreach \x/\y/\z in {1/1/0, 1/2/1, 2/1/1}{
\atoms{z2}{z\x\y\z/p={$0.5*\x*(x)+0.5*\y*(y)+0.5*\z*(z)$}}
}
\foreach \x/\y/\z in {1/1/2, 1/0/1, 0/1/1}{
\atoms{z2,astyle=gray}{z\x\y\z/p={$0.5*\x*(x)+0.5*\y*(y)+0.5*\z*(z)$}}
}
\draw (z110)--(d100) (z110)--(d210) (z110)--(d120) (z110)--(d010) (z211)--(d201) (z211)--(d210) (z211)--(d212) (z211)--(d221) (z121)--(d221) (z121)--(d120) (z121)--(d021) (z121)--(d122);
\draw[gray] (z112)--(d102) (z112)--(d212) (z112)--(d122) (z112)--(d012) (z011)--(d001) (z011)--(d010) (z011)--(d012) (z011)--(d021) (z101)--(d201) (z101)--(d100) (z101)--(d001) (z101)--(d102);
\foreach \x/\y/\z in {0/1/0, 0/2/1, 1/2/2, 1/0/0, 2/0/1, 1/0/2, 0/0/1}{
\atoms{delta,bdastyle=red}{d\x\y\z/p={$0.5*\x*(x)+0.5*\y*(y)+0.5*\z*(z)$}}
}
\end{tikzpicture}\;.
\end{multlined}
\end{equation}
In other words, these equations impose the topological invariance of the anyon worldline, in addition to that of the bulk itself.
More precisely, Eq.~\eqref{eq:anyon_invariance} allows us to change the $e$ 1-cycle by adding $1$ (mod 2) to the edges in the boundary of any face.
That is, we can arbitrarily deform the pattern of anyon worldlines as long as the \emph{homology class} does not change.
The same holds for the $m$ anyon worldlines on the dual lattice.

In general, one could use different types of defects as the syndrome defects, whose configurations obey other local constraints than the cocycle condition, and have other equivalences than homology.
Examples for this are topological phases with non-abelian anyons that are only topologically (not homologically) invariant, or fracton phases where syndrome defects are to some extent restricted to rigid submanifolds.
Also, homological syndrome defects can have a higher dimension, for example they could be located on \emph{2-cycles}, which are closed-membrane configurations as used in Section~\ref{sec:3d_toric_code}.
All other examples in this paper just use closed-loop homological syndrome defects such as the $e$ and $m$ anyons in the toric code.

There is a natural equivalence relation for path integrals.
Namely, two tensor-network path integrals $X$ and $Y$ are equivalent if they are related by applying local tensor-network equations.
Applying such an equation means to remove the left-hand side from the tensor network and insert the right-hand side, or vice versa.
More precisely, we apply such equations in parallel everywhere in the tensor network, for a constant number of rounds.
Alternatively, applying the tensor-network equations only inside some region $A$ yields a domain wall between $Y$ on $A$ and $X$ on the complement $\overline A$.
By applying the tensor-network equations we can make $A$ larger or smaller, which freely moves around the domain wall, making it a topological domain wall.
Since we can also remove, fuse, or generate new $A$ islands, the domain wall is also \emph{invertible}.
Equivalence classes under invertible domain walls will be called \emph{fixed-point phases}.
Fixed-point phases are the natural notion of phases of matter for fixed-point path integrals, analogous to how local unitary circuits can be used to define phases in fixed-point Hamiltonians.
Exact tensor-network equations still provide an interesting equivalence relation for general (non-fixed-point) path integrals.
However, for them to capture phases of matter in this general context, some notion of approximation will be necessary.
This is because applying exact tensor-network equations cannot change the correlation length of a path integral.
In this light, consider tensor-network equations imposing topological invariance such as Eq.~\eqref{eq:pachner_move_equation} or Eq.~\eqref{eq:bialgebra_move}:
They imply that the path integral on one lattice is in the same fixed-point phase as that same path integral on another lattice, for any way of superimposing the two lattices.
For more detail and examples we refer the reader to Ref.~\cite{liquid_intro}.

\subsection{Dynamic codes}
\label{sec:dynamic_codes}
In this paper we are thinking of QEC as a dynamic processes, or more technically, as a circuit executed in spacetime.
An error-correcting process needs to be able to filter out noise introduced into the system by extracting entropy.
Thus, the corresponding circuits are circuits of quantum channels rather than unitaries.
It is useful to consider channels that simultaneously act on classical and quantum degrees of freedom, even though these can always be embedded into purely quantum channels.
Mathematically, such a quantum/classical hybrid channel is a tensor where every classical or quantum degree of freedom, either at the input or the output of the channel, corresponds to one index.
More precisely, a qu-$d$-it is represented by a pair of $d$-dimensional indices, one for the ket and one for the bra part, whereas a classical $d$-it is just a $d$-dimensional index.
For example, a channel with one quantum input and one classical input, and two quantum outputs and one classical output can look like:
\begin{equation}
\begin{tikzpicture}
\atoms{square,xscale=4}{0/}
\draw ([sx=-0.5]0-b)--++(-90:0.5) ([sx=-0.3]0-b)--++(-90:0.5) ([sx=-0.6]0-t)--++(90:0.5) ([sx=-0.4]0-t)--++(90:0.5) ([sx=0.1]0-t)--++(90:0.5) ([sx=-0.1]0-t)--++(90:0.5);
\draw[classical] ([sx=0.4]0-b)--++(-90:0.5) ([sx=0.4]0-t)--++(90:0.5);
\end{tikzpicture}\;,
\end{equation}
where the time direction is from bottom to top, like everywhere in this paper.

A proper channel needs to fulfill two conditions:
First, it needs to be \emph{completely positive}:
For every fixed value of the classical indices, block all ket indices and all bra indices such that the tensor becomes a matrix.
This matrix has to be non-negative, for example,
\begin{equation}
\begin{tikzpicture}
\atoms{square,xscale=4}{0/}
\draw ([sx=-0.5]0-b)edge[ind=$a$]++(-90:0.5) ([sx=-0.3]0-b)edge[ind=$d$]++(-90:0.5) ([sx=-0.6]0-t)edge[ind=$b$]++(90:0.5) ([sx=-0.4]0-t)edge[ind=$e$]++(90:0.5) ([sx=0.1]0-t)edge[ind=$f$]++(90:0.5) ([sx=-0.1]0-t)edge[ind=$c$]++(90:0.5);
\draw[classical] ([sx=0.4]0-b)edge[ind=$i$]++(-90:0.5) ([sx=0.4]0-t)edge[ind=$j$]++(90:0.5);
\end{tikzpicture}
\rightarrow [M^{ij}]_{(abc),(def)} \geq 0\quad\forall i,j\;.
\end{equation}
In this context, the matrix $M$ is also known as the \emph{Choi matrix}.
Second, it needs to be \emph{trace preserving}:
When closing all quantum output (double-)indices with a trace, and all classical output indices with a sum, we obtain a trace and sum at all classical and quantum input indices, for example,
\begin{equation}
\label{eq:channel_normalization}
\begin{tikzpicture}
\atoms{square,xscale=4}{0/}
\atoms{deltatens}{d/p={0.4,0.6}}
\draw ([sx=-0.5]0-b)--++(-90:0.5) ([sx=-0.3]0-b)--++(-90:0.5) ([sx=-0.6]0-t)--++(90:0.5)to[out=90,in=90]++(0:0.2)--([sx=-0.4]0-t) ([sx=0.1]0-t)--++(90:0.5)to[out=90,in=90]++(180:0.2)--([sx=-0.1]0-t);
\draw[classical] ([sx=0.4]0-b)--++(-90:0.5) ([sx=0.4]0-t)--(d);
\end{tikzpicture}
=
\begin{tikzpicture}
\atoms{deltatens}{d/}
\draw[classical] (d)--++(-90:0.5);
\draw (-0.5,-0.5)--++(90:0.5)to[out=90,in=90]++(180:0.2)--(-0.7,-0.5);
\end{tikzpicture}\;.
\end{equation}
Here the black dot is the $\delta$-tensor in Eq.~\eqref{eq:delta_definition} with one index, that is, a vector with all entries equal to $1$.
For more discussion on representing quantum-classical hybrid channels as tensors, see Ref.~\cite{cstar_qmech}.

In \emph{topological} QEC, we demand the circuit to be geometrically local.
Only then it is fair to assume that also the noise occurring in the process is local.
The great achievement of topological QEC is fault tolerance with respect to arbitrary local noise, and thereby any noise that is possible.
We will refer to this type of QEC, where the complete circuit of quantum/classical channels is geometrically local, as \emph{fully local QEC}.
Topological QEC has the additional property of being uniform in spacetime, or at least to scale in a uniform way.
Fully local topological QEC is not only of practical but also of fundamental physical interest since it might provide a model for the process of cooling a topologically ordered material.
Fully local QEC can also be viewed as self-correction using engineered dissipation, formulated in discrete time.
Examples for fully local QEC circuits are given by cellular automaton decoders \cite{Kubica2018}.
While fault tolerant fully local decoders are known exist in $4+1$ dimensions, the situation is unclear in $3+1$ and $2+1$ dimensions.

Since the feasibility of fully local topological QEC in low dimensions is an open question, we consider \emph{quantum-local QEC} as a second type of topological QEC, where only the quantum part of the circuit is assumed to be local.
Quantum-local QEC consists of a geometrically local circuit of channels with additional open classical inputs and outputs.
These inputs and outputs are then coupled to a purely classical \emph{decoder} that is not implemented by a classical circuit in the same spacetime, but treated as a black box that can be evaluated instantly and without noise.
In practice, the efficiency of this decoder is of course still of great importance, and any reasonable decoder should be executable in at most a polynomially larger spacetime.
An example for this is minimum-weight-perfect-matching decoding of the toric code:
The quantum parts, namely the stabilizer measurements and corrections, are local, while the classical decoding algorithm has more-than-constant runtime even if we allowed for instant non-local communication.
From a fundamental point of view, quantum-local QEC is not scalable with a fault tolerant threshold.
This is because for large enough system sizes the quantum circuit has to wait for the results of the decoder, and during this waiting time additional errors accumulate.
Nonetheless, quantum-local QEC might have a practical impact, since current implementations of qubits are by orders of magnitude larger, slower, and noisier, than classical information technology.
A toy example for quantum-local QEC in a $1+1$-dimensional spacetime looks like
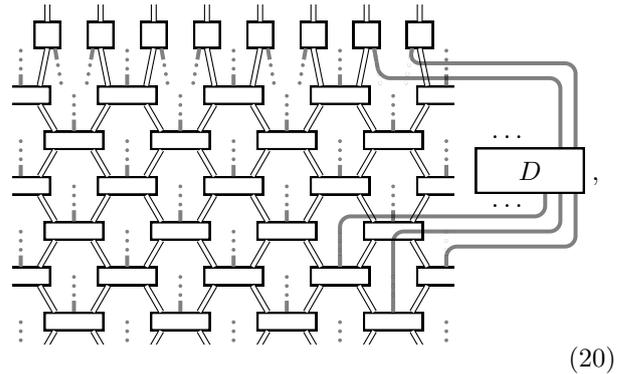
\begin{equation}
\label{eq:qec_circuit}
\begin{tikzpicture}
\begin{scope}
\clip (0.6,0.9) rectangle (6.4,5.4);
\foreach \y in {0,1,2,3}{
\foreach \x in {0,1,2,3,4,5,6}{
\atoms{square,yscale=0.7,xscale=2.2}{{a\x\y/p={\x*1.4,\y*1.2}}}
\draw[classical,black!50] (a\x\y-t)edge[mark={three dots,a}]++(90:0.15);
}};
\foreach \y in {0,1,2,3}{
\foreach \x in {0,1,2,3,4,5,6}{
\atoms{square,yscale=0.7,xscale=2.2}{{b\x\y/p={\x*1.4+0.7,\y*1.2+0.6}}}
\draw[classical,black!50] (b\x\y-t)edge[mark={three dots,a}]++(90:0.15);
}};
\foreach[count=\yy from 1] \y in {0,1,2,3}{
\foreach[count=\xx from 0] \x in {1,2,3,4,5,6}{
\draw[quantum] ([sx=0.25]a\x\y-t)--([sx=-0.25]b\x\y-b);
\draw[quantum] ([sx=-0.25]a\x\y-t)--([sx=0.25]b\xx\y-b);
}};
\foreach[count=\yy from 1] \y in {0,1,2}{
\foreach[count=\xx from 0] \x in {1,2,3,4,5,6}{
\draw[quantum] ([sx=-0.25]b\xx\y-t)--([sx=0.25]a\xx\yy-b);
\draw[quantum] ([sx=0.25]b\xx\y-t)--([sx=-0.25]a\x\yy-b);
}};
\foreach \x in {0,1,2,3,4,5}{
\atoms{square}{a\x/p={\x*1.4+0.35,5}, b\x/p={\x*1.4+1.05,5}}
\draw[classical,black!50] ([sx=-0.1]a\x-b)edge[mark={three dots,a}]++(-100:0.15) ([sx=0.1]b\x-b)edge[mark={three dots,a}]++(-80:0.15);
\draw[quantum] ([sx=-0.25]b\x3-t)--(a\x-b) ([sx=0.25]b\x3-t)--(b\x-b) (a\x-t)--++(90:0.25) (b\x-t)--++(90:0.25);
};
\end{scope}
\atoms{square,xscale=4,yscale=1.7,lab={t=$D$,p={0,0}}}{c/p={7.4,3.2}}
\draw[classical,white] (b31-t)edge[mark={three dots,a}]++(90:0.15) (a41-t)edge[mark={three dots,a}]++(90:0.15) (b41-t)edge[mark={three dots,a}]++(90:0.15) ([sx=-0.1]a4-b)edge[mark={three dots,a}]++(-100:0.15) ([sx=0.1]b3-b)edge[mark={three dots,a}]++(-80:0.15);
\draw[classical,opacity=0.5,rc] ([sx=0.2]c-b)--++(-90:0.3)-|(b31-t) ([sx=0.4]c-b)--++(-90:0.5)-|(a41-t) ([sx=0.6]c-b)--++(-90:0.7)-|(b41-t) ([sx=-0.1]a4-b)--++(-100:0.2)-|([sx=0.6]c-t) ([sx=0.1]b3-b)--++(-80:0.4)-|([sx=0.4]c-t);
\path (c-b)--++(-0.3,-0.15)node{$\ldots$};
\path (c-t)--++(-0.3,0.15)node{$\ldots$};
\end{tikzpicture}\;,
\end{equation}
where we have semi-transparently drawn some of the classical bonds connecting the circuit and the decoder $D$, and omitted the remaining ones.
Note that for a real topological error-correcting circuit we would need at least $2+1$ spacetime dimensions.
The shown example has a special layout where we first apply only hybrid channels without classical inputs for a time $T\sim L$.
Physically, such hybrid channels are known as \emph{instruments} in quantum information theory and describe measurements, which are 2-qu-$d$-it measurements in the example above.
We will refer to the recorded classical outputs/measurement results as \emph{spacetime syndrome}.
Then, at time $T$, we perform a constant-depth \emph{correction} layer of quantum channels with an additional classical input, which are single-qu-$d$-it operators in the example above.
The inputs to these correction channels are obtained from applying the decoding algorithm $D$ to the spacetime syndrome.

Note that in general, corrections could also be applied in every time step like measurements, and not only after a time $T\sim L$.
This might be necessary for example for topological error correction based on non-abelian phases.
However, for all examples considered in this paper, a layout as in Eq.~\eqref{eq:qec_circuit} works.

\subsection{Imaginary versus real time}
In topological quantum computation, we store information in the ground space of topologically ordered models defined on spatial configurations of non-trivial topology.
In order to perform logical operations we change the topology, either by adiabatic variation of the model parameters or by code deformation.
It is in principle possible to perform logical operations by only changing the topology of some bare spatial manifold, for example by physically performing a Dehn twist on a torus and using Pachner moves to undo the associated distortion of the lattice.
However, the set of accessible logic gates becomes much richer if we introduce defects, such as boundaries, anyons, domain walls, twist defects, and so on.
We refer to such defects as \emph{computational defect} to stress that they serve a very different purpose from the syndrome defects introduced in Section~\ref{sec:fixed_points}.
Computational defects are also necessary for implementing computation in practice where we need to faithfully embed the topological manifolds into the Euclidean space we happen to live in.
Consider the following two examples of processes involving defects, with time flowing from bottom to top,
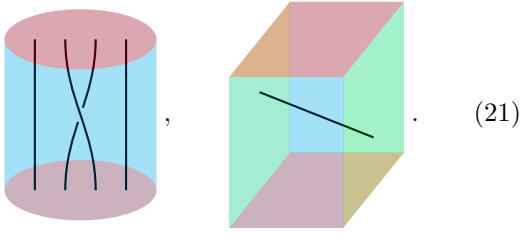
\begin{equation}
\label{eq:spacetime_processes}
\begin{tikzpicture}
\fill[cyan,opacity=0.2] (0,0)arc (180:0:1cm and 0.4cm)--++(0,2) arc(0:180:1cm and 0.4cm)--cycle;
\fill[red,opacity=0.3] (0,0)arc (-180:180:1cm and 0.4cm);
\draw[thick] (0.4,0)--(0.4,2) (0.8,0)to[out=90,in=-90](1.2,2) (1.6,0)--(1.6,2);
\fill[cyan!20] (1,1)circle(0.1);
\draw[thick] (1.2,0)to[out=90,in=-90](0.8,2);
\fill[red,opacity=0.3] (0,2)arc (-180:180:1cm and 0.4cm);
\fill[cyan,opacity=0.2] (0,0)arc (-180:0:1cm and 0.4cm)--++(0,2) arc(0:-180:1cm and 0.4cm)--cycle;
\end{tikzpicture}
\;,
\qquad
\begin{tikzpicture}
\fill[cyan,opacity=0.2] (0.8,1)rectangle++(1.5,2);
\fill[green,opacity=0.2] (0,0)--++(0.8,1)--++(0,2)--++(-0.8,-1)--cycle;
\fill[red,opacity=0.3] (0,0)--++(1.5,0)--++(0.8,1)--++(-1.5,0)--cycle;
\draw[thick] (1.9,1.2)--(0.4,1.8);
\fill[red,opacity=0.3] (0,2)--++(1.5,0)--++(0.8,1)--++(-1.5,0)--cycle;
\fill[green,opacity=0.2] (1.5,0)--++(0.8,1)--++(0,2)--++(-0.8,-1)--cycle;
\fill[cyan,opacity=0.2] (0,0)rectangle++(1.5,2);
\end{tikzpicture}\;.
\end{equation}
The left side shows a process where two of four anyons on a disk are exchanged.
\footnote{Note that for this to define a non-trivial logical operation, the anyons have to be non-Abelian, or we have to replace anyon worldlines by tube-like holes of extensive diameter.}
The right side shows a code on a rectangle with two different types of boundary like the surface code, blue at the back and front and green on the left and right.
An anyon is moved from the green boundary on the right to the green boundary on the left.

The operations implemented on the logical quantum degrees of freedom only depend on the topological phase of the bulk, boundaries, anyons, etc.
Since the phase is a ground state property, it is captured by a path integral in a spacetime with an imaginary time direction, that is, with a Euclidean signature.
Now assume we are given a Euclidean fixed-point path integral for the bulk, boundary, anyons, and other computational defects involved.
The logical operations corresponding to a spacetime process are obtained by simply evaluating this path integral, which can be done for a minimal cellulation.
Note that at the blue and green boundaries in Eq.~\eqref{eq:spacetime_processes}, the tensor-network path integral is terminated without open indices, possibly by introducing some extra tensors and bonds.
There are different choices for such a termination corresponding to different physical boundary conditions, and we will thus refer to these boundaries as \emph{physical boundary}.
In contrast, at the gray-red boundaries at the bottom and top we simply cut the tensor network resulting in open indices such that we get open indices, and evaluating the tensor network yields a state on this boundary.
We call these boundaries \emph{spatial boundaries}, as discussed in Section~\ref{sec:fixed_points}.
So the evaluation yields a linear operator from the bottom to the top open indices of the path integral.
This operator is only non-zero inside the ground state subspace at both input and output, and restricted to this ground state subspace yields the logical operation.

In other words, performing topological quantum computation is the same as executing the imaginary time evolution of some topological phase on some spacetime manifold, possibly including computational defects.
However, in the real world, we can only perform real time evolution.
Real time evolution can be described by a tensor-network path integral as well, namely as a unitary circuit.
However, the tensors of the imaginary time path integrals are not at all unitaries, and therefore it is impossible to execute the Euclidean path integral in the real world.
In this paper, we will understand how topological QEC is precisely a way to circumvent this impossibility by enriching the Euclidean path integral with syndrome defects.
Namely, a topological QEC protocol together with the according corrections effectively performs a fixed-point imaginary-time evolution through a real-time evolution, when restricted to the ground-state subspace.
As argued in Section~\ref{sec:dynamic_codes}, the resulting real-time path integrals will in fact not be unitary but circuits of quantum/classical hybrid channels.

For fully local QEC, the overall QEC protocol defines a dissipative circuit consisting of CPTP maps, and defines a real-time dissipative path integral.
We will now propose a very direct way in which a dissipative circuit representing a topological QEC protocol should be related to an imaginary-time fixed-point path integral of the underlying topological phase.
More precisely, we conjecture that this relation holds for known fully local QEC protocols in $4+1$ dimensions \cite{Kubica2018}, and propose it as a criterion for hypothetical fully local protocols in lower dimensions.
Note that this is not the relation that we use to construct and analyze codes in this paper.
To relate the real and imaginary-time path integrals in this direct way, we look at the spectrum of the transfer operator.
The transfer operator is the operator consisting of one time slice of the path integral, or one time period of the circuit.
When viewing the path integral as a tensor network, the transfer operator is a projected entangled-pair operator (PEPO).
In an imaginary-time fixed-point path integral, this transfer operator is a projector onto the $l$-dimensional ground state space.
Away from the fixed-point, the set of the highest $l$ eigenvalues
\footnote{
Note that it is common in physics to consider the spectral behavior of the \emph{lowest} eigenvalues of a Hamiltonian $H$, corresponding to its ground states.
However, as we discussed in Section~\ref{sec:fixed_points}, the transfer operator has the same qualitative properties as the imaginary-time evolution $e^{-\beta H}$, whose highest eigenvalues correspond to the lowest eigenvalues of $H$.
}
is contained inside an interval that shrinks exponentially with the system size $L$, and the remaining eigenvalues are separated by a gap that is lower bounded by a constant independent of $L$.
The operator consisting of the path integral on a patch of both spatial and temporal size $\sim L$, corresponding to the product of $\sim L$ times the transfer operator, thus converges to the ground state projector exponentially quickly in $L$.
More generally, the operator corresponding to the path integral on some topologically non-trivial spacetime converges to the ground-state projector followed by some logical gate acting on the ground-state space.

We propose that part of this spectral behavior directly carries over to a real-time dissipative circuit corresponding to a fully-local QEC dissipative circuit.
Namely, the highest-magnitude eigenvalues
\footnote{
Again, it is common to consider the lowest-real-part eigenvalues of a Lindbladian $\mathrm L$.
The transfer operator of a real-time dissipative circuit has the same qualitative properties as the continuous-time Lindbladian evolution $e^{-tL}$, whose highest-magnitude eigenvalues correspond to the lowest-real-part eigenvalues of $L$.
}
are still contained within an interval shrinking exponentially with $L$, and the operator corresponding to a circuit with non-trivial spacetime topology equals that of the imaginary-time path integral inside the ground state eigenspace and in the thermodynamic limit.
Outside of the ground state space, however, the real-time path integral has a different spectral behavior from the imaginary-time path integral:
There, the gap of a real-time QEC circuit must shrink at least like $~L^{-1}$ due to the the causality restriction of a circuit, or in other words, the finite propagation speed of information.
Namely, if we insert an ``error'' operation of size $~L$ into the circuit, then it takes time $~L$ to correct this error and return to the steady state.
In contrast, a gapped operator returns to the steady state from any starting point at a system-size independent rate.
Note that despite the closing gap, the ground state space is still well distinguished from the rest of the spectrum, since the shrinking of the ground state interval is exponential instead of polynomial.
We conjecture that fully local QEC protocols in $4+1$ dimensions such as in Ref.~\cite{Kubica2018} indeed have the spectral behavior described above, even after we add arbitrary perturbations below some threshold.
In contrast, imagine a local decoder without threshold, such as locally matching nearest-neighbor syndromes in the $2+1$-dimensional toric code.
Here we expect that the exponential degeneracy of highest-magnitude eigenvalues in the transfer operator breaks after adding arbitrarily small perturbations.

The direct relation suggested above only applies to full-local QEC which so far is only known to exist in unphysical dimensions, and also does help much with constructing such QEC protocols.
Thus, in this paper, we will use a more specific relation that does not take into account the classical decoder:
Namely, for a QEC circuit containing ``syndrome'' measurements, we consider this circuit post-selected on a configuration of measurement outcomes.
This yields a tensor-network path integral that equals an imaginary-time path integral.
The configuration of measurement outcomes corresponds to a pattern of defects inside this path integral.

\subsection{From path integrals to circuits}
In this section we describe an explicit general method to construct topological QEC circuits from topological fixed-point path integrals.
In fact, we will only construct the quantum part of the QEC circuit.
Which classical decoder works depends on the nature of the syndrome defects used.
However, for all examples in this paper, the syndrome defects are of a homological nature.
We propose that in these cases a decoder based on minimum-weight matching works.

We start by putting the path integral on some regular lattice and choosing a time direction.
Then we interpret the tensor network as a geometrically local circuit of operators, where each operator corresponds to a single tensor, or a patch of a few tensors.
The indices of each tensor or patch are divided into input and output in accordance with the chosen time direction.
This can always be done, however the resulting operators, like
\begin{equation}
\begin{tikzpicture}
\atoms{square}{{0/p={-0.3,0},lab={t=$T_1$,p=180:0.4}}}
\draw (0-tl)--++(135:0.6) (0-bl)--++(-135:0.6) (0-tr)--++(45:0.6) (0-br)--++(-45:0.6);
\end{tikzpicture}\;,
\end{equation}
are not in general unitaries, or equivalently, stacking two copies does not result in a channel that is normalized as in Eq.~\eqref{eq:channel_normalization},
\begin{equation}
\begin{tikzpicture}
\atoms{square}{{0/p={-0.3,0},lab={t=$T_1$,p=180:0.4}}, {1/p={0.3,0},lab={t=$T_1^*$,p=0:0.4}}}
\draw[rc] (0-tl)--++(135:0.6)--++(0:0.6)--(1-tl) (0-tr)--++(45:0.6)--++(0:0.6)--(1-tr);
\draw (0-bl)--++(-135:0.6) (0-br)--++(-45:0.6);
\draw (1-bl)--++(-135:0.6) (1-br)--++(-45:0.6);
\end{tikzpicture}
\neq
\begin{tikzpicture}
\draw[rc] (0,0)--++(0,0.5)-|++(0.3,-0.5) (0.8,0)--++(0,0.5)-|++(0.3,-0.5);
\end{tikzpicture}\;.
\end{equation}
In fact, it does never happen that all operators are unitary, since the operator corresponding to a full layer of imaginary-time evolution is a projector of low rank, and thus not a unitary.

Even though $T_1$ does not define a channel, it can always occur as part of an instrument.
To this end, we choose further tensors $T_2,T_3,\ldots$, that we combine into one single tensor using an additional classical output index,
\begin{equation}
\begin{tikzpicture}
\atoms{square}{{0/p={-0.3,0},lab={t=$\mathbf T$,p=180:0.4}}}
\draw (0-tl)--++(135:0.4) (0-bl)--++(-135:0.4) (0-tr)--++(45:0.4) (0-br)--++(-45:0.4);
\draw[classical] (0-t)--++(90:0.4);
\end{tikzpicture}
\coloneqq
\bigl(
\begin{tikzpicture}
\atoms{square}{{0/p={-0.3,0},lab={t=$T_1$,p=180:0.4}}}
\draw (0-tl)--++(135:0.4) (0-bl)--++(-135:0.4) (0-tr)--++(45:0.4) (0-br)--++(-45:0.4);
\end{tikzpicture}\;,
\begin{tikzpicture}
\atoms{square}{{0/p={-0.3,0},lab={t=$T_2$,p=180:0.4}}}
\draw (0-tl)--++(135:0.4) (0-bl)--++(-135:0.4) (0-tr)--++(45:0.4) (0-br)--++(-45:0.4);
\end{tikzpicture}\;,
\begin{tikzpicture}
\atoms{square}{{0/p={-0.3,0},lab={t=$T_3$,p=180:0.4}}}
\draw (0-tl)--++(135:0.4) (0-bl)--++(-135:0.4) (0-tr)--++(45:0.4) (0-br)--++(-45:0.4);
\end{tikzpicture}\;,
\ldots
\bigr)\;.
\end{equation}
We then use this tensor to define an instrument,
\begin{equation}
\label{eq:fixed_point_instrument}
I[\mathbf T]=
\begin{tikzpicture}
\atoms{square,xscale=4}{0/}
\draw[quantum] ([sx=-0.5]0-b)--++(-90:0.4) ([sx=0.5]0-b)--++(-90:0.4) ([sx=-0.5]0-t)--++(90:0.4) ([sx=0.5]0-t)--++(90:0.4);
\draw[classical] (0-t)--++(90:0.6);
\end{tikzpicture}
\coloneqq
\begin{tikzpicture}
\atoms{square}{{0/p={-0.3,0},lab={t=$\mathbf T$,p=180:0.4}}, {1/p={0.3,0},lab={t=$\mathbf T^*$,p=0:0.4}}}
\atoms{deltatens}{d/p={0,0.6}}
\draw (0-tl)--++(135:0.6) (0-bl)--++(-135:0.6) (0-tr)--++(45:0.6) (0-br)--++(-45:0.6);
\draw (1-tl)--++(135:0.6) (1-bl)--++(-135:0.6) (1-tr)--++(45:0.6) (1-br)--++(-45:0.6);
\draw[rc,classical] (0-t)--++(90:0.3)--(d) (1-t)--++(90:0.3)--(d) (d)--++(90:0.3);
\end{tikzpicture}\;.
\end{equation}
The small dot on the right denotes a $\delta$-tensor as defined in Eq.~\eqref{eq:delta_definition}, though here it serves a different function and the bond dimension can be different from $2$.
The normalization condition in Eq.~\eqref{eq:channel_normalization} of this instrument reduces to the following condition for $T$:
\begin{equation}
\begin{tikzpicture}
\atoms{square}{{0/p={-0.3,0},lab={t=$\mathbf T$,p=180:0.4}}, {1/p={0.3,0},lab={t=$\mathbf T$,p=0:0.4}}}
\draw[rc] (0-tl)--++(135:0.6)--++(0:0.6)--(1-tl) (0-tr)--++(45:0.6)--++(0:0.6)--(1-tr);
\draw (0-bl)--++(-135:0.6) (0-br)--++(-45:0.6);
\draw (1-bl)--++(-135:0.6) (1-br)--++(-45:0.6);
\draw[rc,classical] (0-t)--++(90:0.3)-|(1-t);
\end{tikzpicture}
=
\begin{tikzpicture}
\draw[rc] (0,0)--++(0,0.5)-|++(0.3,-0.5) (0.8,0)--++(0,0.5)-|++(0.3,-0.5);
\end{tikzpicture}\;.
\end{equation}
In other words, we are looking for tensors $T_2,T_3,\ldots$, such that the collection $\mathbf T=(T_1,T_2,T_3,\ldots)$ forms an isometry.

We now turn the fixed-point path integral into a circuit of instruments as in Eq.~\eqref{eq:fixed_point_instrument}.
If we happen to always get the trivial measurement outcome corresponding to $T_1$, then we have successfully executed the imaginary-time fixed-point path integral.
However, if some of the outcomes are non-trivial, we have performed another path integral including some tensors $T_2,T_3,\ldots$.
In this case we need to apply corrections such that the corrected time evolution is equivalent to that with only $T_1$.
In order to be able to do this, also the non-trivial outcomes must correspond to an exactly solvable fixed-point path integral of some sort.
This is where we use the syndrome defects such as anyons:
We choose $T_2,T_3,\ldots$ such that each of these tensors corresponds to a piece of fixed-point path integral that includes one or more segments of syndrome defect.
Then every configuration of classical outputs corresponds to a topological path integral with a pattern of syndrome defects.
The corrections are then implemented by classically controlled operations in the circuit that insert additional segments of syndrome defects depending on the classical control.
This motivates the following informal definition.
\begin{mydef}
\label{def:path_integral_code}
A \emph{fixed-point path integral code} is a uniform geometrically local circuit of quantum channels with additional classical inputs and outputs, such that the following holds:
\begin{itemize}
\item When fixing a configuration of classical inputs and outputs, the circuit becomes a mixed-state tensor-network path integral.
This path integral is a stack of two copies of the same (pure-state) path integral, with one of them complex conjugated.
\item This path integral is (in the same fixed-point phase as) a fixed-point path integral for a topological phase, including a pattern of syndrome defects.
This pattern only depends locally on the classical inputs and outputs.
\end{itemize}
The \emph{phase} of a fixed-point path integral code is the phase of the underlying topological path integral, with the trivial pattern of syndrome defects.
Two codes are considered equivalent if they are in the same phase, that is, the underlying path integrals (without defects) are related by local tensor-network equations.
\end{mydef}

In order to turn a fixed-point path integral code into a completely specified process, we have to couple the classical inputs and outputs to a classical decoder $D$.
Very vaguely speaking, the resulting process is error correcting if $D$ yields a total defect pattern (formed by the outputs and inputs) that is equivalent to the trivial one.
If there is noise, the total defect pattern does not fulfill the local constraints, so instead we take the closest defect pattern that does.
More concretely, let us give a decoder that works if the defect pattern form (co-)cycles, which is the case for all examples given in this paper.
This can be viewed as a generalization of decoding the toric code in the presence of measurement errors \cite{Dennis2001}.
\begin{myprop}
\label{prop:path_integral_decoding}
A fixed-point path-integral code whose syndrome defects are (co-)cycles can be turned into a complete fault tolerant process as follows.
The overall circuit layout is that of Eq.~\eqref{eq:qec_circuit}, where we first record measurement outcomes for a time $T\sim L$, and then perform corrections at time $T$.
Here, $T\sim L$ denotes that $T$ scales proportional to the linear system size $L$, for example if we put the code on an $L\times L$ torus.
Thereby, we need to insert enough controlled operations at time $T$ to be able to close off any measured defect pattern.
The classical decoder $D$ is given as follows:
\begin{enumerate}
\item Consider the (co-)chain(s) corresponding to the recorded spacetime syndrome by definition of the fixed-point path integral code.
Choose a minimum-weight fix turning the (co-)chain(s) into (co-)cycle(s).
Thereby, treat the time-like boundary at time $T$ as ``open'', such that (co-)cycles can freely terminate there.
In contrast, treat the initial time-like boundary at time $0$ as ``closed'', such that (co-)cycles are not allowed to terminate there.
\item Consider the endpoints of the (co-)cycle(s) at time $T$.
Choose a set of defect segments at time $T$ that together with the fixed (co-)cycle(s) in the spacetime forms homologically trivial (co-)cycle(s).
This set of defects determines the input to the classically controlled correction operations.
\end{enumerate}
\end{myprop}
Let us give a rough argument for why this process has a fault tolerant threshold under local noise, a detailed proof will appear in future work.
If we perform the circuit without noise, then the classical outputs correspond to a defect pattern consisting of (co-)cycle(s).
Otherwise the path integral evaluates to zero as in Eq.~\eqref{eq:cocycle_constraint}, and the corresponding configuration of outcomes is measured with probability zero.
However, if we perturb the circuit by adding (weak) noise, the (co-)cycle(s) are (slightly) broken.
We find that (1) the probability that they are broken everywhere inside a connected region is exponentially small in the size of that region,
(2) two cycles of different homology classes differ inside a region of at least size $\sim L$, and
(3) the number of connected regions of size $L$ is at most exponential in $L$.
Thus for weak enough noise, the probability for the minimum-weight fix to yield the wrong cohomology class is exponentially small in $L$.
This loop-counting argument is analogous to the one in Ref.~\cite{Dennis2001}.

For all examples in this paper except for Section~\ref{sec:3d_toric_code}, the syndrome defects are anyon worldlines.
In this case, the correction operators closing the spacetime string net pattern at time $T$ are known as \emph{string operators}.
Fixing the string net pattern in the presence of noise means pairing up the string endpoints in spacetime.
A polynomial-time algorithm solving this problem is known as \emph{minimum weight perfect matching} \cite{Edmonds1965}.

As we have seen, it is the task of the decoder to perform corrections that close off the syndrome defects in such a way that it is equivalent to the empty syndrome.
Thus we see that the phase of a code alone determines the operation that the code performs on the logical space.
This holds whenever we use the code as building block in some larger protocol of topological quantum computation, which may involve defects, boundaries, interfaces to other codes, etc.

\section{Known codes in terms of path integrals}
\label{sec:examples}
In this section, we consider four different examples of fixed-point path-integral codes, which we all find to be equivalent to existing codes, namely the stabilizer toric code, subsystem toric code, CSS honeycomb Floquet code, and honeycomb Floquet code.
The first three examples are all based on the toric-code path integral introduced in Section~\ref{sec:fixed_points}, which we put on different spacetime lattices with different choices of time direction.
The fourth example differs from the previous ones by only a change of basis of the tensor-network path integral.

\subsection{Stabilizer toric code}
\label{sec:stabilizer_toric_code}
As a first example let us consider the first of all topological error-correcting codes, namely the toric code on a square lattice \cite{Kitaev1997,Dennis2001}.
The underlying tensor-network path integral is the toric-code path integral from Section~\ref{sec:fixed_points} on a cubic lattice, whose unit vectors we call $x$, $y$, and $z$.
The time direction $t$ is coincident with $z$,
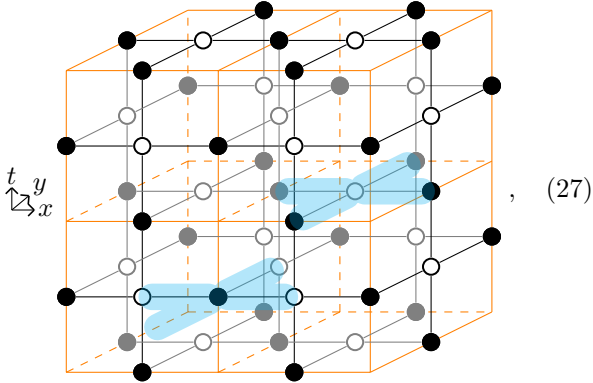
\begin{equation}
\label{eq:toric_code_marked_stabilizer}
\begin{tikzpicture}
\draw (-0.6,0.3)edge[mark={arr,e},ind=$x$]++(0:0.3) (-0.6,0.3)edge[mark={arr,e},ind=$y$]++(40:0.3) (-0.6,0.3)edge[mark={arr,e},ind=$t$]++(90:0.3);
\end{tikzpicture}
\begin{tikzpicture}
\atoms{void}{x/p={2,0}, y/p={0,2}, z/p={1.6,0.8}}
\foreach \x in {0,1,2}{
\foreach \y in {0,1,2}{
\foreach \z in {0,1}{
\atoms{void}{\x\y\z/p={$\x*(x)+\y*(y)+\z*(z)$}}
}}}
\draw[orange] (000)--(200) (010)--(210) (020)--(220) (000)--(020) (100)--(120) (200)--(220) (201)--(221) (021)--(221);
\draw[orange,dashed] (001)--(201) (011)--(211) (001)--(021) (101)--(121);
\draw[orange, dashed] (000)--(001) (100)--(101) (010)--(011) (110)--(111);
\draw[orange] (200)--(201) (210)--(211) (220)--(221) (020)--(021) (120)--(121);
\foreach \x/\y/\z in {1/0/0, 3/0/0, 1/2/0, 3/2/0, 1/4/0, 3/4/0, 0/1/0, 0/3/0, 2/1/0, 2/3/0, 4/1/0, 4/3/0, 4/0/1, 4/2/1, 0/4/1, 2/4/1, 4/4/1, 4/1/2, 4/3/2, 1/4/2, 3/4/2}{
\atoms{delta}{d\x\y\z/p={$0.5*\x*(x)+0.5*\y*(y)+0.5*\z*(z)$}}
}
\foreach \x/\y/\z in {1/0/2, 3/0/2, 1/2/2, 3/2/2, 0/1/2, 0/3/2, 2/1/2, 2/3/2, 2/2/1, 0/2/1, 0/0/1, 2/0/1}{
\atoms{delta,astyle=gray}{d\x\y\z/p={$0.5*\x*(x)+0.5*\y*(y)+0.5*\z*(z)$}}
}
\foreach \x/\y/\z in {1/1/0, 3/1/0, 1/3/0, 3/3/0, 1/4/1, 3/4/1, 4/1/1, 4/3/1}{
\atoms{z2}{z\x\y\z/p={$0.5*\x*(x)+0.5*\y*(y)+0.5*\z*(z)$}}
}
\foreach \x/\y/\z in {1/1/2, 3/1/2, 1/3/2, 3/3/2, 1/2/1, 3/2/1, 2/1/1, 2/3/1, 0/1/1, 0/3/1, 1/0/1, 3/0/1}{
\atoms{z2,astyle=gray}{z\x\y\z/p={$0.5*\x*(x)+0.5*\y*(y)+0.5*\z*(z)$}}
}
\foreach \x/\xx/\xxx in {0/1/2,2/3/4}{
\foreach \y/\yy/\yyy in {0/1/2,2/3/4}{
\foreach \z in {0}{
\draw (z\xx\yy\z)--(d\x\yy\z) (z\xx\yy\z)--(d\xxx\yy\z) (z\xx\yy\z)--(d\xx\y\z) (z\xx\yy\z)--(d\xx\yyy\z);
}
\foreach \z in {2}{
\draw[gray] (z\xx\yy\z)--(d\x\yy\z) (z\xx\yy\z)--(d\xxx\yy\z) (z\xx\yy\z)--(d\xx\y\z) (z\xx\yy\z)--(d\xx\yyy\z);
}}}
\foreach \x/\xx/\xxx in {0/1/2,2/3/4}{
\foreach \y in {0,2}{
\draw[gray] (z\xx\y1)--(d\x\y1) (z\xx\y1)--(d\xxx\y1) (z\xx\y1)--(d\xx\y0) (z\xx\y1)--(d\xx\y2);
}
\foreach \y in {4}{
\draw (z\xx\y1)--(d\x\y1) (z\xx\y1)--(d\xxx\y1) (z\xx\y1)--(d\xx\y0) (z\xx\y1)--(d\xx\y2);
}}
\foreach \y/\yy/\yyy in {0/1/2,2/3/4}{
\foreach \x in {0,2}{
\draw[gray] (z\x\yy1)--(d\x\y1) (z\x\yy1)--(d\x\yyy1) (z\x\yy1)--(d\x\yy0) (z\x\yy1)--(d\x\yy2);
}
\foreach \x in {4}{
\draw (z\x\yy1)--(d\x\y1) (z\x\yy1)--(d\x\yyy1) (z\x\yy1)--(d\x\yy0) (z\x\yy1)--(d\x\yy2);
}}

\draw[cyan,opacity=0.3,line width=10,line cap=round] (z321)--(d221) (z321)--(d421) (z321)--(d320) (z321)--(d322);
\draw[cyan,opacity=0.3,line width=10,line cap=round] (d210)--(z110) (d210)--(z310) (d210)--(z211) (d210)--++($-0.5*(z)$);
\end{tikzpicture}\;,
\end{equation}
where the background cubic lattice is in orange and the tensor-network diagram is in black.
We now view the tensor-network path integral as a circuit of operators, where each operator corresponds to one or a few tensors.
There are two types of operators, as marked above in semi-transparent blue.
Both operators act on 4 qubits that correspond to $t$-directed bonds in the tensor-network diagram.
Specifically, there is an operator $T_1$ at each $xy$ face, and an operator $V_1$ at every $t$ edge,
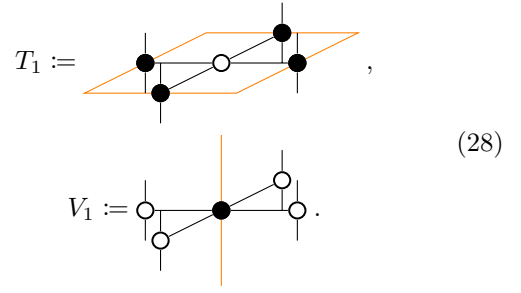
\begin{equation}
\begin{gathered}
T_1\coloneqq
\begin{tikzpicture}
\draw[orange] (-1.8,-0.4)--++(2,0)--++(1.6,0.8)--++(-2,0)--cycle;
\atoms{z2}{0/}
\atoms{delta}{1/p={1,0}, 2/p={-1,0}, 3/p={0.8,0.4}, 4/p={-0.8,-0.4}}
\draw (0)--(1) (0)--(2) (0)--(3) (0)--(4) (1)--++(-90:0.4) (1)--++(90:0.4) (2)--++(-90:0.4) (2)--++(90:0.4) (3)--++(-90:0.4) (3)--++(90:0.4) (4)--++(-90:0.4) (4)--++(90:0.4);
\end{tikzpicture}\;,\\
V_1\coloneqq
\begin{tikzpicture}
\draw[orange] (0,-1)--(0,1);
\atoms{delta}{0/}
\atoms{z2}{1/p={1,0}, 2/p={-1,0}, 3/p={0.8,0.4}, 4/p={-0.8,-0.4}}
\draw (0)--(1) (0)--(2) (0)--(3) (0)--(4) (1)--++(-90:0.4) (1)--++(90:0.4) (2)--++(-90:0.4) (2)--++(90:0.4) (3)--++(-90:0.4) (3)--++(90:0.4) (4)--++(-90:0.4) (4)--++(90:0.4);
\end{tikzpicture}\;.
\end{gathered}
\end{equation}
Note that these diagrams are identical to well-known $ZX$ diagrams for the vertex and plaquette terms of the toric code \cite{Gidney2022,Kissinger2022,Bombin2023}.
In order to get the decomposition, we need to split up all the 4-index $\zz_2$ tensors at the $xt$ and $yt$ faces into two 3-index $\zz_2$-tensors,
\begin{equation}
\begin{tikzpicture}
\atoms{void}{0/, 1/p={1,0}, 2/p={0,1}, 3/p={1,1}}
\draw[orange] (3)--(2) (3)--(1) (0)--(1) (0)--(2);
\atoms{z2}{a/p={0.5,0.5}}
\draw (a)--++(90:0.6) (a)--++(0:0.6) (a)--++(-90:0.6) (a)--++(180:0.6);
\end{tikzpicture}
=
\begin{tikzpicture}
\atoms{void}{0/, 1/p={1,0}, 2/p={0,1}, 3/p={1,1}}
\draw[orange] (1)--(2) (3)--(2) (3)--(1) (0)--(1) (0)--(2);
\atoms{z2}{a/p={0.4,0.3}, b/p={0.6,0.7}}
\draw (a)--(b) (b)--++(90:0.4) (b)--++(0:0.4) (a)--++(-90:0.4) (a)--++(180:0.4);
\end{tikzpicture}
=
\begin{tikzpicture}
\atoms{void}{0/, 1/p={1,0}, 2/p={0,1}, 3/p={1,1}}
\draw[orange] (0)--(3) (3)--(2) (3)--(1) (0)--(1) (0)--(2);
\atoms{z2}{a/p={0.4,0.7}, b/p={0.6,0.3}}
\draw (a)--(b) (b)--++(-90:0.4) (b)--++(0:0.4) (a)--++(90:0.4) (a)--++(180:0.4);
\end{tikzpicture}
\;.
\end{equation}
As shown, this splitting up can be represented geometrically as dividing each plaquette into two triangles.
After this, $V_1$ corresponds to a $t$ edge together with the adjacent triangles.
As shown (see also Eq.~\eqref{eq:22pachner}), there are two different ways to split up the plaquette/tensor.
As we will discuss more later, these correspond to different orderings in which $V_1$ at neighboring $t$ edges act on the same qubit.
Dually, we need to split each 4-index $\delta$-tensor at a $x$ or $y$ edge into two 3-index $\delta$-tensors.
Geometrically, this corresponds to splitting a 4-valent edge into two 3-valent edges separated by a 2-gon face yielding a configuration as shown in Eq.~\eqref{eq:2gon_trivial}.
After this, $T_1$ corresponds to a $xy$ face together with the adjacent 3-valent edges.
Neither $T_1$ nor $V_1$ are unitary, which is not a surprise given that the path integral represents an imaginary, and not a real time evolution.
In fact, $T_1$ is the projector onto the $+1$ eigenspace of the Pauli operator $Z_0Z_1Z_2Z_3$, and $V_1$ the projector onto the $+1$ eigenspace of $X_0X_1X_2X_3$.
To fix this, we define a second projector $T_m$ corresponding to a $xy$ face carrying a segment of $m$ worldline,
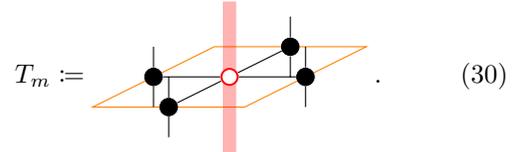
\begin{equation}
\label{eq:stabilizer_plaquette_operator}
T_m\coloneqq
\begin{tikzpicture}
\draw[orange] (-1.8,-0.4)--++(2,0)--++(1.6,0.8)--++(-2,0)--cycle;
\draw[worldline] (0,-1)--(0,1);
\atoms{z2,bdastyle=red}{0/}
\atoms{delta}{1/p={1,0}, 2/p={-1,0}, 3/p={0.8,0.4}, 4/p={-0.8,-0.4}}
\draw (0)--(1) (0)--(2) (0)--(3) (0)--(4) (1)--++(-90:0.4) (1)--++(90:0.4) (2)--++(-90:0.4) (2)--++(90:0.4) (3)--++(-90:0.4) (3)--++(90:0.4) (4)--++(-90:0.4) (4)--++(90:0.4);
\end{tikzpicture}\;.
\end{equation}
This way, $T_1$ is extended to an isometry $\mathbf T$,
\begin{equation}
\mathbf T\coloneqq (T_1,T_m)
=
\begin{tikzpicture}
\atoms{z2}{0/}
\atoms{delta}{1/p={0.8,0}, 2/p={-0.8,0}, 3/p={0.4,0.5}, 4/p={-0.4,-0.5}}
\draw (0)--(1) (0)--(2) (0)--(3) (0)--(4) (1)--++(-90:0.4) (1)--++(90:0.4) (2)--++(-90:0.4) (2)--++(90:0.4) (3)--++(-90:0.4) (3)--++(90:0.4) (4)--++(-90:0.4) (4)--++(90:0.4);
\draw[classical](0)--++(90:0.5);
\end{tikzpicture}\;.
\end{equation}
$\mathbf T$ defines an instrument $I[\mathbf T]$ via Eq.~\eqref{eq:fixed_point_instrument}, which is in fact just a projective $Z_0Z_1Z_2Z_3$ measurement.
Dually, we can define an operator $V_e$ carrying an $e$ anyon segment along a $t$ edge,
\begin{equation}
V_e\coloneqq
\begin{tikzpicture}
\draw[orange] (0,-1)--(0,1);
\draw[worldline] (0,-1)--(0,1);
\atoms{delta,bdastyle=red}{0/}
\atoms{z2}{1/p={1,0}, 2/p={-1,0}, 3/p={0.8,0.4}, 4/p={-0.8,-0.4}}
\draw (0)--(1) (0)--(2) (0)--(3) (0)--(4) (1)--++(-90:0.4) (1)--++(90:0.4) (2)--++(-90:0.4) (2)--++(90:0.4) (3)--++(-90:0.4) (3)--++(90:0.4) (4)--++(-90:0.4) (4)--++(90:0.4);
\end{tikzpicture}\;.
\end{equation}
This gives rise to an isometry $\mathbf V$,
\begin{equation}
\label{eq:stabilizer_vertex_operator}
\mathbf V\coloneqq (V_1,V_e)
=
\begin{tikzpicture}
\atoms{hadamard}{h/p={0,0.5}}
\atoms{delta}{0/}
\atoms{z2}{1/p={0.8,0}, 2/p={-0.8,0}, 3/p={0.4,0.5}, 4/p={-0.4,-0.5}}
\draw (0)--(1) (0)--(2) (0)--(3) (0)--(4) (1)--++(-90:0.4) (1)--++(90:0.4) (2)--++(-90:0.4) (2)--++(90:0.4) (3)--++(-90:0.4) (3)--++(90:0.4) (4)--++(-90:0.4) (4)--++(90:0.4);
\draw[classical] (0)--(h) (h-t)--++(90:0.4);
\end{tikzpicture}\;.
\end{equation}
$\mathbf V$ yields an instrument $I[\mathbf V]$, which is just a projective $X_0X_1X_2X_3$ measurement.
The presence of the \emph{Hadamard} matrix,
\begin{equation}
\label{eq:hadamard}
\begin{tikzpicture}
\atoms{hadamard}{0/}
\draw (0-l)--++(180:0.4) (0-r)--++(0:0.4);
\end{tikzpicture}
\coloneqq
H
\coloneqq
\frac1{\sqrt2}
\begin{pmatrix}
1&1\\1&-1
\end{pmatrix}\;,
\end{equation}
seemingly spoils the duality between Eq.~\eqref{eq:stabilizer_vertex_operator} and Eq.~\eqref{eq:stabilizer_plaquette_operator}.
This is explained by the fact that in Eq.~\eqref{eq:fixed_point_instrument} we use a $\delta$-tensor for both $I[\mathbf T]$ and $I[\mathbf V]$.

As mentioned earlier, the way in which we divide each plaquette into two triangles and each 4-valent edge into two 3-valent edges determines the ordering in which we act on the qubits.
A straight-forward way to choose an ordering is to checkerboard-number each vertex and $t$ edge with $x/y$-coordinate $ax+by$ by $a+b\mod 2$.
Then we first act with all the $0$-labeled $V$ operators (which act on mutually disjoint quadruples of qubits) and then with all the $1$-labeled ones.
Dually, we can checkerboard-number cubes and $xy$ faces according to their $x/y$ coordinate and act with $0$-labeled $T$ operators first.
One full period of the circuit then consists of four rounds of instruments,
\begin{equation}
\rightarrow I[\mathbf T]_0\rightarrow I[\mathbf T]_1\rightarrow I[\mathbf V]_0\rightarrow I[\mathbf V]_1\rightarrow\;.
\end{equation}
Geometrically, this corresponds to dividing each $xt$ and $yt$ plaquette along a diagonal edge connecting a $0$ vertex at time $T$ to a $1$ vertex at time $T-1$.
Dually, we also split each 4-valent edge by inserting a 2-gon that is adjacent to a $0$ cube at time $T$ and a $1$ cube at time $T-1$.
The following shows a section of this modified cubic lattice (this time in black instead of orange),
\begin{equation}
\begin{tikzpicture}
\draw (-0.6,0.3)edge[mark={arr,e},ind=$x$]++(0:0.3) (-0.6,0.3)edge[mark={arr,e},ind=$y$]++(40:0.3) (-0.6,0.3)edge[mark={arr,e},ind=$t$]++(90:0.3);
\end{tikzpicture}
\begin{tikzpicture}
\atoms{vertex}{0/, {1/p={1.5,0}}, {2/p={3,0}}, {3/p={1.2,1},astyle=gray}, {4/p={2.7,1},astyle=gray}, {5/p={4.2,1}}, {6/p={2.4,2},astyle=gray}, {7/p={3.9,2},astyle=gray}, {8/p={5.4,2}}}
\atoms{vertex}{0x/p={0,0.8}, 1x/p={1.5,0.8}, 2x/p={3,0.8}, 3x/p={1.2,1.8}, 4x/p={2.7,1.8}, 5x/p={4.2,1.8}, 6x/p={2.4,2.8}, 7x/p={3.9,2.8}, 8x/p={5.4,2.8}}
\draw (0)edge[mark={slab=$\scriptstyle 0$}](0x) (1)edge[mark={slab=$\scriptstyle 0$}](1x) (2)edge[mark={slab=$\scriptstyle 0$}](2x) (3)edge[dashed,gray,mark={slab=$\scriptstyle 0$}](3x) (4)edge[dashed,gray,mark={slab=$\scriptstyle 0$}](4x) (5)edge[mark={slab=$\scriptstyle 0$}](5x) (6)edge[dashed,gray,mark={slab=$\scriptstyle 0$}](6x) (7)edge[dashed,gray,mark={slab=$\scriptstyle 0$}](7x) (8)edge[mark={slab=$\scriptstyle 0$}](8x);
\draw (0)edge[bend left=10](1) (1)edge[bend left=20](2) (3)edge[bend left=20,dashed,gray](4) (4)edge[bend left=10,dashed,gray](5) (6)edge[bend left=10,dashed,gray](7) (7)edge[bend left=20,dashed,gray](8);
\draw (0)edge[bend right=10](1)(1)edge[bend right=20](2) (3)edge[bend right=20,dashed,gray](4)(4)edge[bend right=10,dashed,gray](5) (6)edge[bend right=10,dashed,gray](7)(7)edge[bend right=20,dashed,gray](8);
\draw (0)edge[bend left=20,dashed,gray](3)(3)edge[bend left=10,dashed,gray](6) (1)edge[bend left=10,dashed,gray](4)(4)edge[bend left=20,dashed,gray](7) (2)edge[bend left=20](5)(5)edge[bend left=10](8);
\draw (0)edge[bend right=20,dashed,gray](3)(3)edge[bend right=10,dashed,gray](6) (1)edge[bend right=10,dashed,gray](4)(4)edge[bend right=20,dashed,gray](7) (2)edge[bend right=20](5)(5)edge[bend right=10](8);
\draw (0x)edge[bend left=10](1x) (1x)edge[bend left=20](2x) (3x)edge[bend left=20](4x) (4x)edge[bend left=10](5x) (6x)edge[bend left=10](7x) (7x)edge[bend left=20](8x);
\draw (0x)edge[bend right=10](1x)(1x)edge[bend right=20](2x) (3x)edge[bend right=20](4x)(4x)edge[bend right=10](5x) (6x)edge[bend right=10](7x)(7x)edge[bend right=20](8x);
\draw (0x)edge[bend left=20](3x)(3x)edge[bend left=10](6x) (1x)edge[bend left=10](4x)(4x)edge[bend left=20](7x) (2x)edge[bend left=20](5x)(5x)edge[bend left=10](8x);
\draw (0x)edge[bend right=20](3x)(3x)edge[bend right=10](6x) (1x)edge[bend right=10](4x)(4x)edge[bend right=20](7x) (2x)edge[bend right=20](5x)(5x)edge[bend right=10](8x);
\draw (0)--(1x) (0)edge[dashed,gray](3x) (2)--(1x) (2)--(5x) (4)edge[dashed,gray](1x) (4)edge[dashed,gray](5x) (4)edge[dashed,gray](7x) (4)edge[dashed,gray](3x) (6)edge[dashed,gray](3x) (6)edge[dashed,gray](7x) (8)edge[dashed,gray](7x) (8)--(5x);
\node at (1.4,1.3){$\scriptstyle 0$};
\node at (2.9,1.3){$\scriptstyle 1$};
\node at (2.6,2.3){$\scriptstyle 1$};
\node at (4.1,2.3){$\scriptstyle 0$};
\end{tikzpicture}\;.
\end{equation}
Here we should imagine the 1-labeled $xy$ faces and the adjacent edges being bent slightly towards the positive $t$ direction.
The fact that all microscopic details of the protocol can be captured by a modified or refined cellulation also has at least one practical application:
When decoding the syndrome, it is natural to perform minimum-weight matching on precisely this spacetime lattice or its Poincar\'e dual:
If we assume single-qubit circuit-level noise acting with a Pauli operator on each qubit at each time step in the protocol with a constant probability, then the matching weight in this lattice is proportional to the probability of the corresponding error configuration.

Note that, since all the operators commute, applying them in any order defines a valid spacetime cellulation.
For example, consider the alternative ordering
\begin{equation}
\rightarrow I[\mathbf T]_0\rightarrow I[\mathbf V]_0\rightarrow I[\mathbf T]_1\rightarrow I[\mathbf V]_1\rightarrow\;.
\end{equation}
This yields a modified cellulation where the $1$ $xy$ faces are replaced by faces formed by four diagonals.
That is, the $0$ cubes remain the same, but the $1$ cubes have their $0$ vertices shifted by $+1$ in $t$ direction.
The following shows the original (gray) and modified (black) 1-cubes:
\begin{equation}
\begin{tikzpicture}
\atoms{vertex}{{0/p={0,0.8}}, {1/p={1.5,0}}, {2/p={1.2,1}}, {3/p={2.7,1.8}}, {0x/p={0,1.6}}, {1x/p={1.5,0.8}}, {2x/p={1.2,1.8}}, 3x/p={2.7,2.6}}
\atoms{vertex,style=gray}{0y/p={0,0}, 1y/p={1.5,1.6}, 2y/p={1.2,2.6}, 3y/p={2.7,1}}
\draw[line width=1.3] (0)--(1) (1)--(3) (0)edge[dashed](2) (2)edge[dashed](3) (0x)--(1x) (1x)--(3x) (0x)edge[dashed](2x) (2x)edge[dashed](3x) (0)edge[mark={slab=$\scriptstyle 0$}](0x) (1)edge[mark={slab=$\scriptstyle 1$}](1x) (2)edge[mark={slab=$\scriptstyle 1$},dashed](2x) (3)edge[mark={slab=$\scriptstyle 0$,r}](3x);
\draw[gray] (0)--(0y) (3)--(3y) (1y)--(1x) (2y)edge[dashed](2x) (0y)--(1) (1)--(3y) (0y)edge[dashed](2) (2)edge[dashed](3y) (0)--(1x) (1x)--(3) (0)edge[dashed](2x) (2x)edge[dashed](3) (0x)--(1y) (1y)--(3x) (0x)--(2y) (2y)--(3x);
\node at (1.4,2.1){$\scriptstyle 1$};
\end{tikzpicture}\;.
\end{equation}
In addition to changing the ordering of measurements, we may put the toric code on different spatial cellulations.
In this case the spacetime cellulation is based on the cartesian product of the spatial cellulation with the regular 1-dimensional lattice.
Further, different decompositions of a stabilizer measurement into CX gates acting on an ancilla also yield different cellulations.
As an intermediate step to obtain these cellulations, we may translate the measurement circuit into a ZX diagram, which is a standard technique in the literature \cite{Gidney2022}.
Then, we replace every $X$-type tensor with a face and every $Z$-type tensor with an edge.
The representation of the circuit as a spacetime cellulation also has a practical application:
The spacetime cellulation provides a natural lattice on which the classical decoder performs minimum-weight matching, since individual Pauli-$X$ or Pauli-$Z$ errors at specific moments in time correspond to different faces and edges in this lattice.

Let us briefly describe the general decoder of Proposition~\ref{prop:path_integral_decoding} for the present code.
We first record the spacetime syndrome until the time $T\sim L$ that we assume to be after the plaquette measurements, where the associated spatial slice of the lattice is a square lattice.
This syndrome is a subset of $t$ edges and $xy$ faces forming an $e$ 1-chain and a $m$ 2-cochain inside the (modified) cubic spacetime lattice.
The classical decoder $D$ now finds a minimum-weight set of edges (faces), such that flipping these edges (faces) fixes the $e$ 1-chain ($m$ 2-cochain) to a 1-cycle (2-cocycle).
It is important to note that the flipped edges (faces) can be any edges (faces) of the modified cubic lattices and not only $t$ edges ($xy$ faces).
Thereby the fixed $e$ 1-cycle ($m$ 2-cocycle) is allowed to terminate at the square-lattice spatial slice at time $T$.
To perform the corrections, we insert a 2-gon and a 2-valent edge in between every time-$T$ $x$ and $y$ edge and the adjacent time-$(T+\frac12)$ $xt$ and $yt$ face.
Then $D$ chooses any subset of the inserted 2-valent edges (2-gon faces), such that these edges (faces) together with the fixed spacetime $e$ 1-cycle ($m$ 2-cocycle) forms a homologically trivial 1-cycle (2-cocycle) that does not terminate on the spatial slice at time $T$.
At every chosen 2-valent edge (2-gon face), we put an $e$ ($m$) worldline, such that the overall defect configuration is equivalent to the trivial one, for example,
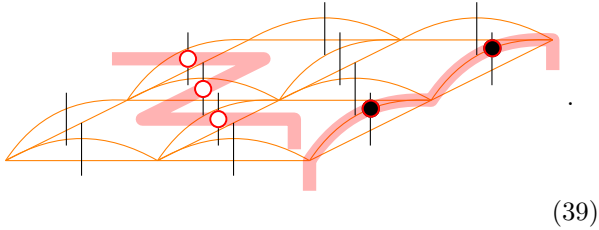
\begin{equation}
\label{eq:toric_code_correction}
\begin{tikzpicture}
\atoms{void}{0/, {1/p={2,0}}, 2/p={4,0}, 3/p={1.6,0.8}, 4/p={3.6,0.8}, 5/p={5.6,0.8}, 6/p={3.2,1.6}, 7/p={5.2,1.6}, {8/p={7.2,1.6}}}
\draw[orange] (0)--(1) (1)--(2) (3)--(4) (4)--(5) (6)--(7) (7)--(8);
\draw[orange] (0)--(3) (3)--(6) (1)--(4) (4)--(7) (2)--(5) (5)--(8);
\draw[orange] (0)edge[bend left](1) (1)edge[bend left](2) (3)edge[bend left](4) (4)edge[bend left](5) (6)edge[bend left](7) (7)edge[bend left](8);
\draw[orange] (0)edge[bend left](3) (3)edge[bend left](6) (1)edge[bend left](4) (4)edge[bend left](7) (2)edge[bend left](5) (5)edge[bend left](8);
\draw[worldline] ($(2-c)+(-90:0.4)$)--(2-c)to[bend left]coordinate[pos=0.576](m0)(5-c)to[bend left]coordinate[pos=0.576](m1)(8-c)--++(-90:0.4);
\draw[worldline] ($(3.8,0.55)+(-90:0.4)$)--++(90:0.4)--coordinate[midway](m2)++(-2,0)--coordinate[midway](m3)++(1.6,0.8)--coordinate[midway](m4)++(-2,0);
\draw (1,-0.2)--++(90:0.7) (3,-0.2)--++(90:0.7) (0.8,0.2)--++(90:0.7) (2.8,0.2)--++(90:0.7) (4.8,0.2)--++(90:0.7) (2.6,0.6)--++(90:0.7) (4.6,0.6)--++(90:0.7) (2.4,1)--++(90:0.7) (4.4,1)--++(90:0.7) (6.4,1)--++(90:0.7) (4.2,1.4)--++(90:0.7) (6.2,1.4)--++(90:0.7);
\atoms{delta,bdastyle=red}{xm0/p=m0, xm1/p=m1}
\atoms{z2,bdastyle=red}{xm2/p=m2, xm3/p=m3, xm4/p=m4}
\end{tikzpicture}\;.
\end{equation}
As shown, adding these the worldlines corresponds to inserting 2-index $\delta$ and $\zz_2$-tensors into the circuit, which are single qubit Pauli operators,
\begin{equation}
\begin{tikzpicture}
\atoms{z2,bdastyle=red}{0/}
\draw (0)--++(90:0.3) (0)--++(-90:0.3);
\end{tikzpicture}
=
X\;,\qquad
\begin{tikzpicture}
\atoms{delta,bdastyle=red}{0/}
\draw (0)--++(90:0.3) (0)--++(-90:0.3);
\end{tikzpicture}
=
Z\;.
\end{equation}
Note that in the absence of noise, the measured $e$ 1-chain ($m$ 2-cochain) is a 1-cycle (2-cocycle) with probability $1$.
Since $e$ ($m$) is only supported on $t$ edges ($xy$ faces), the measured syndrome consists of a subset of infinite (dual) lines in $t$ direction.
Thus, future measurement outcomes are determined by the past history.
In particular, if we start with the ground state, we deterministically measure the trivial syndrome.
This property is not necessary for fault tolerance and is the key qualitative difference between the stabilizer toric code and the subsystem and Floquet versions thereof that we will look at in the following.

\subsection{Subsystem toric code}
\label{sec:subsystem}
The \emph{subsystem toric code} is a topological subsystem code developed in Ref.~\cite{Bravyi2012}, that only involves 3-body measurements.
The code has qubits located on both vertices and edges of a square lattice.
The gauge checks consist of $X_0X_1X_2$ measurements involving either the three qubits near the top-right or the three qubits near the bottom-left corner of each square, and $Z_0Z_1Z_2$ checks involving the three qubits near the top-left or bottom-right corner.
The subsystem code is defined by these gauge checks alone, but in order to obtain a concrete QEC protocol we need to choose a periodic schedule in which we measure them.
The QEC protocol that we reconstruct in this section corresponds to a schedule where we alternatingly measure first all $X$ and then all $Z$ stabilizers.
The underlying path integral is still the toric-code path integral an a specific spacetime cellulation.
This cellulation is similar to that of stabilizer toric code on a regular triangular spatial lattice, and obtained from the latter by a simple modification.
For the stabilizer toric code we would consider a spacetime cellulation consisting of triangle-prism volumes,
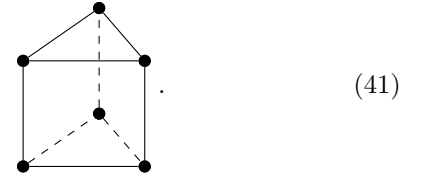
\begin{equation}
\label{eq:triangle_prism}
\begin{tikzpicture}
\atoms{vertex}{{0/}, {2/p={1.6,0}}, 1/p={1,0.7}}
\atoms{vertex}{0x/p={0,1.4}, 2x/p={1.6,1.4}, 1x/p={1,2.1}}
\draw[dashed] (0)--(1) (1)--(2) (1)--(1x);
\draw (0)--(2) (0x)--(1x) (1x)--(2x) (0x)--(2x) (0)--(0x) (2)--(2x);
\end{tikzpicture}\;.
\end{equation}
This way, the operators $T_1$ at the $xy$ faces are already 3-qubit, but the operators $V_1$ at the $t$ edges are 6-qubit.
However, $V_1$ can be split up into two 3-qubit operators $V_1^a$ and $V_1^b$ by the following trick.
We choose one spatial direction $x$ aligned with one third of the edges, and refer to the orthogonal direction as $y$.
Then at every spatial vertex, we pair up the two adjacent triangles whose centers are located in the positive and in the negative $y$ direction,
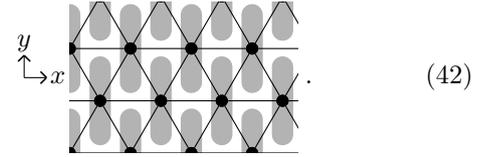
\begin{equation}
\begin{tikzpicture}
\draw (-0.6,1)edge[mark={arr,e},ind=$x$]++(0:0.3) (-0.6,1)edge[mark={arr,e},ind=$y$]++(90:0.3);
\clip (0,0)rectangle (3,2);
\foreach \y in {0,...,6}{
\foreach \x in {0,...,9}{
\draw ($\x*(0:0.8)+\y*(120:0.8)$)--(${\x+1}*(0:0.8)+\y*(120:0.8)$);
\draw ($\x*(0:0.8)+\y*(120:0.8)$)--(${\x+1}*(0:0.8)+{\y+1}*(120:0.8)$);
\draw (${\x+1}*(0:0.8)+\y*(120:0.8)$)--(${\x+1}*(0:0.8)+{\y+1}*(120:0.8)$);
\atoms{vertex}{\x\y/p={$\x*(0:0.8)+\y*(120:0.8)$}}
\draw[opacity=0.3,line width=8,line cap=round] ($\x*(0:0.8)+\y*(120:0.8)+(-90:0.45)$)--($\x*(0:0.8)+\y*(120:0.8)+(90:0.45)$);
}
}
\end{tikzpicture}\;.
\end{equation}
In the spacetime cellulation, we get a pair of triangle prisms adjacent to each $t$ edge.
We simply split up each 6-valent $t$ edge into two 3-valent edges $a$ and $b$, such that the two adjacent prisms become one single volume,
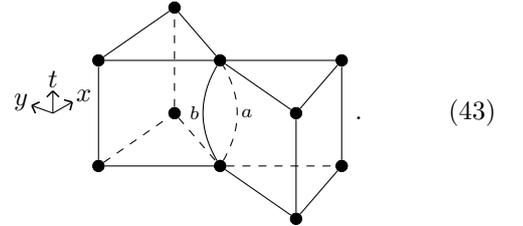
\begin{equation}
\label{eq:subsys_volume}
\begin{tikzpicture}
\draw (-0.6,0.7)edge[mark={arr,e},ind=$x$]++(30:0.3) (-0.6,0.7)edge[mark={arr,e},ind=$y$]++(160:0.3) (-0.6,0.7)edge[mark={arr,e},ind=$t$]++(90:0.3);
\atoms{vertex}{{0/}, {2/p={1.6,0}}, 1/p={1,0.7}, 3/p={3.2,0}, 4/p={2.6,-0.7}}
\atoms{vertex}{0x/p={0,1.4}, 2x/p={1.6,1.4}, 1x/p={1,2.1}, 3x/p={3.2,1.4}, 4x/p={2.6,0.7}}
\draw[dashed] (0)--(1) (1)--(2) (2)--(3);
\draw[dashed]  (1)--(1x) (2)edge[dashed,bend right,mark={slab=$\scriptstyle a$,r}](2x);
\draw (0)--(2) (0x)--(1x) (1x)--(2x) (0x)--(2x) (2)--(4)--(3) (2x)--(4x)--(3x) (2x)--(3x);
\draw (3)--(3x) (4)--(4x) (0)--(0x) (2)edge[bend left,mark={slab=$\scriptstyle b$}](2x);
\end{tikzpicture}\;.
\end{equation}
As for the stabilizer toric code we split each rectangle face into two triangles.
Then we define $V_1^a$ and $V_1^b$ as the operators corresponding to $a$ and $b$ together with the three adjacent triangles, for example
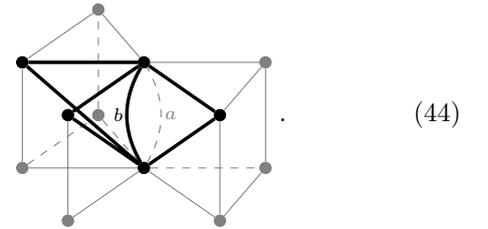
\begin{equation}
\begin{tikzpicture}
\atoms{vertex}{{0/style=gray}, {2/p={1.6,0}}, {1/p={1,0.7},style=gray}, {3/p={3.2,0},style=gray}, {4/p={2.6,-0.7},style=gray}, {5/p={0.6,-0.7},style=gray}}
\atoms{vertex}{0x/p={0,1.4}, 2x/p={1.6,1.4}, {1x/p={1,2.1},style=gray}, {3x/p={3.2,1.4},style=gray}, 4x/p={2.6,0.7}, 5x/p={0.6,0.7}}
\draw[gray,dashed]  (1)--(1x) (2)edge[dashed,bend right,mark={slab=$\scriptstyle a$,r}](2x) (0)--(1) (1)--(2) (2)--(3);
\draw[gray] (0)--(2) (0x)--(1x) (1x)--(2x) (2)--(4)--(3) (4x)--(3x) (2x)--(3x)  (3)--(3x) (4)--(4x) (0)--(0x) (2)--(5) (5)--(5x);
\draw[line width=1.3] (0x)--(2x) (2x)--(4x) (2)--(0x) (2)--(4x) (2)edge[bend left,mark={slab=$\scriptstyle b$}](2x) (2)--(5x) (2x)--(5x);
\end{tikzpicture}\;.
\end{equation}
We also introduce versions $V^a_e$ or $V^b_e$ where $a$ or $b$ carries an $e$ anyon, and the according instruments $I[\mathbf V^a]$, $I[\mathbf V^b]$.

In total, the QEC circuit consists of $X_0X_1X_2$ and $Z_0Z_1Z_2$ measurements on different triples of qubits, which are located at the edges of a triangular lattice.
After drawing a square lattice (black) over the triangular lattice (gray) as follows,
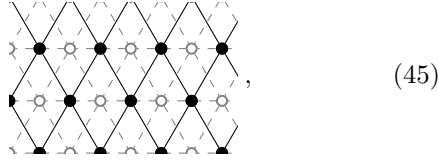
\begin{equation}
\begin{tikzpicture}
\clip (0,0)rectangle (3,2);
\foreach \y in {0,...,6}{
\foreach \x in {0,...,9}{
\draw[gray,dashed] ($\x*(0:0.8)+\y*(120:0.8)$)--(${\x+1}*(0:0.8)+\y*(120:0.8)$);
\draw[gray,dashed] ($\x*(0:0.8)+\y*(120:0.8)$)--(${\x+1}*(0:0.8)+{\y+1}*(120:0.8)$);
\draw[gray,dashed] (${\x+1}*(0:0.8)+\y*(120:0.8)$)--(${\x+1}*(0:0.8)+{\y+1}*(120:0.8)$);
\atoms{circ,tiny,bdastyle=gray}{\x\y/p={$\x*(0:0.8)+\y*(120:0.8)$}}
\atoms{vertex}{\x\y0/p={$\x*(0:0.8)+(180:0.4)+\y*(120:0.8)$}}
\draw ($\x*(0:0.8)+(180:0.4)+\y*(120:0.8)$)--++(60:0.8) ($\x*(0:0.8)+(180:0.4)+\y*(120:0.8)$)--++(120:0.8);
}
}
\end{tikzpicture}\;,
\end{equation}
we recover the subsystem code as presented in Ref.~\cite{Bravyi2012} with qubits on the edges and vertices, and measurements at the corners.

The general decoding procedure of Proposition~\ref{prop:path_integral_decoding} is similar to the stabilizer toric code.
The spacetime lattice in which we find a minimum-weight fix of the $e$ 1-chain ($m$ 2-cochain) is now the modified lattice with volumes as in Eq.~\eqref{eq:subsys_volume}.
The spatial slice of the lattice is a triangular lattice, and the correction works in the same way.
The crucial qualitative difference to the toric code is that the edges where $V_e$ measurements are performed come in $a,b$ pairs forming little loops that support small 1-cycles.
Thus even in the absence of noise measurements are non-deterministic and the results $(x,y)$ and $(x+1,y+1)$ (mod 2) at an $a,b$ pair both occur with probability $\frac12$.
If we start with the ground state, then the measured $e$ 1-cycle is a random subset of $a,b$-loops.

\subsection{CSS honeycomb Floquet code}
\label{sec:css_floquet}
As a third example, we consider the recently discovered CSS honeycomb Floquet code \cite{Kesselring2022,Davydova2022,Aasen2022}.
This code is defined on a hexagonal lattice where each face is colored such that each vertex is adjacent to one red (r), one green (g), and one blue (b) face.
There is one qubit at every vertex.
The code is defined by a dynamic protocol consisting of 6 rounds of measurements that we label rX, gZ, bX, rZ, gX, bZ.
In the round labelled ``fG'', we measure the operator $G_0G_1$ on the two qubits at the vertices of each edge whose two adjacent faces are not colored ``f''.
Note that in every round, each qubit is involved in one 2-body measurement.
To obtain this code from our path-integral picture, we start with the toric code path integral on a cubic spacetime lattice, just as for the stabilizer toric code.
The only difference is that instead of $z$, we choose $t=x+y+z$ as the time direction,
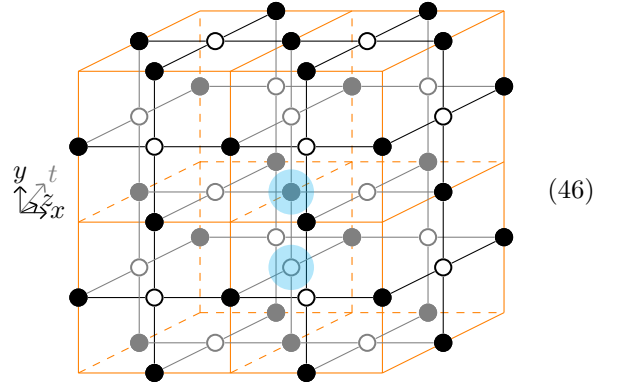
\begin{equation}
\begin{tikzpicture}
\draw (-0.8,0.3)edge[mark={arr,e},ind=$x$]++(0:0.35) (-0.8,0.3)edge[mark={arr,e},ind=$z$]++(25:0.25) (-0.8,0.3)edge[mark={arr,e},ind=$y$]++(90:0.35) (-0.8,0.3)edge[gray,mark={arr,e},ind=$t$]++(50:0.5);
\end{tikzpicture}
\begin{tikzpicture}
\atoms{void}{x/p={2,0}, y/p={0,2}, z/p={1.6,0.8}}
\foreach \x in {0,1,2}{
\foreach \y in {0,1,2}{
\foreach \z in {0,1}{
\atoms{void}{\x\y\z/p={$\x*(x)+\y*(y)+\z*(z)$}}
}}}
\draw[orange] (000)--(200) (010)--(210) (020)--(220) (000)--(020) (100)--(120) (200)--(220) (201)--(221) (021)--(221);
\draw[orange,dashed] (001)--(201) (011)--(211) (001)--(021) (101)--(121);
\draw[orange, dashed] (000)--(001) (100)--(101) (010)--(011) (110)--(111);
\draw[orange] (200)--(201) (210)--(211) (220)--(221) (020)--(021) (120)--(121);
\foreach \x/\y/\z in {1/0/0, 3/0/0, 1/2/0, 3/2/0, 1/4/0, 3/4/0, 0/1/0, 0/3/0, 2/1/0, 2/3/0, 4/1/0, 4/3/0, 4/0/1, 4/2/1, 0/4/1, 2/4/1, 4/4/1, 4/1/2, 4/3/2, 1/4/2, 3/4/2}{
\atoms{delta}{d\x\y\z/p={$0.5*\x*(x)+0.5*\y*(y)+0.5*\z*(z)$}}
}
\foreach \x/\y/\z in {1/0/2, 3/0/2, 1/2/2, 3/2/2, 0/1/2, 0/3/2, 2/1/2, 2/3/2, 2/2/1, 0/2/1, 0/0/1, 2/0/1}{
\atoms{delta,astyle=gray}{d\x\y\z/p={$0.5*\x*(x)+0.5*\y*(y)+0.5*\z*(z)$}}
}
\foreach \x/\y/\z in {1/1/0, 3/1/0, 1/3/0, 3/3/0, 1/4/1, 3/4/1, 4/1/1, 4/3/1}{
\atoms{z2}{z\x\y\z/p={$0.5*\x*(x)+0.5*\y*(y)+0.5*\z*(z)$}}
}
\foreach \x/\y/\z in {1/1/2, 3/1/2, 1/3/2, 3/3/2, 1/2/1, 3/2/1, 2/1/1, 2/3/1, 0/1/1, 0/3/1, 1/0/1, 3/0/1}{
\atoms{z2,astyle=gray}{z\x\y\z/p={$0.5*\x*(x)+0.5*\y*(y)+0.5*\z*(z)$}}
}
\foreach \x/\xx/\xxx in {0/1/2,2/3/4}{
\foreach \y/\yy/\yyy in {0/1/2,2/3/4}{
\foreach \z in {0}{
\draw (z\xx\yy\z)--(d\x\yy\z) (z\xx\yy\z)--(d\xxx\yy\z) (z\xx\yy\z)--(d\xx\y\z) (z\xx\yy\z)--(d\xx\yyy\z);
}
\foreach \z in {2}{
\draw[gray] (z\xx\yy\z)--(d\x\yy\z) (z\xx\yy\z)--(d\xxx\yy\z) (z\xx\yy\z)--(d\xx\y\z) (z\xx\yy\z)--(d\xx\yyy\z);
}}}
\foreach \x/\xx/\xxx in {0/1/2,2/3/4}{
\foreach \y in {0,2}{
\draw[gray] (z\xx\y1)--(d\x\y1) (z\xx\y1)--(d\xxx\y1) (z\xx\y1)--(d\xx\y0) (z\xx\y1)--(d\xx\y2);
}
\foreach \y in {4}{
\draw (z\xx\y1)--(d\x\y1) (z\xx\y1)--(d\xxx\y1) (z\xx\y1)--(d\xx\y0) (z\xx\y1)--(d\xx\y2);
}}
\foreach \y/\yy/\yyy in {0/1/2,2/3/4}{
\foreach \x in {0,2}{
\draw[gray] (z\x\yy1)--(d\x\y1) (z\x\yy1)--(d\x\yyy1) (z\x\yy1)--(d\x\yy0) (z\x\yy1)--(d\x\yy2);
}
\foreach \x in {4}{
\draw (z\x\yy1)--(d\x\y1) (z\x\yy1)--(d\x\yyy1) (z\x\yy1)--(d\x\yy0) (z\x\yy1)--(d\x\yy2);
}}

\fill[cyan,opacity=0.3] (d221)circle(0.3) (z211)circle(0.3);
\end{tikzpicture}
\end{equation}
The operators of the circuit are now individual tensors at the edges and faces as marked in blue above.
Traversing the path integral in the $t$ direction gives a natural direction to each tensor, acting as 2-qubit operators
\begin{equation}
\begin{tikzpicture}
\end{tikzpicture}
T_1\coloneqq
\begin{tikzpicture}
\atoms{void}{0/, 1/p={0,1}}
\draw[orange] (0)--(1);
\fill[orange,opacity=0.2] (0-c)--(1-c)--++(-45:0.4)--++(-90:1)--cycle;
\fill[orange,opacity=0.2] (0-c)--(1-c)--++(-135:0.4)--++(-90:1)--cycle;
\fill[orange,opacity=0.2] (0-c)--(1-c)--++(45:0.6)--++(-90:1)--cycle;
\fill[orange,opacity=0.2] (0-c)--(1-c)--++(135:0.6)--++(-90:1)--cycle;
\atoms{delta}{d/p={0,0.5}}
\draw (d)--++(-45:0.6) (d)--++(-135:0.6) (d)--++(45:0.6) (d)--++(135:0.6);
\end{tikzpicture}
,\qquad 
V_1\coloneqq
\begin{tikzpicture}
\atoms{void}{0/, 1/p={1.4,0}, 2/p={0.7,-0.4}, 3/p={0.7,0.4}}
\draw[orange] (0)--(2)--(1)--(3)--(0);
\atoms{z2}{z/p={0.7,0}}
\draw (z)--++(-45:0.6) (z)--++(-135:0.6) (z)--++(45:0.6) (z)--++(135:0.6);
\end{tikzpicture}\;.
\end{equation}
Neither $T_1$ nor $V_1$ are unitaries.
In fact they are projectors onto the $+1$ subspace of 2-qubit operators $Z_0Z_1$ and $X_0X_1$.
Our construction proceeds by defining versions $T_m$ and $V_e$ of these operators including an anyon worldline segment.
To this end, we slightly modify the cubic lattice.
We split each face into two triangles by diagonal 2-valent edges along the $x+y$, $x+z$, or $y+z$ direction, respectively.
Dually, we split each 4-valent edge into two 3-valent edges separated by a 2-gon, such that the 2-gons are perpendicular to the $x+y$, $x+z$, and $y+z$ directions.
A volume of this slightly modified cubic lattice thus looks like
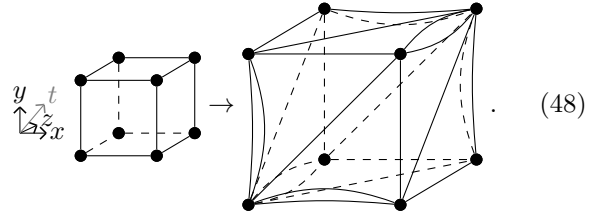
\begin{equation}
\begin{tikzpicture}
\draw (-0.8,0.3)edge[mark={arr,e},ind=$x$]++(0:0.35) (-0.8,0.3)edge[mark={arr,e},ind=$z$]++(25:0.25) (-0.8,0.3)edge[mark={arr,e},ind=$y$]++(90:0.35) (-0.8,0.3)edge[gray,mark={arr,e},ind=$t$]++(50:0.5);
\atoms{vertex}{{0/}, {1/p={1,0}}, {2/p={0.5,0.3}}, {3/p={1.5,0.3}}}
\atoms{vertex}{{0x/p={0,1}}, {1x/p={1,1}}, {2x/p={0.5,1.3}}, {3x/p={1.5,1.3}}}
\draw[dashed] (0)--(2) (2x)--(2)--(3);
\draw (0)--(1) (0)--(0x) (0x)--(1x)--(1)--(3) (3)--(3x) (2x)--(3x) (1x)--(3x) (0x)--(2x);
\end{tikzpicture}
\rightarrow
\begin{tikzpicture}
\atoms{vertex}{{0/}, {1/p={2,0}}, {2/p={1,0.6}}, {3/p={3,0.6}}}
\atoms{vertex}{{0x/p={0,2}}, {1x/p={2,2}}, {2x/p={1,2.6}}, {3x/p={3,2.6}}}
\draw[dashed] (0)to[bend left=20](2) (0)to[bend right=20](2) (2x)--(2)--(3) (3)to[bend left=20](3x) (2x)to[bend right=20](3x);
\draw[dashed] (0)--(3) (0)--(2x) (2)--(3x);
\draw (0)to[bend left=5](1) (0)to[bend left=20](1) (0)to[bend right=5](0x) (0)to[bend right=20](0x) (2x)--(0x)--(1x)--(1)--(3) (3)to[bend left=5](3x) (2x)to[bend right=5](3x) (1x)to[bend left=20](3x) (1x)to[bend right=20](3x);
\draw (0)--(1x) (0x)--(3x) (1)--(3x);
\end{tikzpicture}\;.
\end{equation}
Note that each edge of the cube gives rise to a 2-gon face, but this 2-gon is not always part of the boundary of the modified cube.
In this modified lattice, $T_1$ now corresponds to a 2-gon together with the two adjacent 3-valent edges.
$T_m$ is the same with an $m$ anyon worldline perpendicular to the 2-gon,
\begin{equation}
\label{eq:css_tm}
T_m\coloneqq
\begin{tikzpicture}
\atoms{void}{a/p={0.1,-0.8}, b/p={-0.1,0.8}}
\draw[orange] (a)to[bend left=50](b) (a)to[bend right=50](b);
\draw[worldline] (-100:0.4)--(80:0.4);
\atoms{z2,bdastyle=red}{0/}
\atoms{delta}{1/p={0.35,0}, 2/p={-0.35,0}}
\draw (0)--(1) (0)--(2) (1)--++(-60:0.4) (1)--++(60:0.4) (2)--++(-120:0.4) (2)--++(120:0.4);
\end{tikzpicture}\;.
\end{equation}
Together with $T_1$, we obtain an isometry $\mathbf T$,
\begin{equation}
\label{eq:floquet_t}
\mathbf T
\coloneqq
(T_1,T_m)
=
\begin{tikzpicture}
\atoms{z2}{0/}
\atoms{delta}{1/p={0.5,0}, 2/p={-0.5,0}}
\draw (0)--(1) (0)--(2) (1)--++(-60:0.4) (1)--++(60:0.4) (2)--++(-120:0.4) (2)--++(120:0.4);
\draw[classical] (0)--++(90:0.8);
\end{tikzpicture}\;.
\end{equation}
The according instrument $I[\mathbf T]$ is just a $Z_0Z_1$ measurement. 

Dually, $V_1$ now consists of a 2-valent diagonal edge together with the two adjacent triangles.
$V_e$ is the same with an $e$ anyon worldline segment along the diagonal edge,
\begin{equation}
\label{eq:css_ve}
V_e
\coloneqq
\begin{tikzpicture}
\atoms{void}{a/, b/p={2,0}, c/p={1,-0.7}, d/p={1,0.7}}
\draw[orange] (c)--(d) (a)--(c)--(b)--(d)--(a);
\draw[worldline] (c)--(d);
\atoms{delta,bdastyle=red}{0/p={1,0}}
\atoms{z2}{1/p={1.5,0}, 2/p={0.5,0}}
\draw (0)--(1) (0)--(2) (1)--++(-60:0.4) (1)--++(60:0.4) (2)--++(-120:0.4) (2)--++(120:0.4);
\end{tikzpicture}\;,
\end{equation}
yielding an isometry
\begin{equation}
\label{eq:floquet_v}
\mathbf V
\coloneqq
(V_1,V_e)
=
\begin{tikzpicture}
\atoms{hadamard}{h/p={0,0.5}}
\atoms{delta}{0/}
\atoms{z2}{1/p={0.5,0}, 2/p={-0.5,0}}
\draw (0)--(1) (0)--(2) (1)--++(-60:0.4) (1)--++(60:0.4) (2)--++(-120:0.4) (2)--++(120:0.4);
\draw[classical] (0)--(h) (h)--++(90:0.4);
\end{tikzpicture}\;.
\end{equation}
The according instrument $I[\mathbf V]$ is a $X_0X_1$ measurement. 

In principle, the spacetime cellulation fully specifies the combinatorics of the circuit formed by the instruments defined above.
However, it is instructive to express the circuit in a more conventional form as a sequence of measurements acting on qubits located on a fixed spatial lattice.
We start by decomposing the circuit into rounds of operators acting in parallel.
Within one $t=x+y+z$ period there are three different levels of vertices, which we will label $\vca0$/red, $\vcb1$/green, and $\vcc2$/blue, respectively,
\begin{equation}
\begin{tikzpicture}
\draw (-0.8,0.3)edge[mark={arr,e},ind=$x$]++(0:0.35) (-0.8,0.3)edge[mark={arr,e},ind=$z$]++(25:0.25) (-0.8,0.3)edge[mark={arr,e},ind=$y$]++(90:0.35) (-0.8,0.3)edge[gray,mark={arr,e},ind=$t$]++(50:0.5);
\atoms{vertex}{{0/,vca}, {1/p={1,0},vcb}, {2/p={0.5,0.3},vcb}, {3/p={1.5,0.3},vcc}}
\atoms{vertex}{{0x/p={0,1},vcb}, {1x/p={1,1},vcc}, {2x/p={0.5,1.3},vcc}, {3x/p={1.5,1.3},vca}}
\draw[dashed] (0)--(2) (2x)--(2)--(3);
\draw (0)--(1) (0)--(0x) (0x)--(1x)--(1)--(3) (3)--(3x) (2x)--(3x) (1x)--(3x) (0x)--(2x);
\end{tikzpicture}
\end{equation}
Accordingly, there are three levels of edges, $\vca0\vcb1$, $\vcb1\vcc2$, and $\vcc2\vca0$, and three levels of faces, $\vca0\vcb1\vcc2$, $\vcb1\vcc2\vca0$, and $\vcc2\vca0\vcb1$ (though the cube above only contains faces from two of the levels).
So one $t$ period of the circuit consists of 6 rounds of instruments:
\begin{equation}
\label{eq:floquet_circuit}
\begin{multlined}
\rightarrow I[\mathbf T]_{\vca0\vcb1} \rightarrow I[\mathbf V]_{\vca0\vcb1\vcc2} \rightarrow I[\mathbf T]_{\vcb1\vcc2}\\
\rightarrow I[\mathbf V]_{\vcb1\vcc2\vca0} \rightarrow I[\mathbf T]_{\vcc2\vca0} \rightarrow I[\mathbf V]_{\vcc2\vca0\vcb1}\rightarrow\;.
\end{multlined}
\end{equation}
An appropriate spatial lattice on which the circuit acts can be obtained by projecting the 3-dimensional cubic lattice along the $t$ direction.
This yields a 2-dimensional regular triangular lattice such that the vertices of each triangle have different numbers/colors,
\begin{equation}
\label{eq:triangular_projection}
\begin{tikzpicture}
\clip (0,0)rectangle (4.5,3);
\foreach \y in {0,...,6}{
\foreach \x in {0,...,9}{
\draw ($\x*(0:0.8)+\y*(120:0.8)$)--(${\x+1}*(0:0.8)+\y*(120:0.8)$);
\draw ($\x*(0:0.8)+\y*(120:0.8)$)--(${\x+1}*(0:0.8)+{\y+1}*(120:0.8)$);
\draw (${\x+1}*(0:0.8)+\y*(120:0.8)$)--(${\x+1}*(0:0.8)+{\y+1}*(120:0.8)$);
}
}
\foreach \su in {0,3,6,9}{
\foreach \x in {0,...,9}{
\atoms{vertex,vca}{\x\su/p={$(0.8*\x,0)+{\su-\x}*(120:0.8)$}}
}
}
\foreach \su in {1,4,7,10}{
\foreach \x in {0,...,9}{
\atoms{vertex,vcb}{\x\su/p={$(0.8*\x,0)+{\su-\x}*(120:0.8)$}}
}
}
\foreach \su in {2,5,8,11}{
\foreach \x in {0,...,9}{
\atoms{vertex,vcc}{\x\su/p={$(0.8*\x,0)+{\su-\x}*(120:0.8)$}}
}
}
\end{tikzpicture}\;.
\end{equation}
The spacetime faces become rhombi in this spatial lattice, consisting of two triangles.

Each qubit corresponds to a time-like continued string of bonds in the tensor network/circuit diagram.
The goal is to arrive at a circuit that consists only of 2-body measurements without any swap operations.
This fully determines the time-like strings by the way the inputs and outputs are paired in Eq.~\eqref{eq:floquet_t} and Eq.~\eqref{eq:floquet_v}.
Geometrically, these time-like strings are sequences of adjacent faces and edges.
In the space projection, there is one such sequence for every triangle $F$ as follows,
\begin{equation}
\begin{tikzpicture}
\atoms{void}{x/p={0:1.2}, y/p={60:1.2}}
\atoms{vertex}{0/vca, {1/vcb,p=x}, {2/vcc,p=y}, {3/vcc,p={$(x)-(y)$}}, {4/vca,p={$(x)+(y)$}}, {5/vcb,p={$(y)-(x)$}}}
\draw (0)edge[mark={lab=$a$}](1) (0)edge[mark={lab=$c$}](2) (1)edge[mark={lab=$e$}](2);
\draw (0)--(3)--(1)--(4)--(2)--(5)--(0);
\node at ($0.5*(x)+(-90:0.3)$){$d$};
\node at ($(x)+0.5*(y)-0.5*(x)+(30:0.3)$){$b$};
\node at ($0.5*(y)+(150:0.3)$){$f$};
\node at ($1/3*(x)+1/3*(y)$){$F$};
\end{tikzpicture}\;.
\end{equation}
Here the labels $a$, $c$, $e$ correspond to projections of edges, and the labels $b$, $d$, $f$ at triangles correspond to projections of faces formed by this triangle together with $F$.
Then the sequence $a-b-c-d-e-f$ is the time-like string within one $t$ period.

As we have seen, there is one qubit associated to each triangle.
For each edge, the instrument $I[\mathbf T]$ acts on the qubits at the two triangles adjacent to its projection.
For each face, $I[\mathbf V]$ acts on the qubits at the two triangles contained in its projection.
Note that the instruments $I[\mathbf T]_{\vca0\vcb1}$ act on the same pairs of qubits as the instruments $I[\mathbf V]_{\vcb1\vcc2\vca0}$, and analogous for cyclic permutation of the numbers/colors.
Taking into account that $I[\mathbf T]$ is a $Z_0Z_1$ measurement and $I[\mathbf V]$ is a $X_0X_1$ measurement, we can rewrite Eq.~\eqref{eq:floquet_circuit} as
\begin{equation}
\begin{multlined}
\rightarrow ZZ_{\vca0\vcb1} \rightarrow XX_{\vcc2\vca0}\rightarrow ZZ_{\vca1\vcb2}
\\
\rightarrow XX_{\vcc0\vca1} \rightarrow ZZ_{\vca2\vcb0} \rightarrow XX_{\vcc1\vca2} \rightarrow\;.
\end{multlined}
\end{equation}
After going to the dual hexagonal lattice, we recover the CSS honeycomb Floquet code as introduced in Refs.~\cite{Kesselring2022,Davydova2022,Aasen2022}.

Let us briefly describe the general decoding procedure in Proposition~\ref{prop:path_integral_decoding} for the present code.
The spacetime lattice in which we fix the measured $e$ 1-chain ($m$ 2-cochain) is the rotated modified cubic lattice.
If we choose the correction time $T$ after the $I[\mathbf T]_{\vca0\vcb1}$ measurements, the spatial slice of the lattice looks like
\begin{equation}
\label{eq:css_spatial_slice}
\begin{tikzpicture}
\clip (0.4,0)rectangle (4, 2*0.8*1.732);
\foreach \su in {0,3,6,9}{
\foreach \x in {0,...,9}{
\atoms{vertex,vca}{\x\su/p={$(0.8*\x,0)+{\su-\x}*(120:0.8)$}}
}
}
\foreach \su in {1,4,7,10}{
\foreach \x in {0,...,9}{
\atoms{vertex,vcb}{\x\su/p={$(0.8*\x,0)+{\su-\x}*(120:0.8)$}}
}
}
\draw (00)to[bend left](11)(11)to[bend left](23)(23)to[bend left](24)(24)to[bend left](13) (23)to[bend left](34)(34)to[bend left](33) (34)to[bend left](46)(46)to[bend left](57)(57)to[bend left](56)(56)to[bend left](44)(44)to[bend left](33) (56)to[bend left](67) (57)to[bend left](69)(69)to[bend left](710) (46)to[bend left](47)(47)to[bend left](59)(59)to[bend left](610)(610)to[bend left](69) (24)to[bend left](36)(36)to[bend left](37)(37)to[bend left](26) (36)to[bend left](47);
\draw (00)to[bend right](11)(11)to[bend right](23)(23)to[bend right](24)(24)to[bend right](13) (23)to[bend right](34)(34)to[bend right](33) (34)to[bend right](46)(46)to[bend right](57)(57)to[bend right](56)(56)to[bend right](44)(44)to[bend right](33) (56)to[bend right](67) (57)to[bend right](69)(69)to[bend right](710) (46)to[bend right](47)(47)to[bend right](59)(59)to[bend right](610)(610)to[bend right](69) (24)to[bend right](36)(36)to[bend right](37)(37)to[bend right](26) (36)to[bend right](47);
\end{tikzpicture}\;.
\end{equation}
The fixed spacetime $e$ 1-cycle ($m$ 2-cocycle) terminates at a 0-cycle (2-cocycle) on this spatial lattice.
It is closed in a homologically trivial way by inserting 2-valent edges (2-gon faces) potentially carrying $e$ ($m$) anyon worldlines similar to Eq.~\eqref{eq:toric_code_correction}.
The static toric code on this spatial lattice also coincides with the \emph{instantaneous stabilizer group} of the code at time $T$.
In contrast to the stabilizer and subsystem toric code, the edges (dual edges) where measurements potentially yield $e$ ($m$) anyon worldlines are not aligned with the $t$ direction.
Furthermore, the graph formed by these edges (dual edges) is much more connected.
In the absence of noise, any 1-cycle (2-cocycle) supported these edges (dual edges) is measured with equal probability.
So as for the subsystem toric code the measurement results are non-deterministic, but now they are even more fluctuating and may include homologically non-trivial loops.
This is not a problem for decoding though, since these homologically non-trivial loops are recorded and can be corrected.
The following three pictures illustrate a typical syndrome in the absence of noise on a spacetime patch for the toric code, subsystem toric code, and CSS honeycomb Floquet code,
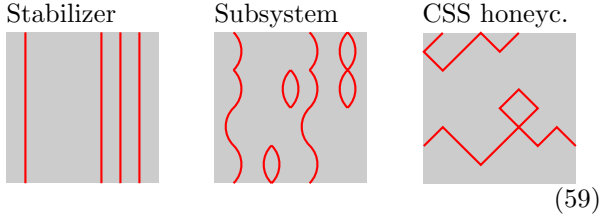
\begin{equation}
\begin{tabular}{lll}
Stabilizer & Subsystem & CSS honeyc.\\
\begin{tikzpicture}
\fill[black!20] (0,0)rectangle(2,2);
\draw[red,thick] (0.25,0)--++(90:2) (1.25,0)--++(90:2) (1.5,0)--++(90:2) (1.75,0)--++(90:2);
\end{tikzpicture}
\hspace{0.2cm}
&
\begin{tikzpicture}
\fill[black!20] (0,0)rectangle(2,2);
\draw[red,thick] (0.25,0)to[bend right=45]++(90:0.5) (0.25,0.5)to[bend left=45]++(90:0.5) (0.25,1)to[bend right=45]++(90:0.5) (0.25,1.5)to[bend right=45]++(90:0.5);
\draw[red,thick] (1.25,0)to[bend right=45]++(90:0.5) (1.25,0.5)to[bend left=45]++(90:0.5) (1.25,1)to[bend right=45]++(90:0.5) (1.25,1.5)to[bend right=45]++(90:0.5);
\draw[red,thick] (1,1)to[bend right=45]++(90:0.5) (1,1)to[bend left=45]++(90:0.5) (0.75,0)to[bend right=45]++(90:0.5) (0.75,0)to[bend left=45]++(90:0.5) (1.75,1.5)to[bend right=45]++(90:0.5) (1.75,1.5)to[bend left=45]++(90:0.5) (1.75,1)to[bend right=45]++(90:0.5) (1.75,1)to[bend left=45]++(90:0.5);
\end{tikzpicture}
\hspace{0.2cm}
&
\begin{tikzpicture}
\fill[black!20] (0,0)rectangle(2,2);
\draw[red,thick] (0,0.5)--++(0.25,0.25)--++(0.5,-0.5)--++(0.5,0.5)--++(0.25,-0.25)--++(0.25,0.25)--++(0.25,-0.25);
\draw[red,thick] (1.25,0.75)--++(0.25,0.25)--++(-0.25,0.25)--++(-0.25,-0.25)--++(0.25,-0.25);
\draw[red,thick] (0.25,2)--++(-0.25,-0.25)--++(0.25,-0.25)--++(0.5,0.5)--++(0.25,-0.25)--++(0.25,0.25);
\end{tikzpicture}
\end{tabular}
\end{equation}
The syndrome is in red and boundary conditions are periodic with left and right identified.
The illustrations differ from actual syndromes in that the spacetime dimension is 2 instead of 3, the underlying lattice is not shown, and we only show one part (either $e$ or $m$) of the syndrome.
Note that if we start in the code space (with empty syndrome at the bottom boundary), the stabilizer toric code syndrome will be empty, the subsystem toric code syndrome consists of a random set of ``bubbles'', and the CSS honeycomb Floquet code syndrome can contain homologically non-trivial loops as shown.

We have seen that the CSS honeycomb Floquet code and the stabilizer toric code are both based on the cubic-lattice toric code path integral, but with different time directions.
If we superimpose the two cubic lattices such that the time directions align, the path integrals are different.
Nonetheless, they are in the same fixed-point phase as defined at the end of Section~\ref{sec:fixed_points}.
The tensor-network equations applied to get from the cubic lattice to the rotated cubic lattice are the equations imposing topological invariance, for example, Eq.~\eqref{eq:22pachner} and Eq.~\eqref{eq:bialgebra_move}.
So the time evolutions of the two codes postselected to the trivial $0$ (or $+1\in \{\pm 1\}$) measurement outcomes are locally equivalent.
In both codes, the non-trivial $1$ (or $-1\in \{\pm 1\}$) outcomes correspond to $e$ or $m$ anyon worldline segments.
However, the positions of these segments in spacetime are different for the two codes.
The subsystem toric code, as well as the honeycomb Floquet code discussed in the following section, are related in the same way.

\subsection{Honeycomb Floquet code}
\label{sec:honeycomb_floquet}
In this section, we consider the honeycomb Floquet code introduced in Ref.~\cite{Hastings2021}.
This is a code defined on a hexagonal lattice whose faces are rgb-colored like for the CSS honeycomb Floquet code.
The code is defined by a QEC protocol consisting of three rounds of measurements that we label r, g, and b.
In the round labeled ``f'', we apply a 2-body measurement involving the two qubits at the vertices of every edge whose two adjacent faces are not colored ``f''.
The type of measurement depends on the direction of the edge:
The edges of a regular hexagonal lattice point in three different directions, and the performed measurement is $XX$, $YY$, or $ZZ$ depending on this direction.
The underlying tensor-network path integral will be referred to as the \emph{honeycomb path integral}.
It has the same geometry as the cubic-lattice toric-code path integral used in previous sections.
However, it involves a third kind of tensor,
\begin{equation}
\begin{multlined}
\begin{tikzpicture}
\atoms{cc}{0/}
\draw (0)edge[mark={arr,s,f}, ind=$a$]++(0:0.5) (0)edge[mark={arr,s,f}, ind=$b$]++(90:0.5) (0)edge[ind=$c$]++(180:0.5);
\node at (-90:0.5){$\ldots$};
\end{tikzpicture}
\\
=
\begin{cases}
(-1)^{\frac{a+b-c+\ldots}{2}} & \text{if } a+b+c+\ldots=0\mod 2\\
0 & \text{otherwise}
\end{cases}\;.
\end{multlined}
\end{equation}
We will refer to this tensor as \emph{$\cc$-tensor} since it is related to the two-dimensional real algebra of complex numbers.
Note that the tensor depends on a choice of arrow direction at each index, which we indicate by an arrow at the incoming indices.
The honeycomb path integral has $\delta$-tensors at every $z$ edge and every $xy$ face, $\zz_2$-tensors at every $x$ edge and $yz$ face, and $\cc$-tensors at every $y$ edge and $xz$ face,
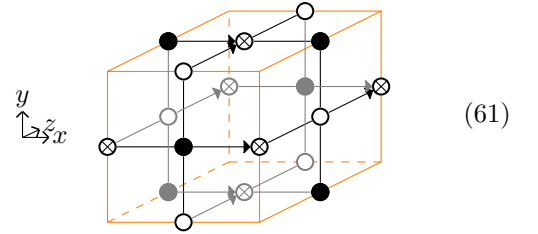
\begin{equation}
\begin{tikzpicture}
\draw (-0.8,0.3)edge[mark={arr,e},ind=$x$]++(0:0.35) (-0.8,0.3)edge[mark={arr,e},ind=$z$]++(25:0.25) (-0.8,0.3)edge[mark={arr,e},ind=$y$]++(90:0.35);
\end{tikzpicture}
\quad
\begin{tikzpicture}
\atoms{void}{x/p={2,0}, y/p={0,2}, z/p={1.6,0.8}}
\foreach \x in {0,1}{
\foreach \y in {0,1}{
\foreach \z in {0,1}{
\atoms{void}{\x\y\z/p={$\x*(x)+\y*(y)+\z*(z)$}}
}}}
\draw[orange] (000)--(100)--(110)--(010)--(000) (100)--(101)--(111)--(110) (010)--(011)--(111);
\draw[orange,dashed] (000)--(001)--(011) (001)--(101);
\foreach \x/\y/\z/\type in {1/0/0/z2, 1/2/0/z2, 0/1/0/cc, 2/1/0/cc, 2/0/1/delta, 0/2/1/delta, 2/2/1/delta, 2/1/2/cc, 1/1/0/delta, 1/2/1/cc, 1/2/2/z2, 2/1/1/z2}{
\atoms{\type}{d\x\y\z/p={$0.5*\x*(x)+0.5*\y*(y)+0.5*\z*(z)$}}
}
\foreach \x/\y/\z/\type in {1/0/2/z2, 0/1/2/cc, 1/1/2/delta, 0/0/1/delta, 1/0/1/cc, 0/1/1/z2}{
\atoms{\type,astyle=gray}{d\x\y\z/p={$0.5*\x*(x)+0.5*\y*(y)+0.5*\z*(z)$}}
}
\draw (d110)--(d100) (d110)edge[mark={arr,f,e}](d210) (d110)--(d120) (d110)--(d010) (d211)--(d201) (d211)--(d210) (d211)edge[mark={arr,f,e}](d212) (d211)--(d221) (d121)--(d221) (d121)edge[mark={arr,f,s}](d120) (d121)edge[mark={arr,f,s}](d021) (d121)--(d122);
\draw[gray] (d112)--(d102) (d112)edge[mark={arr,f,e}](d212) (d112)--(d122) (d112)--(d012) (d011)--(d001) (d011)--(d010) (d011)edge[mark={arr,f,e}](d012) (d011)--(d021) (d101)--(d201) (d101)edge[mark={arr,f,s}](d100) (d101)edge[mark={arr,f,s}](d001) (d101)--(d102);
\end{tikzpicture}
\end{equation}
The arrow directions of the $\cc$-tensors are chosen to point in the positive $x$ and $z$ direction, respectively.

The honeycomb path integral can be decorated with two types of topological syndrome defects, located on 1-cycles and 2-cocycles.
To this end, we define a charged version of the $\cc$-tensor,
\begin{equation}
\begin{multlined}
\begin{tikzpicture}
\atoms{cc,bdastyle=red}{0/}
\draw (0)edge[mark={arr,s,f}, ind=$a$]++(0:0.5) (0)edge[mark={arr,s,f}, ind=$b$]++(90:0.5) (0)edge[ind=$c$]++(180:0.5);
\node at (-90:0.5){$\ldots$};
\end{tikzpicture}
\\
=
\begin{cases}
i^{a+b-c+\ldots} & \text{if } a+b+c+\ldots=1\mod 2\\
0 & \text{otherwise}
\end{cases}\;.
\end{multlined}
\end{equation}
Then, at every edge of the 1-cycle, we replace the associated tensor with the charged version thereof.
We do the same for the faces of the 2-cocycle.
The tensors satisfy tensor-network equations such as:
\begin{equation}
\begin{gathered}
\begin{tikzpicture}
\atoms{cc,bdastyle=red}{0/}
\draw (0)--++(0:0.5) (0)edge[mark={arr,s,f}]++(90:0.5) (0)edge[mark={arr,s,f}]++(180:0.5) (0)--++(-90:0.5);
\end{tikzpicture}
=
\begin{tikzpicture}
\atoms{cc}{0/}
\atoms{cc,bdastyle=red}{1/p=180:0.4}
\draw (0)--++(0:0.5) (0)edge[mark={arr,s,f}]++(90:0.5) (0)--(1) (1)edge[mark={arr,s,f}]++(180:0.4) (0)--++(-90:0.5);
\end{tikzpicture}\;,
\begin{tikzpicture}
\atoms{cc}{0/}
\draw (0)--++(0:0.5) (0)edge[mark={arr,s,f}]++(90:0.5) (0)--++(180:0.4) (0)--++(-90:0.5);
\end{tikzpicture}
=
\begin{tikzpicture}
\atoms{cc}{0/}
\atoms{delta,bdastyle=red}{4/p=180:0.4,1/p=90:0.4, 2/p=-90:0.4, 3/p=0:0.4}
\draw (0)--(3) (3)--++(0:0.4) (0)edge[mark={arr,s,f}](1) (1)--++(90:0.4) (0)--(4) (4)--++(180:0.5) (0)--(2) (2)--++(-90:0.4);
\end{tikzpicture}\;,\\
\begin{tikzpicture}
\atoms{cc,bdastyle=red}{0/}
\atoms{delta,bdastyle=red}{1/p={0.5,0}}
\draw (0)--(1) (1)--++(0:0.5) (0)edge[mark={arr,s,f}]++(180:0.5);
\end{tikzpicture}
=
i\cdot
\begin{tikzpicture}
\atoms{z2,bdastyle=red}{0/}
\draw (0)--++(0:0.5) (0)--++(180:0.5);
\end{tikzpicture}\;,
\begin{tikzpicture}
\atoms{cc,bdastyle=red}{0/, 1/p={0.5,0}}
\draw (0)edge[mark={arr,e,f}](1) (1)--++(0:0.5) (0)edge[mark={arr,s,f}]++(180:0.5);
\end{tikzpicture}
=
\begin{tikzpicture}
\draw (0,0)--++(0:0.8);
\end{tikzpicture}\;.
\end{gathered}
\end{equation}
Using these equations, we can locally change the 1-cycle and 2-cocycle.
For example, we can add the boundary of a $xz$ face to the 1-cocycle as follows,
\begin{equation}
\begin{multlined}
\begin{tikzpicture}
\draw[orange] (-0.6,-0.6)--(-0.6,0.6)--(0.6,0.6)--(0.6,-0.6)--cycle;
\atoms{cc}{0/}
\atoms{z2}{1/p=90:0.6, 2/p=-90:0.6}
\atoms{delta}{3/p=0:0.6,4/p=180:0.6}
\draw (0)--(3) (0)--(1) (0)edge[mark={arr,s,f}](4) (0)edge[mark={arr,s,f}](2);
\draw (1)--++(90:0.3) (1)--++(-140:0.3) (1)--++(40:0.3) (2)--++(-90:0.3) (2)--++(-140:0.3) (2)--++(40:0.3) (3)--++(40:0.3) (3)--++(-140:0.3) (3)--++(0:0.3) (4)--++(40:0.3) (4)--++(-140:0.3) (4)--++(180:0.3);
\end{tikzpicture}
=
\begin{tikzpicture}
\atoms{cc}{0/}
\atoms{z2}{1/p=90:0.8, 2/p=-90:0.8}
\atoms{delta}{3/p=0:0.8, 4/p=180:0.8}
\atoms{delta,bdastyle=red}{z1/p={90:0.4}, z2/p={-90:0.4}, z3/p={0:0.4}, z4/p={180:0.4}}
\draw (0)--(z3)(z3)--(3) (0)--(z1) (z1)--(1) (0)edge[mark={arr,s,f}](z4) (z4)--(4) (0)edge[mark={arr,s,f}](z2) (z2)--(2);
\draw (1)--++(90:0.3) (1)--++(-140:0.3) (1)--++(40:0.3) (2)--++(-90:0.3) (2)--++(-140:0.3) (2)--++(40:0.3) (3)--++(40:0.3) (3)--++(-140:0.3) (3)--++(0:0.3) (4)--++(40:0.3) (4)--++(-140:0.3) (4)--++(180:0.3);
\end{tikzpicture}
=
\begin{tikzpicture}
\atoms{cc}{0/}
\atoms{z2}{1/p=90:1.2, 2/p=-90:1.2}
\atoms{delta}{3/p=0:0.8, 4/p=180:0.8}
\atoms{delta,bdastyle=red}{z1/p={90:0.8}, z2/p={-90:0.8}, z3/p={0:0.4}, z4/p={180:0.4}}
\atoms{cc,bdastyle=red}{c1/p={90:0.4}, c2/p={-90:0.4}}
\draw (0)--(z3) (z3)--(3) (0)edge[mark={arr,e,f}](c1) (c1)--(z1) (z1)--(1) (0)edge[mark={arr,s,f}](z4) (z4)--(4) (0)edge[mark={arr,s,f}](c2) (c2)edge[mark={arr,s,f}](z2) (z2)--(2);
\draw (1)--++(90:0.3) (1)--++(-140:0.3) (1)--++(40:0.3) (2)--++(-90:0.3) (2)--++(-140:0.3) (2)--++(40:0.3) (3)--++(40:0.3) (3)--++(-140:0.3) (3)--++(0:0.3) (4)--++(40:0.3) (4)--++(-140:0.3) (4)--++(180:0.3);
\end{tikzpicture}\\
\propto
\begin{tikzpicture}
\atoms{cc}{0/}
\atoms{z2}{1/p=90:0.8, 2/p=-90:0.8}
\atoms{delta}{3/p=0:0.8, 4/p=180:0.8}
\atoms{z2,bdastyle=red}{z1/p={90:0.4}, z2/p={-90:0.4}}
\atoms{delta,bdastyle=red}{z3/p={0:0.4}, z4/p={180:0.4}}
\draw (0)--(z3)(z3)--(3) (0)--(z1) (z1)--(1) (0)edge[mark={arr,s,f}](z4) (z4)--(4) (0)edge[mark={arr,s,f}](z2) (z2)--(2);
\draw (1)--++(90:0.3) (1)--++(-140:0.3) (1)--++(40:0.3) (2)--++(-90:0.3) (2)--++(-140:0.3) (2)--++(40:0.3) (3)--++(40:0.3) (3)--++(-140:0.3) (3)--++(0:0.3) (4)--++(40:0.3) (4)--++(-140:0.3) (4)--++(180:0.3);
\end{tikzpicture}
=
\begin{tikzpicture}
\draw[orange] (-0.6,-0.6)--(-0.6,0.6)--(0.6,0.6)--(0.6,-0.6)--cycle;
\draw[worldline] (-0.6,-0.6)--(-0.6,0.6)--(0.6,0.6)--(0.6,-0.6)--cycle;
\atoms{cc}{0/}
\atoms{z2,bdastyle=red}{1/p=90:0.6, 2/p=-90:0.6}
\atoms{delta,bdastyle=red}{3/p=0:0.6,4/p=180:0.6}
\draw (0)--(3) (0)--(1) (0)edge[mark={arr,s,f}](4) (0)edge[mark={arr,s,f}](2);
\draw (1)--++(90:0.3) (1)--++(-140:0.3) (1)--++(40:0.3) (2)--++(-90:0.3) (2)--++(-140:0.3) (2)--++(40:0.3) (3)--++(40:0.3) (3)--++(-140:0.3) (3)--++(0:0.3) (4)--++(40:0.3) (4)--++(-140:0.3) (4)--++(180:0.3);
\end{tikzpicture}\;.
\end{multlined}
\end{equation}
Note that global prefactors are irrelevant due to the quantum mechanical interpretation of the path integral.
Also note that the honeycomb path integral and its defect configurations are only defined on a fixed cubic background lattice.

We will now turn the honeycomb path integral into a circuit of instruments using the same time direction as in Section~\ref{sec:css_floquet} when discussing the CSS honeycomb Floquet code.
$\delta$-tensors and $\zz$-tensors become operators $T_1$ and $V_1$ yielding instruments $I[\mathbf T]$ and $I[\mathbf V]$, which are $Z_0Z_1$ and $X_0X_1$ measurements as before.
The operator $W_1$ corresponding to a single $\cc$-tensor,
\begin{equation}
W_1\coloneqq
\begin{tikzpicture}
\atoms{cc}{0/}
\draw (0)edge[mark={arr,s,f}]++(-45:0.6) (0)edge[mark={arr,s,f}]++(-135:0.6) (0)--++(45:0.6) (0)--++(135:0.6);
\end{tikzpicture}
=
\frac12 (1+Y_0Y_1)
\;,
\end{equation}
will be complemented by another tensor
\begin{equation}
\label{eq:yy_measurement}
W_x\coloneqq
\begin{tikzpicture}
\atoms{cc}{0/, 1/p={0.8,0}}
\atoms{delta,bdastyle=red}{d/p={0.4,0}}
\draw (1)edge[mark={arr,s,f}]++(-45:0.6) (0)edge[mark={arr,s,f}]++(-135:0.6) (1)--++(45:0.6) (0)--++(135:0.6) (0)edge[mark={arr,s,f}](d) (d)--(1);
\end{tikzpicture}
=
\frac12 (1-Y_0Y_1)\;,
\end{equation}
such that the instrument $I[\mathbf W]$ defined by $(W_1,W_x)$ is a projective $Y_0Y_1$ measurement.
We now need to turn a configuration of measurement outcomes into a configuration of defects on the cubic lattice.
To this end, we apply tensor-network equations to move the ``charge'' from the middle 2-index tensors in Eq.~\eqref{eq:css_tm}, Eq.~\eqref{eq:css_ve}, or Eq.~\eqref{eq:yy_measurement} to the tensors of the original honeycomb path integral.
For example, for the operator $W_x$, this results in a ``charge'' at the neighboring $\delta$ and $\zz_2$-tensors, but also at the $\cc$-tensor itself,
\begin{equation}
\begin{tikzpicture}
\atoms{cc}{0/p={-0.8,0}, 1/}
\atoms{delta,bdastyle=red}{d/p={-0.4,0}}
\atoms{delta}{a0/p={-45:0.5}}
\atoms{z2}{a1/p={45:0.5}}
\draw (1)edge[mark={arr,s,f}](a0) (0)edge[mark={arr,s,f}]++(-135:0.6) (1)--(a1) (0)--++(135:0.6) (0)edge[mark={arr,s,f}](d) (d)--(1);
\draw (a0)--++(45:0.3) (a0)--++(-45:0.3) (a0)--++(-135:0.3) (a1)--++(45:0.3) (a1)--++(-45:0.3) (a1)--++(135:0.3);
\end{tikzpicture}
=
\begin{tikzpicture}
\atoms{cc}{1/}
\atoms{delta,bdastyle=red}{d0/p=-45:0.4, d1/p={45:0.4}}
\atoms{delta}{a0/p={-45:0.8}}
\atoms{z2}{a1/p={45:0.8}}
\draw (1)edge[mark={arr,s,f}](d0) (d0)--(a0) (1)edge[mark={arr,s,f}]++(-135:0.6) (1)--(d1) (d1)--(a1) (1)--++(135:0.6);
\draw (a0)--++(45:0.3) (a0)--++(-45:0.3) (a0)--++(-135:0.3) (a1)--++(45:0.3) (a1)--++(-45:0.3) (a1)--++(135:0.3);
\end{tikzpicture}
=
\begin{tikzpicture}
\atoms{cc,bdastyle=red}{1/}
\atoms{delta,bdastyle=red}{d1/p={45:0.8}}
\atoms{cc,bdastyle=red}{c1/p={45:0.4}}
\atoms{delta,bdastyle=red}{a0/p={-45:0.5}}
\atoms{z2}{a1/p={45:1.2}}
\draw (1)edge[mark={arr,s,f}](a0) (1)edge[mark={arr,s,f}]++(-135:0.6) (1)edge[mark={arr,e,f}](c1) (c1)--(d1) (d1)--(a1) (1)--++(135:0.6);
\draw (a0)--++(45:0.3) (a0)--++(-45:0.3) (a0)--++(-135:0.3) (a1)--++(45:0.3) (a1)--++(-45:0.3) (a1)--++(135:0.3);
\end{tikzpicture}
\propto
\begin{tikzpicture}
\atoms{cc,bdastyle=red}{1/}
\atoms{delta,bdastyle=red}{a0/p={-45:0.5}}
\atoms{z2,bdastyle=red}{a1/p={45:0.5}}
\draw (1)edge[mark={arr,s,f}](a0) (1)edge[mark={arr,s,f}]++(-135:0.6) (1)--(a1) (1)--++(135:0.6);
\draw (a0)--++(45:0.3) (a0)--++(-45:0.3) (a0)--++(-135:0.3) (a1)--++(45:0.3) (a1)--++(-45:0.3) (a1)--++(135:0.3);
\end{tikzpicture}\;.
\end{equation}
So the $W_x$ measurement outcome corresponds to three different syndrome defect segments being present.
For a $W_x$ operator located at a $xy$ face, we get 1-cycle defect segments on two adjacent edges, but also a 2-cocycle defect at the $xy$ face itself,
\begin{equation}
\begin{tikzpicture}
\draw[orange] (-1,-1)--(-1,1)--(1,1)--(1,-1)--cycle;
\atoms{delta,bdastyle=red}{dx/}
\atoms{cc}{c0/p={135:0.35}, c1/p={-45:0.35}}
\atoms{z2}{1/p=90:1, 2/p=-90:1}
\atoms{delta}{3/p=0:1,4/p=180:1}
\draw (c1)--(3) (c0)--(1) (c0)edge[mark={arr,s,f}](4) (c1)edge[mark={arr,s,f}](2) (dx)edge[mark={arr,e,f}](c0) (dx)--(c1);
\draw (1)--++(90:0.3) (1)--++(-140:0.3) (1)--++(40:0.3) (2)--++(-90:0.3) (2)--++(-140:0.3) (2)--++(40:0.3) (3)--++(40:0.3) (3)--++(-140:0.3) (3)--++(0:0.3) (4)--++(40:0.3) (4)--++(-140:0.3) (4)--++(180:0.3);
\end{tikzpicture}
\propto
\begin{tikzpicture}
\draw[orange] (-1,-1)--(-1,1)--(1,1)--(1,-1)--cycle;
\draw[worldline] (-1,-1)--(1,-1)--(1,1) (-0.8,-0.4)--(0.8,0.4);
\atoms{delta,bdastyle=red}{dx/}
\atoms{cc,bdastyle=red}{c0/}
\atoms{z2}{1/p=90:1, {2/p=-90:1,bdastyle=red}}
\atoms{delta}{{3/p=0:1,bdastyle=red},4/p=180:1}
\draw (c0)--(3) (c0)--(1) (c0)edge[mark={arr,s,f}](4) (c0)edge[mark={arr,s,f}](2) (dx)edge[mark={arr,e,f}](c0) (dx)--(c0);
\draw (1)--++(90:0.3) (1)--++(-140:0.3) (1)--++(40:0.3) (2)--++(-90:0.3) (2)--++(-140:0.3) (2)--++(40:0.3) (3)--++(40:0.3) (3)--++(-140:0.3) (3)--++(0:0.3) (4)--++(40:0.3) (4)--++(-140:0.3) (4)--++(180:0.3);
\end{tikzpicture}\;.
\end{equation}
Note that also in the CSS honeycomb Floquet code in Section~\ref{sec:css_floquet}, a $V_e$ operator at a face corresponds to an $e$ defect segment at a diagonal edge, which is equivalent to $e$ defect segments at two boundary edges like above.
However, in this case there is no $m$ defect segment at the face itself.
Dually, for a $W_x$ measurement outcome at an edge, we obtain 2-cocycle defects at two adjacent faces as well as a 1-cycle defect at the edge itself.
The same is also true for $T_m$ and $V_e$ instead of $W_x$,
\begin{equation}
\begin{tikzpicture}
\atoms{delta}{0/p={-0.8,0}, 1/}
\atoms{z2,bdastyle=red}{d/p={-0.4,0}}
\atoms{cc}{a0/p={-45:0.5}}
\atoms{z2}{a1/p={45:0.5}}
\draw (1)edge[mark={arr,s,f}](a0) (0)edge[mark={arr,s,f}]++(-135:0.6) (1)--(a1) (0)--++(135:0.6) (0)edge[mark={arr,s,f}](d) (d)--(1);
\draw (a0)--++(45:0.3) (a0)--++(-45:0.3) (a0)--++(-135:0.3) (a1)--++(45:0.3) (a1)--++(-45:0.3) (a1)--++(135:0.3);
\end{tikzpicture}
\propto
\begin{tikzpicture}
\atoms{delta,bdastyle=red}{1/}
\atoms{cc,bdastyle=red}{a0/p={-45:0.5}}
\atoms{z2,bdastyle=red}{a1/p={45:0.5}}
\draw (1)edge[mark={arr,s,f}](a0) (1)edge[mark={arr,s,f}]++(-135:0.6) (1)--(a1) (1)--++(135:0.6);
\draw (a0)--++(45:0.3) (a0)--++(-45:0.3) (a0)--++(-135:0.3) (a1)--++(45:0.3) (a1)--++(-45:0.3) (a1)--++(135:0.3);
\end{tikzpicture}\;,
\qquad
\begin{tikzpicture}
\atoms{z2}{0/p={-0.8,0}, 1/}
\atoms{delta,bdastyle=red}{d/p={-0.4,0}}
\atoms{delta}{a0/p={-45:0.5}}
\atoms{cc}{a1/p={45:0.5}}
\draw (1)edge[mark={arr,s,f}](a0) (0)edge[mark={arr,s,f}]++(-135:0.6) (1)--(a1) (0)--++(135:0.6) (0)edge[mark={arr,s,f}](d) (d)--(1);
\draw (a0)--++(45:0.3) (a0)--++(-45:0.3) (a0)--++(-135:0.3) (a1)--++(45:0.3) (a1)--++(-45:0.3) (a1)--++(135:0.3);
\end{tikzpicture}
\propto
\begin{tikzpicture}
\atoms{z2,bdastyle=red}{1/}
\atoms{delta,bdastyle=red}{a0/p={-45:0.5}}
\atoms{cc,bdastyle=red}{a1/p={45:0.5}}
\draw (1)edge[mark={arr,s,f}](a0) (1)edge[mark={arr,s,f}]++(-135:0.6) (1)--(a1) (1)--++(135:0.6);
\draw (a0)--++(45:0.3) (a0)--++(-45:0.3) (a0)--++(-135:0.3) (a1)--++(45:0.3) (a1)--++(-45:0.3) (a1)--++(135:0.3);
\end{tikzpicture}\;.
\end{equation}
All in all, we find that the condition of Definition~\ref{def:path_integral_code} still holds, just that now each measurement outcome corresponds to multiple defect segments of different types.

Let us now look at the combinatorics of the resulting circuit.
The overall geometry is as for the CSS honeycomb Floquet code in Eq.~\eqref{eq:floquet_circuit}, just that the type of measurement now depends on the orientation of the edge or face and not on the time step:
\begin{equation}
\begin{multlined}
\rightarrow (I[\mathbf T]_{z\vca0\vcb1},I[\mathbf V]_{x\vca0\vcb1},I[\mathbf W]_{y\vca0\vcb1})\\
\rightarrow (I[\mathbf T]_{xy\vca0\vcb1\vcc2},I[\mathbf V]_{yz\vca0\vcb1\vcc2},I[\mathbf W]_{xz\vca0\vcb1\vcc2})\\
\rightarrow (I[\mathbf T]_{z\vcb1\vcc2},I[\mathbf V]_{x\vcb1\vcc2},I[\mathbf W]_{y\vcb1\vcc2})\\
\rightarrow (I[\mathbf T]_{xy\vcb1\vcc2\vca0},I[\mathbf V]_{yz\vcb1\vcc2\vca0},I[\mathbf W]_{xz\vcb1\vcc2\vca0})\\
\rightarrow (I[\mathbf T]_{z\vcc2\vca0},I[\mathbf V]_{x\vcc2\vca0},I[\mathbf W]_{y\vcc2\vca0})\\
\rightarrow (I[\mathbf T]_{xy\vcc2\vca0\vcb1},I[\mathbf V]_{yz\vcc2\vca0\vcb1},I[\mathbf W]_{xz\vcc2\vca0\vcb1})\rightarrow\;.
\end{multlined}
\end{equation}
After projecting the cubic lattice along time as in Eq.~\eqref{eq:triangular_projection}, $x$, $y$, and $z$ refer to the three different directions of edges in the resulting triangular lattice.
The measurements at $x\vca0\vcb1$ and $yz\vcb1\vcc2\vca0$ (and analogous pairs) act on the same pair of qubits and are in fact the same type of measurement.
We thus find that the circuit repeats already after three rounds, yielding
\begin{equation}
\begin{multlined}
\rightarrow (ZZ_{z\vca0\vcb1},XX_{x\vca0\vcb1},YY_{y\vca0\vcb1})\\
\rightarrow (ZZ_{z\vcc2\vca0},XX_{x\vcc2\vca0},YY_{y\vcc2\vca0})\\
\rightarrow (ZZ_{z\vcb1\vcc2},XX_{x\vcb1\vcc2},YY_{y\vcb1\vcc2})\rightarrow\;.
\end{multlined}
\end{equation}
After going to the dual hexagonal lattice, we obtain the honeycomb code as presented in \cite{Hastings2021}.

It has been argued in Ref.~\cite{Hastings2021} that the honeycomb Floquet code is closely related to the toric code since the instantaneous stabilizer group of the former is equivalent to the latter.
Here we will make this relation precise by showing that the underlying path integrals are in the same fixed-point phase.
The sequence of tensor-network equations transforming the toric-code path integral into the honeycomb path integral (or vice versa) is as follows:
We first insert a resolution of the identity, $\mathbb1=GG^{-1}$ at every bond.
$G$ is an invertible matrix that depends on the bond within a unit cell, but not on the unit cell.
Then we contract each 4-index tensor with the four surrounding matrices $G$ or $G^{-1}$, yielding a new 4-index tensor at that place.
Note that this is just a complicated way of saying that the two tensor networks are equivalent up to a basis change at every bond.
The matrices $G$ are built from the Hadamard matrix $H$ in Eq.~\eqref{eq:hadamard}, together with the following two matrices,
\begin{equation}
\begin{tikzpicture}
\atoms{smat}{0/}
\draw (0-l)--++(180:0.4) (0-r)--++(0:0.4);
\end{tikzpicture}
\coloneqq
S\coloneqq
\begin{pmatrix}
1&0\\0&i
\end{pmatrix}\;,
\quad
\begin{tikzpicture}
\atoms{umat}{0/}
\draw (0-l)--++(180:0.4) (0-r)--++(0:0.4);
\end{tikzpicture}
\coloneqq
U
\coloneqq HSH\;.
\end{equation}
$H$, $S$, and $U$ are all unitary,
\begin{equation}
\label{eq:honeycomb_gauge2}
\begin{aligned}
\begin{tikzpicture}
\atoms{hadamard}{0/p={-0.3,0}, 1/p={0.3,0}}
\draw (0-l)--++(180:0.3) (1-r)--++(0:0.3) (0-r)--(1-l);
\end{tikzpicture}
&=
\begin{tikzpicture}
\draw (0,0)--++(0:0.8);
\end{tikzpicture}\;,
&
\begin{tikzpicture}
\atoms{smat}{0/p={-0.3,0}, {1/p={0.3,0},lab={t=$*$,p={90:0.2}}}}
\draw (0-l)--++(180:0.3) (1-r)--++(0:0.3) (0-r)--(1-l);
\end{tikzpicture}
&=
\begin{tikzpicture}
\draw (0,0)--++(0:0.8);
\end{tikzpicture}\;,
\\
\begin{tikzpicture}
\atoms{umat}{0/p={-0.3,0}, {1/p={0.3,0},lab={t=$*$,p={90:0.2}}}}
\draw (0-l)--++(180:0.3) (1-r)--++(0:0.3) (0-r)--(1-l);
\end{tikzpicture}
&=
\begin{tikzpicture}
\draw (0,0)--++(0:0.8);
\end{tikzpicture}\;,
&&
\end{aligned}
\end{equation}
where the $*$ denotes complex conjugation.
$H$, $S$, and $U$, together with the 4-index $\delta$, $\zz_2$ and $\cc$ tensors satisfy two types of equations.
First, adding $H$ to all indices exchanges $\delta$ and $\zz_2$, and the same holds with $S$, $\zz_2$, and $\cc$, as well as with $U$, $\delta$, and $\cc$,
\begin{equation}
\label{eq:honeycomb_gauge0}
\begin{aligned}
\begin{tikzpicture}
\atoms{z2}{0/}
\atoms{smat}{{x0/p=0:0.5,lab={t=$*$,p=90:0.25}}, {x1/p=90:0.5,lab={t=$*$,p=0:0.25}}, x2/p=180:0.5, x3/p=-90:0.5}
\draw (0)--(x0) (0)--(x1) (0)--(x2) (0)--(x3) (x0)--++(0:0.4) (x1)--++(90:0.4) (x2)--++(180:0.4) (x3)--++(-90:0.4);
\end{tikzpicture}
&=
\begin{tikzpicture}
\atoms{cc}{0/}
\draw (0)edge[mark={arr,s,f}]++(0:0.5) (0)edge[mark={arr,s,f}]++(90:0.5) (0)--++(180:0.5) (0)--++(-90:0.5);
\end{tikzpicture}\;,
&
\begin{tikzpicture}
\atoms{delta}{0/}
\atoms{hadamard}{{x0/p=0:0.5}, {x1/p=90:0.5}, x2/p=180:0.5, x3/p=-90:0.5}
\draw (0)--(x0) (0)--(x1) (0)--(x2) (0)--(x3) (x0)--++(0:0.4) (x1)--++(90:0.4) (x2)--++(180:0.4) (x3)--++(-90:0.4);
\end{tikzpicture}
&=
\begin{tikzpicture}
\atoms{z2}{0/}
\draw (0)--++(0:0.5) (0)--++(90:0.5) (0)--++(180:0.5) (0)--++(-90:0.5);
\end{tikzpicture}\;,
\\
\begin{tikzpicture}
\atoms{delta}{0/}
\atoms{umat}{{x0/p=0:0.5,lab={t=$*$,p=90:0.25}}, {x1/p=90:0.5,lab={t=$*$,p=0:0.25}}, x2/p=180:0.5, x3/p=-90:0.5}
\draw (0)--(x0) (0)--(x1) (0)--(x2) (0)--(x3) (x0)--++(0:0.4) (x1)--++(90:0.4) (x2)--++(180:0.4) (x3)--++(-90:0.4);
\end{tikzpicture}
&=
\begin{tikzpicture}
\atoms{cc}{0/}
\draw (0)edge[mark={arr,s,f}]++(0:0.5) (0)edge[mark={arr,s,f}]++(90:0.5) (0)--++(180:0.5) (0)--++(-90:0.5);
\end{tikzpicture}\;.
&&
\end{aligned}
\end{equation}
Due to Eq.~\eqref{eq:honeycomb_gauge2}, each $H$, $S$, or $U$ matrix can be either on the right or on the left-hand side.
Furthermore, two $S$ matrices adjacent to a $\delta$-tensor can be canceled, and the same for $H$ and $\cc$, as well as for $U$ and $\zz_2$:
\begin{equation}
\label{eq:honeycomb_gauge3}
\begin{aligned}
\begin{tikzpicture}
\atoms{delta}{0/}
\atoms{smat}{{x0/p=0:0.5}, {x1/p=180:0.5,lab={t=$*$,p=90:0.25}}}
\draw (0)--(x0) (0)--(x1) (x0)--++(0:0.4) (x1)--++(180:0.4) (0)--++(90:0.6) (0)--++(-90:0.6);
\end{tikzpicture}
&=
\begin{tikzpicture}
\atoms{delta}{0/}
\draw (0)edge[]++(0:0.5) (0)edge[]++(90:0.5) (0)--++(180:0.5) (0)--++(-90:0.5);
\end{tikzpicture}\;,
&
\begin{tikzpicture}
\atoms{cc}{0/}
\atoms{hadamard}{{x0/p=0:0.5}, {x1/p=180:0.5}}
\draw (0)edge[mark={arr,f,s}](x0) (0)--(x1) (x0)--++(0:0.4) (x1)--++(180:0.4) (0)edge[mark={arr,f,s}]++(90:0.6) (0)--++(-90:0.6);
\end{tikzpicture}
&=
\begin{tikzpicture}
\atoms{cc}{0/}
\draw (0)--++(0:0.5) (0)edge[mark={arr,s,f}]++(90:0.5) (0)edge[mark={arr,s,f}]++(180:0.5) (0)--++(-90:0.5);
\end{tikzpicture}\;,
\\
\begin{tikzpicture}
\atoms{z2}{0/}
\atoms{umat}{{x0/p=0:0.5}, {x1/p=180:0.5,lab={t=$*$,p=90:0.25}}}
\draw (0)--(x0) (0)--(x1) (x0)--++(0:0.4) (x1)--++(180:0.4) (0)--++(90:0.6) (0)--++(-90:0.6);
\end{tikzpicture}
&=
\begin{tikzpicture}
\atoms{z2}{0/}
\draw (0)edge[]++(0:0.5) (0)edge[]++(90:0.5) (0)--++(180:0.5) (0)--++(-90:0.5);
\end{tikzpicture}\;.
&&
\end{aligned}
\end{equation}
With this, we are now ready to find matrices $G$ that transform the toric-code path integral into the honeycomb path integral.
Each bond inside a unit cell can be specified by the involved edge $a$ (either $x$, $y$, or $z$), the involved face $b$ (either $xy$, $xz$, or $yz$), and the direction $\pm$ of the bond $a\rightarrow b$ relative to the $x$, $y$, or $z$ direction.
Thus, we need to specify 12 different matrices $G(a,b,\pm)$.
As an ansatz, we set $G(a,b,-)\coloneqq G(a,b,+)^*$, and impose that every $G$ is some product formed by $H$, $S$, and $U$.
For each edge $a$, there are different choices for the two matrices $G(a,\ldots,+)$, such that the toric-code tensor together with the surrounding $G$ matrices yields the according honeycomb tensor.
For example, for $a=x$, we want to transform a toric-code $\delta$-tensor into a honeycomb $\zz_2$-tensor.
First, any of the two matrices $G(a,\ldots,+)$ may or may not contain $S$, since each $G$ appears at two ($\pm$) indices and can be annihilated using the first of Eq.~\eqref{eq:honeycomb_gauge3}.
Then, both matrices $G(a,\ldots,+)$ need to contain $H$ in order to transform the $\delta$-tensor into a $\zz_2$-tensor via the second of Eq.~\eqref{eq:honeycomb_gauge0}.
Finally, each of $G(a,\ldots,+)$ may or may not contain $U$, due to the third of Eq.~\eqref{eq:honeycomb_gauge3}.
This yields a set of possible choices for either matrix $G(a,\ldots,+)$ which we denote by $(S)H(U)$.
The following table shows all potential $G(a,\ldots,+)$ for edges $a$ or $G(\ldots,b,+)$ for faces $b$,
\begin{equation}
\begin{tabular}{l|l|l|l}
$a/b$ & toric code & honeycomb & \makecell{potential\\$G(a/\ldots,\ldots/b,+)$}\\
\hline
$x$ & $\delta$ & $\zz_2$ & $(S)H(U)$\\
$y$ & $\delta$ & $\cc$ & $(S)U(H)$\\
$z$ & $\delta$ & $\delta$ & $(S)\mathbb{1}(S)$\\
$xy$ & $\zz_2$ & $\delta$ & $(U)H(S)$\\
$xz$ & $\zz_2$ & $\cc$ & $(U)S(H)$\\
$yz$ & $\zz_2$ & $\zz_2$ & $(U)\mathbb{1}(U)$
\end{tabular}\;.
\end{equation}
In order to find $G(a,b,+)$, we write out all potential $G(a,\ldots,+)$ and $G(\ldots,b,+)$ and take any common element.
A solution is given by
\begin{equation}
\label{eq:honeycomb_basis_transformation}
\begin{tabular}{rcl|l}
$a$&$-$&$b$&$G(a,b,+)$\\
\hline
$x$&$-$&$xy$&$H$\\
$x$&$-$&$xz$&$SH$\\
$y$&$-$&$xy$&$UH$\\
$y$&$-$&$yz$&$U$\\
$z$&$-$&$xz$&$S$\\
$z$&$-$&$yz$&$\mathbb{1}$\\
\end{tabular}\;.
\end{equation}
So we have found that the toric code and honeycomb path integrals are in the same fixed-point phase.
Next we notice that Eq.~\eqref{eq:honeycomb_gauge0} and Eq.~\eqref{eq:honeycomb_gauge3} still hold after we replace all $\delta$, $\zz$ and $\cc$-tensors with their charged versions, at least up to a phase prefactor.
Thus, the toric code path integral with a configuration of $e$ and $m$ defects is locally equivalent to the honeycomb path integral with the same configuration of 1-cycle and 2-cocycle defects.
However, the original honeycomb Floquet code and its CSS version differ by the selection of defect segments corresponding to a measurement in the circuit:
A face measurement in the CSS honeycomb Floquet code corresponds to the presence of an $e$ defect segment at a diagonal edge, or an equvialent pair edges of the original lattice.
A face measurement in the honeycomb Floquet code corresponds to the same $e$ defect segments and an additional $m$ defect segment at the face itself.
The analogous holds for the edge measurements.
So in other words, the direct equivalence of the original honeycomb Floquet code and its CSS version by local tensor-network rewriting only holds for the circuits postselected to $+1$ measurement outcomes, and not for the full circuit of instruments with arbitrary measurement outcomes.
For both codes, the circuit postselected to arbitrary measurement outcomes is the toric code path integral with a pattern of $e$ and $m$ defects, but the locations of these defects depending on the measurement outcomes are different.

Note that the reduced time periodicity is closely related to the fact that the honeycomb path integral on a cubic lattice allows for a smaller choice of unit cell, namely one consisting of only one $\delta$, one $\zz_2$, and one $\cc$-tensor.
For example, consider an $xy$ plane containing $xy$ faces, and the honeycomb path integral restricted to this plane including the tensors at all $xy$ faces, $x$ edges, and $y$ edges.
When shifting this plane by $\frac12 z$, the tensors at $z$ edges, $xz$ faces, and $yz$ faces form exactly the same tensor network on the Poincar\'e dual lattice, that is, shifted by $\frac12x+\frac12y$.
So instead of $\{x,y,z\}$, we can choose $\{x,y,\frac12 x+\frac12 y+\frac12 z\}$ as a unit cell.
When we instead consider these two planes shifted by $\frac12 z$ for the toric code, we swap $\delta$ and $\zz_2$ tensors in addition to going to the dual lattice.
This has the same effect as inserting a duality domain wall (exchanging $e$ and $m$ anyons) in between the two planes.
So with the new unit cell, the phase of both path integrals is a toric code with a rigid stack of duality domain walls perpendicular to $z$.
The according exchange of $e$ and $m$ after one code cycle has already been observed in the honeycomb code \cite{Hastings2021}.
The halved unit cell is also responsible for the weak breaking of translation symmetry in the closely related Kitaev honeycomb model Hamiltonian \cite{Kitaev2005}.

Let us briefly discuss the decoding of Proposition~\ref{prop:path_integral_decoding} for the honeycomb Floquet code.
Each spacetime syndrome maps to a 1-chain and 2-cochain in the cubic spacetime lattice.
The novel feature compared to earlier examples is that a single measurement outcome corresponds to both 1-chain and 2-cochain segments.
Apart from this, the decoding proceeds as usual by finding a minimum-weight fix of the spacetime syndrome and then closing the syndrome at a spatial slice at time $T$ such as in Eq.~\eqref{eq:css_spatial_slice}.
Defects along 1-chains or 2-cochains can be introduced by applying Pauli $X$, $Y$, or $Z$ operators to the corresponding places in the circuit.
Let us discuss when the syndrome 1-chain is broken at a specific vertex $v$, by looking at the measurements for which the boundary of the corresponding defect segments contains $v$.
These include the measurements at the 6 faces in the cubic lattice for which $v$ is the temporally first or last vertex, like for the CSS honeycomb Floquet code.
However, they also include the measurements at the 6 edges indicent to $v$ themselves.
This is in accordance with the fact that a detection cell is formed by 12 measurements in the honeycomb Floquet code \cite{Paetznick2022}, but only 6 in the CSS honeycomb Floquet code \cite{Kesselring2022}.

\section{New codes from tensor-network path integrals}
\label{sec:new_examples}
In this section we use our path integral framework to construct two new dynamic error-correcting codes.
First, we introduce a generalization of the CSS honeycomb Floquet code to $3+1$ dimensions.
Then we construct a non-Pauli dynamic code based on the double-semion string-net model.
\subsection{Floquet toric code in 3+1 dimensions}
\label{sec:3d_toric_code}
In this section, we use our method to construct a new Floquet code, namely a Floquet version of the toric code in $3+1$ dimensions.
As a stabilizer code, the 3-dimensional toric code is defined on a cubic lattice with one qubit at every face.
There is one $Z_0Z_1Z_2Z_3Z_4Z_4$ stabilizer acting on the qubits at the 6 faces adjacent to every volume, and a $X_0X_1X_2X_3$ stabilizer acting on the 4 qubits adjacent to every edge.
The ground state of the model is an equal-weight superposition of $\zz_2$-configurations on all faces that obey a Gauss law at every volume.
That is, it is an equal-weight superposition of 2-cocycles, which can be pictured as closed-loop configurations on the Poincar\'e dual lattice.
Analogous to the ground state, the path integral representing the $3+1$-dimensional toric code is given by a sum over all cellular 2-cocycles inside an arbitrary 4-dimensional spacetime cellulation.
Such a cocycle is a configuration of $\zz_2$ variables on all the faces, such that at every volume the sum of variables at its boundary faces is $0$ (mod $2$).
These 2-cocycles can be pictured as closed-membrane configurations inside the 4-dimensional spacetime cellulation, which restrict to a closed-loop configuration at a spatial boundary.
As a tensor network, the path integral consists of one $\delta$-tensor at every face and one $\zz_2$-tensor at every volume, with bonds shared between pairs of adjacent face and volume.

We can also introduce defects inside the path integral.
We will use two types of syndrome defects, namely line-like $m$ defects, as well as membrane-like $e$ defects.
\footnote{
In the literature, a more common convention is to label line-like defects $e$ and the membrane-like defects $m$.
However, for a smoother geometric interpretation, we have represented the $3+1$-dimensional toric code as a 2-form gauge theory instead of a 1-form gauge theory.
In this dual representation, the fluxes of the 2-form gauge field that we label by $m$ are line-like, and the charges that we label by $e$ are membrane-like.
}
The $e$ defects are placed on 2-cycles, and at every face of the 2-cocycle we replace the $\delta$-tensor by the charged $\delta$-tensor in Eq.~\eqref{eq:e_anyon_tensor}.
The $m$ defects are placed on 3-cocyles, and at every volume of the 3-cocycle we replace the $\zz_2$-tensor by the charged $\zz_2$-tensor in Eq.~\eqref{eq:m_anyon_tensor}.

To construct the Floquet code, we take a 4-dimensional hypercubic lattice spanned by the four unit vectors $w$, $x$, $y$, $z$, and choose $t=w+x+y+z$ as the time direction.
The operators of the circuit are individual $4$-index $\delta$-tensors at the faces, and $6$-index $\zz_2$-tensors at the cubes.
The diagonal $t$ direction allows for a natural interpretation of these tensors as operators by dividing their indices into inputs and outputs,
\begin{equation}
\label{eq:3d_floquet_operators}
T_1\coloneqq
\begin{tikzpicture}
\draw[orange,fill=orange,fill opacity=0.2] (-0.2,-1)--++(0.4,0.4)--++(0,1.6)--++(-0.4,-0.4)--cycle;
\atoms{delta}{d/}
\draw (d)--++(-45:0.6) (d)--++(-135:0.6) (d)--++(45:0.6) (d)--++(135:0.6);
\end{tikzpicture}
\;,\qquad
V_1\coloneqq
\begin{tikzpicture}
\atoms{void}{{0/}, {1/p={0.8,0.8}}, {2/p={0,0.5}}, {3/p={0.8,1.3}}}
\atoms{void}{{0x/p={-0.8,0.8}}, {1x/p={0,1.6}}, {2x/p={-0.8,1.3}}, {3x/p={0,2.1}}}
\draw[dashed,orange] (0)--(2) (2x)--(2)--(3) (3)--(3x) (2x)--(3x);
\draw[orange] (0)--(1) (0)--(0x) (2x)--(0x)--(1x)--(1)--(3) (3)--(3x) (2x)--(3x) (1x)--(3x);
\atoms{z2}{z/p={0,1.05}}
\draw (z)--++(-0.45,-0.45) (z)--++(0,-0.4) (z)--++(0.45,-0.45) (z)--++(-0.45,0.45) (z)--++(0,0.4) (z)--++(0.45,0.45);
\end{tikzpicture}
\;.
\end{equation}
The 4 indices of $T_1$ connect it to the four cubes adjacent to the face (in orange), whose position in the drawing should not be taken literal due to the 4-dimensional nature.
The 6 indices of $V_1$ connect it to the 6 faces of the cube.

$T_1$ is a projector onto the $+1$ eigenspace of $Z_0Z_1$, whereas $V_1$ is a projector onto the intersection of the $+1$ eigenspaces of $X_0X_1$ and $X_1X_2$.
Since both $T_1$ and $V_1$ are non-unitary, we turn them into instruments using additional operators that include defect segments.
In addition to $T_1$ at a face we define another operator $T_m$ that includes a line segment of $m$ defect.
To this end, we modify the lattice slightly by replacing each 4-valent face by a pair of 3-valent faces separated by a pillow-like volume, whose boundary is formed by these two faces only.
$T_1$ then corresponds of the pillow-like volume together with the two faces, and $T_m$ is the same with the pillow-like volume carrying an $m$ anyon segment,
\begin{equation}
T_m\coloneqq
\begin{tikzpicture}
\draw[orange] (-0.2,-1)--++(0.4,0.4)--++(0,1.6)--++(-0.4,-0.4)--cycle;
\fill[orange,fill opacity=0.2] (-0.2,-1)to[out=140,in=-160](0.2,1)to[out=-40,in=20]cycle;
\atoms{delta}{d0/p={-0.5,0}, d1/p={0.5,0}}
\draw[worldline] (-70:0.9)--(110:0.9);
\atoms{z2,bdastyle=red}{m/}
\draw (d1)--++(-45:0.6) (d0)--++(-135:0.6) (d1)--++(45:0.6) (d0)--++(135:0.6);
\draw (m)--(d0) (m)--(d1);
\fill[orange,fill opacity=0.2] (-0.2,-1)to[out=140,in=-160](0.2,1)to[out=-40,in=20]cycle;
\end{tikzpicture}\;.
\end{equation}
The drawn $m$ worldline goes perpendicular to the pillow volume connecting the two adjacent 4-cells, and its positioning in the drawing should not be taken literal.
$T_1$ and $T_m$ together yield a $Z_0Z_1$ measurement $I[\mathbf T]$ as usual.

To turn $V_1$ into an instrument, we define three new operators,  $V_{1e}$, $V_{e1}$, and $V_{ee}$, corresponding to the absence or presence of two different $e$ membrane defect segments.
To this end, we divide the cube into three volumes along two internal 2-valent faces $g$ and $f$.
$V_{1e}$, $V_{e1}$, or $V_{ee}$ then correspond to an $e$ membrane segment being present at either $g$, $f$, or at both.
Specifically, in a cube with faces labelled like
\begin{equation}
\label{eq:cube_index_labeling}
\begin{tikzpicture}
\draw (0,0)edge[mark={arr,e},ind=$w$]++(135:0.45) (0,0)edge[mark={arr,e},ind=$y$]++(90:0.25) (0,0)edge[mark={arr,e},ind=$x$]++(45:0.45);
\end{tikzpicture}\quad
\begin{tikzpicture}
\atoms{vertex}{{0/}, {1/p={0.8,0.8}}, {2/p={0,0.5}}, {3/p={0.8,1.3}}}
\atoms{vertex}{{0x/p={-0.8,0.8}}, {1x/p={0,1.6}}, {2x/p={-0.8,1.3}}, {3x/p={0,2.1}}}
\draw (0,1.3)edge[out=60,in=-90,ind=$o_1$](0.2,2.4) (0.4,0.65)edge[out=-60,in=90,ind=$i_1$](1,-0.3) (-0.4,0.65)edge[out=-120,in=90,ind=$i_2$](-1,-0.3);
\draw[dashed,front] (0)--(2) (2x)--(2)--(3) (3)--(3x) (2x)--(3x);
\draw[front] (0)--(1) (0)--(0x) (2x)--(0x)--(1x)--(1)--(3) (3)--(3x) (2x)--(3x) (1x)--(3x);
\draw (0,0.8)edge[front, out=-60,in=90,ind=$i_0$](0.2,-0.3) (0.4,1.45)edge[front, out=60,in=-90,ind=$o_0$](1,2.4) (-0.4,1.45)edge[front, out=120,in=-90,ind=$o_2$](-1,2.4);
\end{tikzpicture}
\;,
\end{equation}
we choose $f$ to have the same boundary as $i_0$ together with $o_0$, and $g$ with the same boundary as $i_2$ and $o_2$.
The isometry resulting from combining the different $V$ operators is given by
\begin{equation}
\label{eq:3d_floquet_volume_measurement}
\begin{multlined}
\mathbf V\coloneqq
(V_1,V_{e1},V_{1e},V_{ee})
=
\begin{tikzpicture}
\atoms{hadamard}{h0/p={-0.5,0.5}, h1/p={0.5,0.5}}
\atoms{delta}{d0/p={-0.5,0},d1/p={0.5,0}}
\atoms{z2}{z0/p={-1,0},z1/,z2/p={1,0}}
\draw (z0)--(z1)--(z2) (z0)edge[ind=$i_0$]++(-90:0.4) (z0)edge[ind=$o_0$]++(90:0.4) (z1)edge[ind=$i_1$]++(-90:0.4) (z1)edge[ind=$o_1$]++(90:0.4) (z2)edge[ind=$i_2$]++(-90:0.4) (z2)edge[ind=$o_2$]++(90:0.4);
\draw[classical] (d0)--(h0) (h0)edge[ind=$f$]++(90:0.4) (d1)--(h1) (h1)edge[ind=$g$]++(90:0.4);
\end{tikzpicture}\\
=
\begin{tikzpicture}
\atoms{hadamard}{h0/p={-0.5,0.5}, h1/p={0.5,1}}
\atoms{delta}{d0/p={-0.5,0},d1/p={0.5,0.5}}
\atoms{z2}{z0/p={-1,0},z1/, zx/p={0,0.5},z2/p={1,0.5}}
\draw (z0)--(d0)--(z1) (z1)edge[mark={slab=$\textcolor{red}{\scriptstyle h}$,r}](zx) (zx)--(d1)--(z2) (z0)edge[ind=$i_0$]++(-90:0.4) (z0)edge[ind=$o_0$]++(90:0.4) (z1)edge[ind=$i_1$]++(-90:0.4) (zx)edge[ind=$o_1$]++(90:0.4) (z2)edge[ind=$i_2$]++(-90:0.4) (z2)edge[ind=$o_2$]++(90:0.4);
\draw[classical] (d0)--(h0) (h0)edge[ind=$f$]++(90:0.4) (d1)--(h1) (h1)edge[ind=$g$]++(90:0.4);
\end{tikzpicture}
\;.
\end{multlined}
\end{equation}
As shown, $\mathbf V$ can be split up, such that $I[\mathbf V]$ consists of two consecutive measurements $X_0X_1$ and $X_1X_2$.
Geometrically, this corresponds to adding another internal 2-valent face $h$ whose boundary is that of $i_0$, $i_1$, and $o_0$ together (or equivalently $i_2$, $o_1$, and $o_2$).
Note that $h$ does not carry a potential $e$ defect segment.

The definition of $\mathbf V$ above depends on a choice of labelling the bottom faces of a cube by $i_0$, $i_1$, and $i_2$, and of the top faces by $o_0$, $o_1$, and $o_2$.
In principle, any choice would yield a valid error-correcting circuit.
However, we aim to obtain a circuit consisting of projective 2-qubit measurements acting on a fixed set of qubits without intermediate swap gates.
The following choice achieves this goal, as will come clear later in this section.
First, we divide the cubes into 8 different sorts labelled by $p_{abc}$, and there will be a different choice for each sort.
$p\in \{0,1\}$ labels whether the cube appears at an even ($0$) or odd ($1$) time step in the circuit.
$abc$ labels the three spanning directions $\{a,b,c\}\subset \{w,x,y,z\}$ of the cube, ordered according to $w,x,y,z$.
With this, $i_0$, $i_1$, and $i_2$ are the bottom faces with spanning directions $ab$, $bc$, and $ca$, respectively.
If $p=0$, then $o_0$, $o_1$, and $o_2$ are the top faces with spanning directions $bc$, $ca$, and $ab$.
If $p=1$, then they are given by $ca$, $ab$, and $bc$.
For example, Eq.~\eqref{eq:cube_index_labeling} shows the labelling for a $1_{wxy}$ cube.

In principle, the combinatorics of the circuit formed by the instruments $I[\mathbf T]$ and $I[\mathbf V]$ is fully specified by the 4-dimensional cellulation.
However, it is instructive to give a more conventional description of this circuit in terms of measurements acting on fixed qubits on a spatial lattice.
We start by decomposing the circuit into rounds of instruments that are applied in parallel.
To this end, we notice that there are four different levels of vertices in the 4-dimensional cubic lattice along the $t$ direction, which we label/color by $\vca0$/red, $\vcb1$/green, $\vcc2$/blue, and $\vcd3$/yellow.
Accordingly, there are four levels of faces, $\vca0\vcb1\vcc2$, $\vcb1\vcc2\vcd3$, $\vcc2\vcd3\vca0$, and $\vcd3\vca0\vcb1$, and four levels of volumes, $\vca0\vcb1\vcc2\vcd3$, $\vcb1\vcc2\vcd3\vca0$, $\vcc2\vcd3\vca0\vcb1$, and $\vcd3\vca0\vcb1\vcc2$.
So one time period of the circuit consists of 8 rounds of instruments,
\begin{equation}
\label{eq:floquet3d_circuit}
\begin{multlined}
\rightarrow I[\mathbf T]_{\vca0\vcb1\vcc2} \rightarrow I[\mathbf V]_{\vca0\vcb1\vcc2\vcd3} \rightarrow I[\mathbf T]_{\vcb1\vcc2\vcd3}\rightarrow I[\mathbf V]_{\vcb1\vcc2\vcd3\vca0}\\
\rightarrow I[\mathbf T]_{\vcc2\vcd3\vca0} \rightarrow I[\mathbf V]_{\vcc2\vcd3\vca0\vcb1}\rightarrow I[\mathbf T]_{\vcd3\vca0\vcb1} \rightarrow I[\mathbf V]_{\vcd3\vca0\vcb1\vcc2}\;.
\end{multlined}
\end{equation}
To get an appropriate spatial lattice, we project the 4-dimensional cubic lattice onto $3$-dimensional space along the $t=w+x+y+z$ axis.
To this end, we choose new basis vectors
\begin{equation}
\begin{gathered}
\overline x=\frac12 w+\frac12 x-\frac12 y-\frac12 z\;,\\
\overline y=\frac12 w-\frac12 x+\frac12 y-\frac12 z\;,\\
\overline z=\frac12 w-\frac12 x-\frac12 y+\frac12 z\;,
\end{gathered}
\end{equation}
orthogonal to $t$.
The projected $\vca0$ and $\vcc2$ vertices then form a cubic lattice $A$ with unit vectors $\overline x$, $\overline y$, and $\overline z$.
The $\vcb1$ and $\vcd3$ vertices form a second cubic lattice $B$ shifted by $\frac12(\overline x+\overline y+\overline z)$, such that the vertices of $A$ are the centers of the cubes of $B$ and vice versa.
Within $A$, $\vca0$ and $\vcc2$ vertices alternate in a checkerboard manner, and the same for $\vcb1$ and $\vcd3$ vertices within $B$.
The projected edges have length $\sqrt{\frac34}$ and connect each $B$ vertex with the $8$ corner vertices of the corresponding $A$ cube, and vice versa.
The edges of the $A$ and $B$ lattice themselves are not projected edges of the 4-dimensional cubic lattice.
The following depicts a section of the lattice with four layers of vertices in $\overline y$ direction, projected edges in gray, edges of $A$ and $B$ in black, and edges connecting vertices of the two back layers dotted:
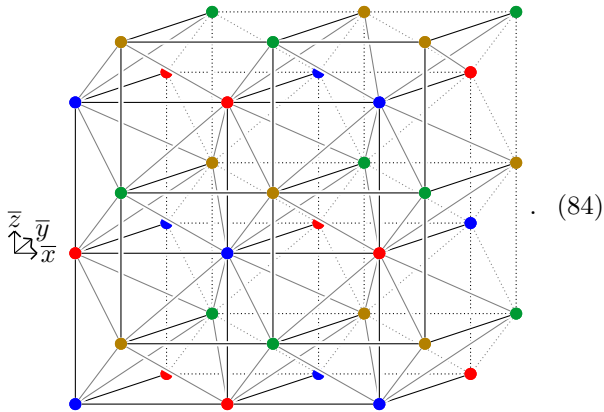
\begin{equation}
\label{eq:4_colorable_triangulation}
\begin{tikzpicture}
\atoms{void}{coordsys/p={-2,1.6},x/p={2,0}, y/p={0,2}, s1/p={0.6,0.8}, s2/p={-1.2,-0.4}}
\draw (coordsys)edge[mark={arr,e},ind=$\overline x$]++(0:0.3) (coordsys)edge[mark={arr,e},ind=$\overline y$]++(40:0.3) (coordsys)edge[mark={arr,e},ind=$\overline z$]++(90:0.3);
\atoms{vertex}{{0/vca}, {1/p={x},vcc}, {2/p={$2*(x)$},vca}, {3/p={y},vcc}, {4/p={$(x)+(y)$},vca}, {5/p={$2*(x)+(y)$},vcc}, {6/p={$2*(y)$},vca}, {7/p={$(x)+2*(y)$},vcc}, {8/p={$2*(x)+2*(y)$},vca}}
\draw[densely dotted] (0)--(1)--(2) (3)--(4)--(5) (6)--(7)--(8) (0)--(3)--(6) (1)--(4)--(7) (2)--(5)--(8);
\atoms{vertex}{{0x/vcb, p={$(0)+(s1)$}}, {1x/vcd, p={$(1)+(s1)$}}, {2x/vcb, p={$(2)+(s1)$}}, {3x/vcd, p={$(3)+(s1)$}}, {4x/vcb, p={$(4)+(s1)$}}, {5x/vcd, p={$(5)+(s1)$}}, {6x/vcb, p={$(6)+(s1)$}}, {7x/vcd, p={$(7)+(s1)$}}, {8x/vcb, p={$(8)+(s1)$}}}
\draw[densely dotted,gray] (0x)--(0) (0x)--(1) (0x)--(3) (0x)--(4) (1x)--(1) (1x)--(2) (1x)--(4) (1x)--(5) (2x)--(2) (2x)--(5) (3x)--(3) (3x)--(4) (3x)--(6) (3x)--(7) (4x)--(4) (4x)--(5) (4x)--(7) (4x)--(8) (5x)--(5) (5x)--(8) (6x)--(6) (6x)--(7) (7x)--(7) (7x)--(8) (8x)--(8);
\draw[front,densely dotted] (0x)--(1x)--(2x) (3x)--(4x)--(5x) (6x)--(7x)--(8x) (0x)--(3x)--(6x) (1x)--(4x)--(7x) (2x)--(5x)--(8x);
\atoms{vertex}{{0y/p={$(0)+(s2)$},vcc}, {1y/p={$(1)+(s2)$},vca}, {2y/p={$(2)+(s2)$},vcc}, {3y/p={$(3)+(s2)$},vca}, {4y/p={$(4)+(s2)$},vcc}, {5y/p={$(5)+(s2)$},vca}, {6y/p={$(6)+(s2)$},vcc}, {7y/p={$(7)+(s2)$},vca}, {8y/p={$(8)+(s2)$},vcc}}
\draw[front] (0)--(0y) (1)--(1y) (2)--(2y) (3)--(3y) (4)--(4y) (5)--(5y) (6)--(6y) (7)--(7y) (8)--(8y);
\draw[gray,front] (0x)--(0y) (0x)--(1y) (0x)--(3y) (0x)--(4y) (1x)--(1y) (1x)--(2y) (1x)--(4y) (1x)--(5y) (2x)--(2y) (2x)--(5y) (3x)--(3y) (3x)--(4y) (3x)--(6y) (3x)--(7y) (4x)--(4y) (4x)--(5y) (4x)--(7y) (4x)--(8y) (5x)--(5y) (5x)--(8y) (6x)--(6y) (6y)--(7y) (7x)--(7y) (7x)--(8y) (8x)--(8y);
\draw[front] (0y)--(1y)--(2y) (3y)--(4y)--(5y) (6y)--(7y)--(8y) (0y)--(3y)--(6y) (1y)--(4y)--(7y) (2y)--(5y)--(8y);
\atoms{vertex}{{0z/vcd, p={$(0x)+(s2)$}}, {1z/vcb, p={$(1x)+(s2)$}}, {2z/vcd, p={$(2x)+(s2)$}}, {3z/vcb, p={$(3x)+(s2)$}}, {4z/vcd, p={$(4x)+(s2)$}}, {5z/vcb, p={$(5x)+(s2)$}}, {6z/vcd, p={$(6x)+(s2)$}}, {7z/vcb, p={$(7x)+(s2)$}}, {8z/vcd, p={$(8x)+(s2)$}}}
\draw[front] (0x)--(0z) (1x)--(1z) (2x)--(2z) (3x)--(3z) (4x)--(4z) (5x)--(5z) (6x)--(6z) (7x)--(7z) (8x)--(8z);
\draw[gray,front] (0z)--(0y) (0z)--(1y) (0z)--(3y) (0z)--(4y) (1z)--(1y) (1z)--(2y) (1z)--(4y) (1z)--(5y) (2z)--(2y) (2z)--(5y) (3z)--(3y) (3z)--(4y) (3z)--(6y) (3z)--(7y) (4z)--(4y) (4z)--(5y) (4z)--(7y) (4z)--(8y) (5z)--(5y) (5z)--(8y) (6z)--(6y) (6z)--(7y) (7z)--(7y) (7z)--(8y) (8z)--(8);
\draw[front] (0z)--(1z)--(2z) (3z)--(4z)--(5z) (6z)--(7z)--(8z) (0z)--(3z)--(6z) (1z)--(4z)--(7z) (2z)--(5z)--(8z);
\end{tikzpicture}\;.
\end{equation}
The edges of $A$ and $B$ together with all the projected edges define a triangulation where each tetrahedron has one $\vca0$, one $\vcb1$, one $\vcc2$, and one $\vcd3$ vertex.
The projections of spacetime faces are rhombi consisting of two triangles.
The projections of the spacetime cubes are (rhombic) cubes consisting of 6 tetrahedra, 3 left-handed and 3 right-handed ones.
If a cube is adjacent to a face, then one of the right-handed tetrahedra contains one of the triangles of the face.

As usual, qubits can be identified by following the timeline of the bonds in the tensor-network/circuit diagram.
There is one such timeline for every tetrahedron $F$ that is right-handed relative to the vertex ordering $\vca0\vcb1\vcc2\vcd3$,
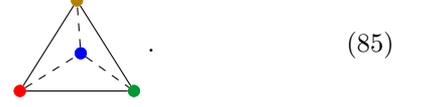
\begin{equation}
\begin{tikzpicture}
\atoms{vertex}{{0/vca}, {1/vcb,p={1.5,0}}, {2/vcc,p={0.8,0.5}}, {3/vcd,p={0.75,1.2}}}
\draw[dashed] (0)--(2)--(1) (2)--(3);
\draw (0)--(1)--(3)--(0);
\end{tikzpicture}\;.
\end{equation}
Let $F_{i,i+1,i+2}$ be the spacetime face whose projection is spanned by the $(i,i+1)$ and $(i+1,i+2)$ edges of the tetrahedron, where all numbers are understood mod 4.
Let $F_{i,i+1,i+2,i+3}$ be the spacetime cube whose projection is spanned by the $(i,i+1)$, $(i+1,i+2)$, and $(i+2,i+3)$ edges of the tetrahedron.
Then, within a fixed $t$-period, the timeline of bonds is given by the following sequence of adjacent faces and cubes,
\begin{equation}
\begin{multlined}
F_{\vca0\vcb1\vcc2} - F_{\vca0\vcb1\vcc2\vcd3} - F_{\vcb1\vcc2\vcd3} - F_{\vcb1\vcc2\vcd3\vca0
} - F_{\vcc2\vcd3\vca0}
\\
 - F_{\vcc2\vcd3\vca0\vcb1} - F_{\vcd3\vca0\vcb1} - F_{\vcd3\vca0\vcb1\vcc2}-\;.
\end{multlined}
\end{equation}
To go from the face $F_{i,i+1,i+2}$ to the face $F_{i+1,i+2,i+3}$ inside the projection of the cube $F_{i,i+1,i+2,i+3}$, we have to either rotate left or right when looking in the direction $i\rightarrow i+3$.
Since the tetrahedron is right-handed relative to the orderings $\vca0\vcb1\vcc2\vcd3$ and $\vcc2\vcd3\vca0\vcb1$ but left-handed for $\vcb1\vcc2\vcd3\vca0$ and $\vcd3\vca0\vcb1\vcc2$, we rotate right for $i=0$ and $i=2$, and right for $i=1$ and $i=3$.
This fits our choice of labeling the faces of each cube by $i_0$, $\ldots$, $o_2$, which we have discussed in the paragraph after Eq.~\eqref{eq:3d_floquet_volume_measurement}:
As can be seen in Eq.~\eqref{eq:cube_index_labeling}, in order to go from $i_x$ to $o_x$ we turn either right or left in the spatial projection of the cube when looking from bottom to top.
We turn left for even time steps ($p=0$ which we identify with $i=1$ or $i=3$), and right for odd time steps ($p=1$, which is $i=0$ or $i=2$).

As we have seen, there is one qubit associated to every right-handed tetrahedron.
Each instrument $I[\mathbf T]$ at a spacetime face acts on the qubits at the two right-handed tetrahedra adjacent to the two triangles that are contained in the projection of the face.
Alternatively, these two right-handed tetrahedra are the ones adjacent to the diagonal $(i,i+2)$ edge of a $(i,i+1,i+2)$ face, which is an edge of the $A$ or $B$ cubic lattice.
Each instrument $I[\mathbf V]$ at a spacetime cube acts on the qubits at the three right-handed tetrahedra contained in the projection of the cube.
Alternatively, these two right-handed tetrahedra are the ones adjacent to the diagonal $(i+3,i)$ edge of a $(i,i+1,i+2,i+3)$ cube.

So in total we obtain the following dynamic code.
Consider two shifted cubic lattices $A$ and $B$ together with all length-$\sqrt{\frac34}$ edges connecting $A$ and $B$, defining a triangulation whose vertices are 4-colorable as $\vca0$, $\vcb1$, $\vcc2$, or $\vcd3$.
There is one qubit at every right-handed tetrahedron.
The sequence of measurements consists in 8 rounds,
\begin{equation}
\label{floquet30_measurements}
\begin{multlined}
ZZ_{\vca0\vcc2}\rightarrow (XX,XX)_{\vcd3\vca0}\rightarrow ZZ_{\vcb1\vcd3}\rightarrow (XX,XX)_{\vca0\vcb1}\\
\rightarrow
ZZ_{\vcc2\vca0}\rightarrow (XX,XX)_{\vcb1\vcc2}
\rightarrow ZZ_{\vcd3\vcb1}\\
\rightarrow (XX,XX)_{\vcc2\vcd3}\rightarrow\;.
\end{multlined}
\end{equation}
In each round we measure either $Z_0Z_1$ on the two adjacent right-handed tetrahedra adjacent to all edges of the specified type, or we measure $X_0X_1$ and $X_1X_2$ on the three right-handed tetrahedra adjacent to all edges of that type.
Note that the rounds 0 and 4 (numbered starting from 0), as well as 2 and 6 in Eq.~\eqref{floquet30_measurements} are identical.

This Floquet code can be generalized to arbitrary triangulations with 4-colored vertices.
In every round, we measure $Z_0Z_1$, $Z_1Z_2$, $\ldots$ $Z_{i-1}Z_i$ on the set of right-handed tetrahedra adjacent to the specified type of edges in the lattice, or the same for $X$ instead of $Z$.
The Poincar\'e dual to such a lattice has 4-colorable volumes and is used in the definition of the 3-dimensional color code \cite{Bombin2006}.
However, our code involves only half of the qubits.
The dual lattice of the triangulation depicted in Eq.~\eqref{eq:4_colorable_triangulation} is known as \emph{bitruncated cubic honeycomb} \cite{Wiki_bitrunc}.
The volumes are bitruncated cubes,
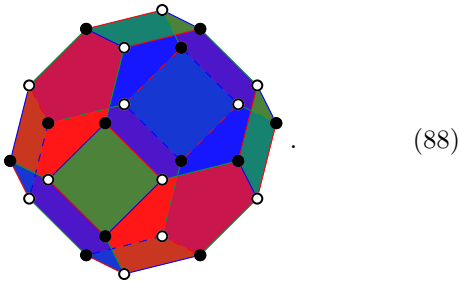
\begin{equation}
\begin{tikzpicture}
\tikzset{opac/.style={opacity=0.7}}
\atoms{void}{x/p={3,0}, y/p={1,1}, z/p={0,3}}
\atoms{void}{0/p={$0.25*(x)+0.5*(y)$}, 1/p={$0.75*(x)+0.5*(y)$}, 2/p={$(x)+0.25*(z)+0.5*(y)$}, 3/p={$(x)+0.75*(z)+0.5*(y)$}, 4/p={$0.75*(x)+0.5*(y)+(z)$}, 5/p={$0.25*(x)+0.5*(y)+(z)$}, 6/p={$0.75*(z)+0.5*(y)$}, 7/p={$0.25*(z)+0.5*(y)$}, 8/p={$0.5*(x)+0.25*(z)$}, 9/p={$0.75*(x)+0.5*(z)$}, 10/p={$0.5*(x)+0.75*(z)$}, 11/p={$0.25*(x)+0.5*(z)$}, 12/p={$0.5*(x)+0.25*(z)+(y)$}, 13/p={$0.75*(x)+0.5*(z)+(y)$}, 14/p={$0.5*(x)+0.75*(z)+(y)$}, 15/p={$0.25*(x)+0.5*(z)+(y)$}, 16/p={$0.5*(x)+0.25*(y)$}, 17/p={$(x)+0.5*(z)+0.25*(y)$}, 18/p={$(z)+0.5*(x)+0.25*(y)$}, 19/p={$0.5*(z)+0.25*(y)$}, 20/p={$0.5*(x)+0.75*(y)$}, 21/p={$(x)+0.5*(z)+0.75*(y)$}, 22/p={$(z)+0.5*(x)+0.75*(y)$}, 23/p={$0.5*(z)+0.75*(y)$}}
\fill[darkgreen,opac] (12)--(13)--(14)--(15)--cycle (7)--(23)--(6)--(19)--cycle (0)--(16)--(1)--(20)--cycle;
\fill[red,opac] (0)--(20)--(12)--(15)--(23)--(7)--cycle (13)--(21)--(3)--(4)--(22)--(14)--cycle;
\fill[blue,opac] (23)--(15)--(14)--(22)--(5)--(6)--cycle (1)--(2)--(21)--(13)--(12)--(20)--cycle;
\fill[darkgreen,opac] (2)--(21)--(3)--(17)--cycle (8)--(9)--(10)--(11)--cycle (5)--(18)--(4)--(22);
\fill[red,opac] (1)--(2)--(17)--(9)--(8)--(16)--cycle (19)--(11)--(10)--(18)--(5)--(6)--cycle;
\fill[blue,opac] (0)--(16)--(8)--(11)--(19)--(7)--cycle (9)--(17)--(3)--(4)--(18)--(10)--cycle;
\atoms{vertex}{0/p=0, 1/p=1, 17/p=17, 21/p=21, 8/p=8, 10/p=10, 12/p=12, 14/p=14, 5/p=5, 4/p=4, 19/p=19, 23/p=23}
\atoms{circ,tiny}{16/p=16, 20/p=20, 2/p=2, 3/p=3, 18/p=18, 22/p=22, 6/p=6, 7/p=7, 11/p=11, 9/p=9, 15/p=15, 13/p=13}
\draw[red] (0)--(16) (1)edge[dashed](20) (2)--(21) (3)--(17) (4)--(18) (5)--(22) (6)edge[dashed](23) (7)--(19) (9)--(10) (11)--(8) (12)edge[dashed](13) (14)edge[dashed](15);
\draw[blue] (16)--(1) (0)edge[dashed](20) (2)--(17) (3)--(21) (4)--(22) (5)--(18) (6)--(19) (7)edge[dashed](23) (8)--(9) (10)--(11) (13)edge[dashed](14) (15)edge[dashed](12);
\draw[darkgreen] (8)--(16) (9)--(17) (10)--(18) (11)--(19) (12)edge[dashed](20) (13)edge[dashed](21) (14)edge[dashed](22) (15)edge[dashed](23) (1)--(2) (3)--(4) (5)--(6) (7)--(0);
\end{tikzpicture}\;.
\end{equation}
The drawn volume is dual to a $\vcd3$ vertex.
The blue shaded 6-gon faces are dual to $\vcc2\vcd3$ edges, and the red shaded 6-gon faces to $\vcd3\vca0$ edges.
The green shaded 4-gon faces are dual to $\vcb1\vcd3$ edges.
The red, green, and blue edges are dual to $\vcb1\vcc2\vcd3$-triangles, $\vcc2\vcd3\vca0$-triangles, and $\vcd3\vca0\vcb1$-triangles, respectively.
The overall lattice also contains faces dual to $\vca0\vcb1$ edges, $\vcb1\vcc2$ edges and $\vca0\vcc2$ edges, as well as edges dual to $\vca0\vcb1\vcc2$ triangles, but none of these are contained in the boundary of the $\vcd3$ volume shown above.
There are qubits on all the full vertices, and none at the empty vertices.
The measurements in the dual lattice take place on the faces and involve the qubits at the vertices.
For example, the $ZZ_{\vcb1\vcd3}$ measurements take place simulteneously on all green 4-gon faces shown above.

Let us briefly look at the decoding procedure from Proposition~\ref{prop:path_integral_decoding} for the present code.
The spacetime syndrome measured over some time $T\sim L$ consists of one outcome at every face and two outcomes at every cube of the hypercubic lattice.
The syndrome yields a $e$ 2-chain and an $m$ 3-cochain inside the 4-dimensional modified hypercubic lattice, supported on the pillow-like volumes and the dividing $f$ and $g$ faces.
The boundary of the $m$ 3-cochain is a (0-dimensional) 4-cocyle, and the boundary of the $e$ 2-chain is a (1-dimensional) 1-cycle.
We then use the classical decoder $D$ to find a low-weight fix that turns $e$ into a 2-cycle and $m$ into a 3-cocycle.
For closing off the $e$ 2-cycle and $m$ 3-cocycle, we choose $T$ to be after a round of $I[\mathbf T]_{\vcb1\vcc2\vcd3}$ instruments.
The corresponding spatial slice of the modified hypercubic lattice at this time is obtained by (1) taking only the $\vcb1\vcc2\vcd3$ faces in the lattice in Eq.~\eqref{eq:4_colorable_triangulation}, and (2) replacing every face by two copies separated by a pillow-like volume.
The non-pillow volumes of this spatial slice are rhombic dodecahedra, each formed by the four $\vca0\vcb1\vcc2\vcd3$ cubes adjacent to a $\vca0$ vertex in Eq.~\eqref{eq:4_colorable_triangulation}.
Each $\vcb1\vcc2\vcd3$ face in Eq.~\eqref{eq:4_colorable_triangulation} has two adjacent qubits, so there is one qubit for every face of the spatial slice.
The $m$ 3-cocycle restricted to this spatial slice is again a 3-cocycle, that is, a collection of rhombic dodecahedra and pillow volumes.
We close this 3-cochain by a 2-cochain, and apply a Pauli-$X$ operator to the qubits at each face of this 2-cochain.
The $e$ 2-cycle restricted to the spatial slice becomes a 1-cycle, that is, a collection of edges.
We close this 1-cycle by a 2-chain, and apply a Pauli-$Z$ operator to the qubits at each face of this 2-cochain.
Following Proposition~\ref{prop:path_integral_decoding}, the error configurations that potentially lead to a wrong correction of the $e$ ($m$) syndrome must contain more than half of a membrane (line) of non-trivial homology.
The number of such configurations grows at most exponentially in the number of errors, whereas the probability for an error configuration shrinks exponentially, with the error probability $p$ in the basis.
Thus, for $p$ below some threshold, the probability of a logical error shrinks exponentially in the size of the smallest membrane (line) configuration of non-trivial homology, so like $\sim e^{-L}$ for the $m$ syndrome, and $\sim e^{-L^2}$ for the $e$ syndrome.
Note that the $e$ part of the syndrome could also be corrected by a local cellular automaton shinking the corresponding 1-cycle in each time step using a mechanism similar to \emph{Toom's rule} \cite{Kubica2018}.

\subsection{Dynamic double-semion string-net code}
\label{sec:double_semion_code}
In this section we will give an example for a non-Pauli fixed-point path integral code, which is based on the double-semion Turav-Viro/Dijkgraaf-Witten model \cite{Dijkgraaf1990,Turaev1992}, the state-sum version of the double-semion string-net model \cite{Levin2004,Hu2012}.
The double-semion model is not a Pauli stabilizer code, but only a commuting-projector Hamiltonian.
The model is defined a triangular lattice with a branching structure, that is, a direction assigned to all edges that is acyclic around every triangle.
The branching structure defines a local ordering of the vertices within each triangle, and accordingly allows us to label the vertices by $0$, $1$, and $2$.
There is one local ground state projector at every vertex $v$, acting on the 12 qubits at the edges of the 6 adjacent triangles.
This ground state projector is given by
\begin{equation}
(\mathbb1 + \prod_{t: v=t_1} Z_{t_{12}} \prod_{e:v\subset e} X_e \prod_{t:v\subset t} CZ_{t_{01},t_{12}})\circ\pcoc\;.
\end{equation}
Here, $e$ runs over all edges adjacent to the vertex $v$, and $t$ over all adjacent triangles.
$CZ$ denotes a controlled-$Z$ operator, $t_{01}$ and $t_{12}$ are the $01$ and the $12$ edge of the triangle $t$, and $t_1$ is the $1$ vertex of $T$.
$\pcoc$ denotes the local projector onto the subspace given by the configurations that form 1-cocycles,
\begin{equation}
\pcoc = \prod_{v\subset t} (\mathbb1 + Z_{t_{01}} Z_{t_{12}} Z_{t_{02}})\;.
\end{equation}
Note that there are in fact non-Pauli as well as Pauli stabilizer codes for this phase (and any Abelian non-chiral anyon model) \cite{Dauphinais2018,Magdalena2020,Ellison2021}.
Here, we present a dynamic non-Pauli and non-stabilizer code.
This code can be seen as somewhere between stabilizer and Floquet codes, since the anyon worldlines forming the spacetime syndrome move in a fixed direction, but this direction does not coincide with the $t$ direction.
Apart from this, our code has some similarities to recent protocols for syndrome extraction for the non-Abelian double-Fibonacci string-net model presented in Ref.~\cite{Schotte2020}.
The goal here is not to produce a particularly practical code, but rather to demonstrate the applicability of our framework beyond the toric-code phase.

We consider a path integral defined on any $2+1$-dimensional triangulation with a branching structure, that is, a direction for all the edges that is acyclic around every triangle.
As for the toric code, the path integral is a sum over $\zz_2$-valued 1-cocycles $A$, but now there is a non-trivial action $(-1)^{A\cup A\cup A}$.
That is, the state sum has an additional weight,
\begin{equation}
\omega(a,b,c)=\omega_{a,b,c}=(-1)^{abc}\;,
\end{equation}
at every tetrahedron,
\begin{equation}
\label{eq:statesum_tetrahedron}
\begin{tikzpicture}
\atoms{vertex}{0/, 1/p={-150:1}, 2/p={-30:1}, 3/p={90:1}}
\draw (0)edge[mark={arr,-}](1) (0)edge[mark={arr,-}, mark={slab=$c$,p=0.4}](2) (0)edge[mark={arr,-}](3) (1)edge[mark=arr](2) (2)edge[mark={arr,-}, mark={slab=$b$,r}](3) (1)edge[mark=arr, mark={slab=$a$}](3);
\end{tikzpicture}\;.
\end{equation}
Note that since the configuration of state-sum variables is restricted to a 1-cocycle, the variables at the three edges of each triangle sum to $0$ mod $2$.
Thus, the variables at two edges of a triangle determine the variable on the third edge.
Hence, all variables on the 6 edges of the tetrahedron are determined by the variables at three ``generating'' edges, which are labeled $a$, $b$, and $c$ above.
The state sum can be written as a tensor network with one $\delta$-tensor at every edge, one 3-index $\zz_2$ tensor at every face, and one 3-index $\omega$ tensor at every tetrahedron.
This path integral is invariant under Pachner moves including the one depicted in Eq.~\eqref{eq:pachner_move_equation}.
The equations corresponding to this invariance are equivalent to the fact that $\omega$ is a $\zz_2$ group 3-cocycle.
The string-net picture of this model is obtained by considering space-only Poncar\'e dual lattices.

We can equip this path integral with anyon worldlines.
Geometrically, these worldlines are represented by sequences of cylinder-like 3-cells or \emph{tube segments} embedded into the triangulation.
The boundary of such a tube segment consists of two \emph{anyon 1-gons} (in red at the bottom and top) and one rectangle (wrapping around the side) which can be divided into two triangles,
\begin{equation}
\label{eq:anyon_segment}
\begin{tikzpicture}
\atoms{vertex}{0/, 1/p={0,1.4}}
\draw[mark={slab=$g$,r,p=0.75},mark={arr,p=0.4},looseness=3,anyon1gon] (0)to[out=0,in=0]++(0,0.5)to[out=180,in=180](0);
\draw[mark={arr,p=0.4},looseness=3,anyon1gon] (1)to[out=0,in=0]++(0,0.5)to[out=180,in=180](1);
\draw[mark={slab=$h$,r,p=0.7},mark={arr,p=0.7}] (0)--(1);
\draw[dashed, mark={arr,p=0.7},looseness=2.5] (0)to[out=30,in=-30]++(0,0.7)to[out=150,in=-150](1);
\end{tikzpicture}\;.
\end{equation}
There are further tube segments attached to the two anyon 1-gons, and ordinary tetrahedra attached to the two triangles.
There are no additional state-sum variables other than the group elements at each edge, but there is an additional state-sum weight
\begin{equation}
\rho_{g,h}
\end{equation}
at each tube segment with $\zz_2$ variables as in Eq.~\eqref{eq:anyon_segment}.
There are four types of anyons, $1$, $s$, $\bar s$, and $s\bar s$, and the associated weights are
\begin{equation}
\begin{gathered}
\rho^1_{g,h}=\delta_{g,0}\;,\\
\rho^s_{g,h}=\delta_{g,1} i^{h}\;,\\
\rho^{\bar s}_{g,h}=\delta_{g,1}(-i)^h\;,\\
\rho^{s\bar s}_{g,h}=\delta_{g,0} (-1)^h\;.
\end{gathered}
\end{equation}
The different $\rho^x$ are irreducible representations of the \emph{tube algebra} defined by $\omega$ \cite{Bullivant2019,Lan2013}.
For a review of defects in the path integral language used here, see Appendix~D of Ref.~\cite{Magdalena2023}.
The string-net analogue of this way of introducing anyons as explicit defects is given in Ref.~\cite{Hu2015}.

We now consider this path integral on a triangulation consisting of two cubic lattices $A$ and $B$ with unit vectors $x$, $y$, and $z$, shifted by relative to each other by $\frac12 x+\frac12 y+\frac12 z$.
Each tetrahedron is formed by one $A$ edge, one nearby $B$ edge, as well as four length-$\sqrt{\frac34}$-edges connecting $A$ vertices with nearby $B$ vertices.
So this is the same as the lattice depicted in Eq.~\eqref{eq:4_colorable_triangulation}, just that we color all $A$ vertices red and all $B$ vertices green.
The branching structure can be chosen such that for every directed edge with associated vector $ax+by+cz$, we have $a+b+c>0$.

We turn the path integral into a circuit of operators choosing $t=z$ as the time direction.
There are two kinds of operators in the circuit which correspond to different volumes as follows.
For every $t$ edge, there is an operator $T_1$ consisting of the four adjacent tetrahedra, acting on 8 qubits (here with coloring for a $t$ edge of $B$),
\begin{equation}
\begin{gathered}
\begin{tikzpicture}
\draw (-0.6,0.3)edge[mark={arr,e},ind=$x$]++(0:0.3) (-0.6,0.3)edge[mark={arr,e},ind=$y$]++(40:0.3) (-0.6,0.3)edge[mark={arr,e},ind=$t$]++(90:0.3);
\end{tikzpicture}\qquad
T_1\coloneqq
\begin{tikzpicture}
\atoms{vertex,vca}{0/, 1/p={1.6,0}, 2/p={0.5,1}, 3/p={2.1,1}}
\atoms{vertex,vcb}{m0/p={1,0.3}, m1/p={1,1.3}}
\draw[dashed] (2)edge[mark=arr](3) (m0)edge[mark=arr](1) (m0)edge[mark=arr](2) (m0)edge[mark=arr](3) (m0)edge[mark={arr,-}](0);
\draw[dotted] (m0)edge[mark=arr](m1);
\draw (0)edge[mark=arr](1) (1)edge[mark=arr](3) (0)edge[mark=arr](2) (m1)edge[mark={arr,-}](1) (m1)edge[mark={arr,-}](2) (m1)edge[mark=arr](3) (m1)edge[mark={arr,-}](0);
\end{tikzpicture}
\;,
\\
\begin{multlined}
T_1 \Ket{
\begin{tikzpicture}
\atoms{vertex,vca}{0/, 1/p={1.6,0}, 2/p={0,1.6}, 3/p={1.6,1.6}}
\atoms{vertex,vcb}{m/p={0.8,0.8}}
\draw (0)edge[mark={slab=$\scriptstyle a$}](2) (2)edge[mark={slab=$\scriptstyle b$}](3) (3)edge[mark={slab=$\scriptstyle c$}](1) (1)edge[mark={slab=$\scriptstyle d$}](0) (m)edge[mark={slab=$\scriptstyle e$}](0) (m)edge[mark={slab=$\scriptstyle f$}](2) (m)edge[mark={slab=$\scriptstyle g$}](3) (m)edge[mark={slab=$\scriptstyle h$}](1);
\end{tikzpicture}
}
\\
=
\sum_y
\omega_{e,f,f+y}\omega_{f,f+y,g+y}\omega_{h,h+y,g+y}\omega_{e,h,h+y}\\
\pcoc \Ket{
\begin{tikzpicture}
\atoms{vertex,vca}{0/, 1/p={1.6,0}, 2/p={0,1.6}, 3/p={1.6,1.6}}
\atoms{vertex,vcb}{m/p={0.8,0.8}}
\draw (0)edge[mark={slab=$\scriptstyle a$}](2) (2)edge[mark={slab=$\scriptstyle b$}](3) (3)edge[mark={slab=$\scriptstyle c$}](1) (1)edge[mark={slab=$\scriptstyle d$}](0) (m)edge[mark={slab=$\scriptstyle e+y$,p=0.4,sideoff=-0.1}](0) (m)edge[mark={slab=$\scriptstyle f+y$,p=0.4,sideoff=-0.1}](2) (m)edge[mark={slab=$\scriptstyle g+y$,p=0.4,sideoff=-0.1}](3) (m)edge[mark={slab=$\scriptstyle h+y$,p=0.4,sideoff=-0.1}](1);
\end{tikzpicture}
}
\;.
\end{multlined}
\end{gathered}
\end{equation}
The large ket around diagram denotes the computational basis state given by the labels, supported on the drawn edges.
$\pcoc$ acting on a triangle with edge labels $a$, $b$, and $c$ is the projector onto the \emph{cocycle subspace}, spanned by the configurations that fulfil $a+b=c$.
Here and in the following, we also use $\pcoc$ for the product of $\pcoc$ on all the triangles that are currently acted on.
As shown, $T_1$ contains the $\omega$-tensors of the involved tetrahedra, and the $\zz_2$-tensors at the internal and bottom faces.
The $\delta$-tensors at the edges of the lattice are split between the adjacent volumes.

For every $x$ or $y$ edge of $A$ or $B$ there is an operator $V_1$ consisting of the tetrahedron spanned by this edge and the $y$ or $x$ edge of $B$ or $A$ whose center is shifted by $\frac12 t$,
\begin{equation}
\label{eq:v1_definition}
\begin{gathered}
V_1\coloneqq
\begin{tikzpicture}
\atoms{vertex,vca}{1/p={1.2,0}, 3/p={1.5,0.8}}
\atoms{vertex,vcb}{m1/p={0.8,1}, m2/p={2,1}}
\draw[dashed] (1)edge[mark=arr](3) (m1)edge[mark=arr](3) (3)edge[mark=arr](m2);
\draw (1)edge[mark=arr](m1) (m1)edge[mark=arr](m2) (1)edge[mark=arr](m2);
\end{tikzpicture}\;,\\
V_1\Ket{
\begin{tikzpicture}
\atoms{vertex,vca}{2/p={0.5,0.5}, 3/p={0.5,-0.5}}
\atoms{vertex,vcb}{0/, 1/p={1,0}}
\draw (0)edge[mark={slab=$\scriptstyle a$}](2) (2)edge[mark={slab=$\scriptstyle b$}](1) (1)edge[mark={slab=$\scriptstyle c$}](3) (3)edge[mark={slab=$\scriptstyle d$}](0) (2)edge[mark={slab=$\scriptstyle e$}](3);
\end{tikzpicture}
}
=
\omega_{d,a,b} \pcoc
\Ket{
\begin{tikzpicture}
\atoms{vertex,vca}{2/p={0.7,0.7}, 3/p={0.7,-0.7}}
\atoms{vertex,vcb}{0/, 1/p={1.4,0}}
\draw (0)edge[mark={slab=$\scriptstyle a$}](2) (2)edge[mark={slab=$\scriptstyle b$}](1) (1)edge[mark={slab=$\scriptstyle c$}](3) (3)edge[mark={slab=$\scriptstyle d$}](0) (0)edge[mark={slab=$\scriptstyle e+d+b$}](1);
\end{tikzpicture}
}\;.
\end{gathered}
\end{equation}
Neither $T_1$ nor $V_1$ are unitary since we have
\begin{equation}
\label{eq:t_cocycle_restriction}
T_1 = \pcoc T_1=T_1\pcoc = \pcoc T_1\pcoc\;,
\end{equation}
and the same for $V_1$ instead of $T_1$.
So the support of $T_1$ and $V_1$ is contained in the cocycle subspace of the involved triangles.
Restricted to this cocycle subspace, $V_1$ is indeed unitary,
\begin{equation}
\label{eq:omega_unitarity}
V_1^\dagger V_1=
\pcoc
=
\begin{tikzpicture}
\atoms{vertex,vcb}{1/, 3/p={2,0}}
\atoms{vertex,vca}{m1/p={1.2,0.6}, m2/p={0.8,-0.6}}
\draw[dashed] (m2)edge[bend right,mark=arr](m1);
\draw (m2)edge[bend left,mark=arr](m1) (1)edge[mark=arr](m1) (1)edge[mark=arr](m2) (m1)edge[mark=arr](3) (m2)edge[mark=arr](3);
\end{tikzpicture}
\;.
\end{equation}
On the right, we have depicted the corresponding volume that arises from gluing the tetrahedron with a reflected copy.
\footnote{This is just the unitarity condition for the unitary fusion category defined by $\omega$.}
This is not the case for $T_1$, whose support is contained in but not equal to the cocycle subspace.
We will now show how to extend $T_1$ to an isometry that is fully supported on the cocycle subspace, and later extend both $T_1$ and $V_1$ to the full Hilbert space using a different method.
To this end, we slightly modify the spacetime lattice to incorporate anyon worldlines running along the $x+y+t$ direction.
We consider all the edges aligned with the $x+y-t$ direction.
We split every such edge into two edges separated by a 2-gon perpendicular to the $x+y+t$ direction.
Then we insert an anyon 1-gon into each such 2-gon, at the vertex with the smaller $t$ component, for example,
\begin{equation}
\begin{tikzpicture}
\atoms{vertex,vca}{0/}
\atoms{vertex,vcb}{1/p=10:1.3}
\draw (0)edge[mark=arr](1);
\end{tikzpicture}
\rightarrow
\begin{tikzpicture}
\atoms{void}{0/,1/p=10:1.3}
\draw (0)edge[bend left=35,mark=arr](1) (0)edge[bend right=35,mark=arr](1);
\draw[anyon1gon,mark=arr] (1-c)to[out=160,in=100]++(-170:0.6)to[out=-80,in=-140]cycle;
\atoms{vertex,vca}{0x/p={0}}
\atoms{vertex,vcb}{1x/p={1}}
\end{tikzpicture}
\;.
\end{equation}
The $T_1$ volume then gets two anyon 1-gons at its boundary, which we connect using an anyon tube along the $x+y+t$ edge,
\begin{equation}
\label{eq:diagonal_tube_segment}
\begin{tikzpicture}
\atoms{vertex,vca}{0/, 1/p={2,0}, 2/p={0.5,1.2}, 3/p={2.5,1.2}}
\atoms{vertex,vcb}{m0/p={1.2,0.4}, m1/p={1.2,1.5}}
\draw[dashed] (2)edge[mark=arr](3) (m0)edge[mark=arr](1) (m0)edge[mark=arr](2) (m0)edge[mark=arr](3) (m0)edge[bend left=20,mark={arr,-}](0) (m0)edge[bend right=40,mark={arr,-}](0);
\draw[dashed,anyon1gon,mark={arr,-}] (m0-c)to[out=-150,in=-80]++(-170:0.6)to[out=100,in=160]cycle;
\draw (0)edge[mark=arr](1) (1)edge[mark=arr](3) (0)edge[mark=arr](2) (m1)edge[mark={arr,-}](1) (m1)edge[mark={arr,-}](2) (m1)edge[mark={arr,-}](0) (m1)edge[out=0,in=140,mark=arr](3) (m1)edge[out=-40,in=-150,mark=arr](3);
\draw[anyon1gon,mark={arr,-}] (3-c)to[out=-155,in=-85]++(175:0.6)to[out=95,in=145]cycle;
\end{tikzpicture}\;.
\end{equation}
With this, we can replace $T_1$ by a collection of partial isometries $\mathbf T=(T_x)_{x\in \{1,s,\bar s,s\bar s\}}$,
\begin{equation}
\label{eq:anyon_isometry}
\begin{gathered}
T_x\coloneqq
\begin{tikzpicture}
\atoms{vertex,vca}{0/, 1/p={2,0}, 2/p={0.5,1.2}, 3/p={2.5,1.2}}
\atoms{vertex,vcb}{m0/p={1.2,0.4}, m1/p={1.2,1.5}}
\draw[dashed] (2)--(3) (m0)--(1) (m0)--(2) (m0)--(3) (m0)to[bend left=20](0) (m0)to[bend right=40](0);
\draw[dashed,anyon1gon] (m0-c)to[out=-150,in=-80]++(-170:0.6)to[out=100,in=160]cycle;
\draw (0)--(1)--(3) (2)--(0) (m1)--(1) (m1)--(2) (m1)--(0) (m1)to[out=0,in=140](3) (m1)to[out=-40,in=-150](3);
\draw[anyon1gon] (3-c)to[out=-155,in=-85]++(175:0.6)to[out=95,in=145]cycle;
\node at (1.4,0.7){\textcolor{red}{$\rho^x$}};
\end{tikzpicture}\;,\\
\begin{multlined}
T_x \Ket{
\begin{tikzpicture}
\atoms{vertex,vca}{0/, 1/p={1.6,0}, 2/p={0,1.6}, 3/p={1.6,1.6}}
\atoms{vertex,vcb}{m/p={0.8,0.8}}
\draw (0)edge[mark={slab=$\scriptstyle a$}](2) (2)edge[mark={slab=$\scriptstyle b$}](3) (3)edge[mark={slab=$\scriptstyle c$}](1) (1)edge[mark={slab=$\scriptstyle d$}](0) (m)edge[mark={slab=$\scriptstyle f$}](2) (m)edge[mark={slab=$\scriptstyle g$}](3) (m)edge[mark={slab=$\scriptstyle h$}](1) (m)edge[bend left,mark={slab=$\scriptstyle i$}](0) (m)edge[bend right,mark={slab=$\scriptstyle e$,r}](0);
\draw[anyon1gon,mark={slab=$\scriptstyle j$,r}] (m-c)to[out=-160,in=135]++(-135:0.5)to[out=-45,in=-110]cycle;
\end{tikzpicture}
}\\
=
\sum_y
\rho^x_{j,g}
\omega_{i,h,h+y} \omega_{h,h+y,g+y}\\
\omega_{e,f,f+j+y}\omega_{f,f+j+y,g+j+y}
\omega_{i+y,g+j+y,j}\omega_{e,g,j} \omega_{e,j,g}
\\
\pcoc
\Ket{
\begin{tikzpicture}
\atoms{vertex,vca}{0/, 1/p={1.6,0}, 2/p={0,1.6}, 3/p={1.6,1.6}}
\atoms{vertex,vcb}{m/p={0.8,0.8}}
\draw (0)edge[mark={slab=$\scriptstyle a$}](2) (2)edge[mark={slab=$\scriptstyle b$}](3) (3)edge[mark={slab=$\scriptstyle c$}](1) (1)edge[mark={slab=$\scriptstyle d$}](0) (m)edge[mark={slab=$\scriptstyle \substack{f+j\\+y}$,p=0.3,sideoff=-0.1}](2) (m)edge[mark={slab=$\scriptstyle i+y$,sideoff=-0.1,p=0.6}](0) (m)edge[mark={slab=$\scriptstyle h+y$,r,sideoff=-0.1,p=0.4}](1) (m)edge[bend left,mark={slab=$\scriptstyle \substack{g+j\\+y}$,sideoff=-0.1,p=0.3}](3) (m)edge[bend right,mark={slab=$\scriptstyle g+y$,r,sideoff=-0.1,p=0.3}](3);
\draw[anyon1gon,mark={slab=$\scriptstyle j$,r}] (3-c)to[out=-160,in=135]++(-135:0.5)to[out=-45,in=-110]cycle;
\end{tikzpicture}
}\;.
\end{multlined}
\end{gathered}
\end{equation}
Here we have used a cellulation of the volume with one anyon tube and 7 tetrahedra.
$\mathbf T$ is indeed an isometry when restricted to the cocycle subspace,
\begin{equation}
\mathbf T^\dagger \mathbf T = \sum_x T_x^\dagger T_x =P_{\text{cocyc}}\;.
\end{equation}
In order to see this, we compute $T_x^\dagger T_x$ by gluing Eq.~\eqref{eq:anyon_isometry} with a time-reflected copy and using the topological invariance, yielding a projector,
\begin{equation}
\label{eq:anyon_projector}
T_x^\dagger T_x=
\begin{tikzpicture}
\atoms{vertex,vca}{0/, 1/p={2,0}, 2/p={0.5,1.2}, 3/p={2.5,1.2}}
\atoms{vertex,vcb}{m0/p={1.2,0.4}, m1/p={1.2,1.5}}
\draw[dashed] (m0)--(1) (m0)--(2) (m0)--(3) (m0)to[bend left=20](0) (m0)to[bend right=40](0);
\draw[dashed,anyon1gon] (m0-c)to[out=-150,in=-80]++(-170:0.6)to[out=100,in=160]cycle;
\draw[dotted] (m0)--(m1);
\draw (0)--(1)--(3)--(2)--(0) (m1)--(1) (m1)--(2) (m1)--(3) (m1)to[bend left=40](0) (m1)to[bend right=20](0);
\draw[anyon1gon] (m1-c)to[out=-90,in=-30]++(-120:0.6)to[out=150,in=-140]cycle;
\end{tikzpicture}\;,
\end{equation}
where the bottom and top 1-gon are connected via a tube segment along the $t$ edge.
\footnote{This is a projector since gluing two copies of this volume stacked on top of each other yields the same volume, which corresponds to the equation $P^2=P$.}
Then we compute the sum over all tube segments,
\begin{equation}
\rho^1_{g,h}+\rho^s_{g,h}+\rho^{\bar s}_{g,h}+\rho^{s\bar s}_{g,h} = \delta_{h,0}\;.
\end{equation}
Setting $h$ to $0$ geometrically corresponds to removing the anyon tube and the $t$ edge in Eq.~\eqref{eq:anyon_projector}, and identifying the loop edges at the top and bottom.
So we obtain the following volume of solid-torus topology:
\begin{equation}
\mathbf T^\dagger \mathbf T
=
\sum_x T_x^\dagger T_x=
\begin{tikzpicture}
\atoms{vertex}{0/, 1/p={2,0}, 2/p={0.5,1.2}, 3/p={2.5,1.2}, m/p={1.2,0.6}}
\draw[dashed] (m)to[bend right=10](1) (m)to[bend left=10](2) (m)to[bend right=10](3) (m)to[out=-110,in=20](0) (m)to[out=180,in=80](0);
\draw (m-c)to[out=-170,in=135]++(-150:0.6)to[out=-45,in=-110]cycle;
\draw (0)--(1)--(3)--(2)--(0) (m)to[bend left=10](1) (m)to[bend right=10](2) (m)to[bend left=10](3) (m)to[out=-90,in=10](0) (m)to[out=-170,in=70](0);
\end{tikzpicture}
=
\pcoc\;.
\end{equation}
For the last equation we have used that this spacetime volume can be obtained from gluing one volume as in Eq.~\eqref{eq:omega_unitarity} for every pair of neighboring triangles.
With this, using $\mathbf T$ in Eq.~\eqref{eq:fixed_point_instrument} defines an instrument $I[\mathbf T]$ restricted to the cocycle subspace.

We will now discuss how to extend $I[\mathbf T]$ and $I[\mathbf V]$ to full instruments also outside the cocycle subspace.
The first step is to choose arbitrary extensions $\widetilde{\mathbf T}$ and $\widetilde{\mathbf V}$ to the full Hilbert space.
\footnote{In general, this might also involve enlarging the output dimension by adding new measurement outcomes. This is not necessary in the present case though.}
However, the circuit consisting of the extended instruments $I[\widetilde{\mathbf T}]$ and $I[\widetilde{\mathbf V}]$ clearly violates Definition~\ref{def:path_integral_code}.
This can be fixed by introducing a new channel $C$ to the circuit, with the following task:
$C$ measures whether the cocycle constraint is violated at any of the triangles, and maps back to the cocycle subspace if yes.
Roughly speaking, this works because (1) $\widetilde{\mathbf T}$ and $\widetilde{\mathbf V}$ still preserve the cocycle subspace,
\begin{equation}
\label{eq:cchannel_commutativity}
\begin{gathered}
\widetilde{T}_x\circ \pcoc = T_x = \pcoc\circ T_x\;,\\
\widetilde{V}_x\circ \pcoc = V_x = \pcoc\circ V_x\;,
\end{gathered}
\end{equation}
and (2) $\pcoc$ consists of the same triangle terms for each isometry.

Concretely, it suffices to apply a channel $C$ before every $I[\widetilde{\mathbf T}]$ instrument.
The space that $\widetilde{\mathbf T}$ acts on is given by
\begin{equation}
\label{eq:triangle_correction}
\begin{tikzpicture}
\atoms{void}{b/}
\atoms{vertex,vca}{a0/p={-1,-1}, a1/p={1,-1}, a2/p={-1,1}, a3/p={1,1}}
\draw (a0)edge[mark={slab=$a$,r}](a1) (a0)edge[mark={slab=$b$}](a2) (a1)edge[mark={slab=$c$,r}](a3) (a2)edge[mark={slab=$d$}](a3) (b)edge[mark={slab=$e$}](a3) (b)edge[mark={slab=$f$,r}](a1) (b)edge[mark={slab=$g$}](a2) (b)edge[bend left=35,mark={slab=$h$,p=0.4}](a0) (b)edge[bend right=35,mark={slab=$i$,r,p=0.4}](a0);
\draw[anyon1gon,mark={slab=$j$,r}] (b-c)to[out=-165,in=135]++(-135:0.8)to[out=-45,in=-105]cycle;
\atoms{vertex,vcb}{bx/p=b}
\end{tikzpicture}\;,
\end{equation}
and $C$ acts on that same space.
$C$ is the product of one 3-qubit channel $C^t$ for each of the 5 different triangles,
\begin{equation}
C^t_{c,e,f}\rightarrow C^t_{d,e,g}\rightarrow C^t_{f,a,h}\rightarrow C^t_{g,b,i}\rightarrow C^t_{h,i,j}\;.
\end{equation}
Each instrument $C^t$ acts on the qubits at the three edges of the triangle, as indicated by the labels which refer to Eq.~\eqref{eq:triangle_correction}.
The 3-qubit instrument $C^t_{a,b,c}$ is defined as follows.
First we measure $x=a+b+c\mod 2$, which is the same as a $Z_0Z_1Z_2$ measurement just that we label the outcome with $x\in \{0,1\}$ instead of $\pm 1$.
Then we apply a classically controlled operation $c\rightarrow c+x$, which is the same as a CNOT after turning the classical bit $x$ into a qubit.
In other words, $C^t_{a,b,c}$ fixes the cocycle condition by flipping the edge $c$, and $C$ pushes potential cocycle constraint violations into the anyon 1-gon.
It is easy to see that $C$ (1) maps everything into the cocycle subspace,
\begin{equation}
\label{eq:pcoc_cchannel1}
C=(\pcoc\otimes\pcoc)\circ C\;,\\
\end{equation}
and (2) acts as the identity inside the cocycle subspace,
\begin{equation}
\label{eq:pcoc_cchannel2}
C\circ(\pcoc\otimes\pcoc) = \pcoc\otimes \pcoc\;.
\end{equation}

With this, the complete QEC circuit consists of 6 rounds of channels/instruments.
First we apply $I[\widetilde{\mathbf T}]$ for every $t$ edge of $A$ whose center is within a fixed $xy$ plane of the $B$ lattice, and apply the according operator $C$ before that.
Then we apply $I[\widetilde{\mathbf V}]$ at all $x$ and all $y$ edges of $B$ inside this $xy$ plane.
We then shift the $xy$ plane by $\frac12 t$ and perform the same instruments with $A$ and $B$ exchanged.
In total we obtain
\begin{equation}
\label{eq:doublesemion_floquet_circuit}
\begin{multlined}
\rightarrow C_{At} \rightarrow  I[\widetilde{\mathbf T}]_{At} \rightarrow (I[\widetilde{\mathbf V}]_{Bx},I[\widetilde{\mathbf V}]_{By})\\
\rightarrow C_{Bt} \rightarrow  I[\widetilde{\mathbf T}]_{Bt} \rightarrow (I[\widetilde{\mathbf V}]_{Ax},I[\widetilde{\mathbf V}]_{Ay})\rightarrow\;.
\end{multlined}
\end{equation}
Let us now show that this circuit defines a valid path-integral QEC circuit according to Definition~\ref{def:path_integral_code}.
To this end, we use the tensor-network equations Eq.~\eqref{eq:pcoc_cchannel1}, Eq.~\eqref{eq:cchannel_commutativity}, and Eq.~\eqref{eq:pcoc_cchannel2} transform the circuit in Eq.~\eqref{eq:doublesemion_floquet_circuit} into the circuit
\begin{equation}
\label{eq:doublesemion_simple_circuit}
\begin{multlined}
\rightarrow I[\mathbf T]_{At} \rightarrow (I[\mathbf V]_{Bx},I[\mathbf V]_{By})\\
\rightarrow I[\mathbf T]_{Bt} \rightarrow (I[\mathbf V]_{Ax},I[\mathbf V]_{Ay})\rightarrow\;.
\end{multlined}
\end{equation}
Specifically, applying Eq.~\eqref{eq:pcoc_cchannel1} to all channels $C_{At}$/$C_{Bt}$ inserts $\pcoc$ on all triangles of the corresponding spatial cut of the lattice (here coloring like before $C_{Bt}$),
\begin{equation}
\begin{tikzpicture}
\clip (0.25,0.25)rectangle (3.25,2.75);
\foreach \x in {0,1,2,3,4}{
\foreach \y in {0,1,2,3}{
\atoms{vertex}{{a\x\y/vca,p={\x,\y}}, {b\x\y/vcb,p={\x+0.5,\y+0.5}}}
\atoms{void}{{a\x\y/p={\x,\y}}, {b\x\y/p={\x+0.5,\y+0.5}}}
}
}
\foreach[count=\xx from 1] \x in {0,1,2,3}{
\foreach[count=\yy from 1] \y in {0,1,2}{
\draw (a\x\y)--(a\xx\y) (a\x\y)--(a\x\yy) (b\x\y)--(a\xx\yy) (b\x\y)--(a\xx\y) (b\x\y)--(a\x\yy) (b\x\y)to[bend left=35](a\x\y) (b\x\y)to[bend right=35](a\x\y);
\draw[anyon1gon] (b\x\y-c)to[out=-165,in=135]++(-135:0.4)to[out=-45,in=-105]cycle;
}
}
\foreach \x in {0,1,2,3,4}{
\foreach \y in {0,1,2,3}{
\atoms{vertex}{{xa\x\y/vca,p={a\x\y}}, {xb\x\y/vcb,p={b\x\y}}}
}
}
\end{tikzpicture}\;.
\end{equation}
Then applying Eq.~\eqref{eq:cchannel_commutativity} moves $\pcoc$ to different spatial cuts.
Finally, applying Eq.~\eqref{eq:pcoc_cchannel2} removes all the channels $C_{Bt}$/$C_{At}$.
The remaining $\pcoc$ can be absorbed into the following $I[\mathbf T]_{Bt}$/$I[\mathbf T]_{At}$ using Eq.~\eqref{eq:t_cocycle_restriction}.
The transformation implies that the circuit in Eq.~\eqref{eq:doublesemion_floquet_circuit} is in the same fixed-point phase as the circuit in Eq.~\eqref{eq:doublesemion_simple_circuit}.
Since for the circuit in Eq.~\eqref{eq:doublesemion_simple_circuit}, every spacetime syndrome corresponds to a fixed-point path integral with anyon worldlines, the circuit in Eq.~\eqref{eq:doublesemion_floquet_circuit} fulfils Definition~\ref{def:path_integral_code} as well.

Depending on how we map the circuit onto a fixed set of qubits, $I[\widetilde{\mathbf T}]$ acts on at least 10 qubits.
So in order to implement it in practice we should decompose it into smaller gates.
Surely, any gate can be written as a circuit using a small fixed universal gate set, but this circuit might be approximate and finding it might be hard for such a large operator.
However, a first decomposition can be obtained by decomposing the volume in Eq.~\eqref{eq:anyon_isometry} into tetrahedra or at least smaller volumes.
Let us give such a decomposition as a sequence of spatial lattices that we get from gluing these smaller volumes step by step,
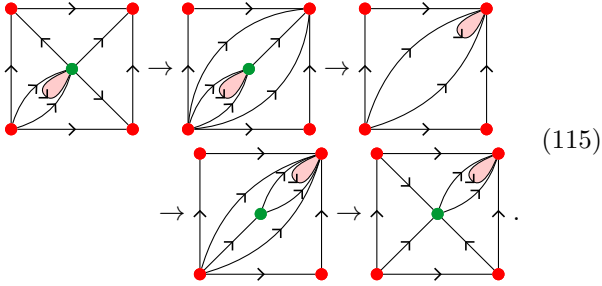
\begin{equation}
\begin{multlined}
\begin{tikzpicture}
\atoms{vertex,vca}{0/, 1/p={1.6,0}, 2/p={0,1.6}, 3/p={1.6,1.6}}
\atoms{void}{m/p={0.8,0.8}}
\draw (0)edge[mark=arr](2) (2)edge[mark=arr](3) (1)edge[mark=arr](3) (0)edge[mark=arr](1) (m)edge[mark=arr](2) (m)edge[mark=arr](3) (m)edge[mark=arr](1) (m)edge[bend left,mark={arr,-}](0) (m)edge[bend right,mark={arr,-}](0);
\draw[anyon1gon,mark=arr] (m-c)to[out=-160,in=135]++(-135:0.5)to[out=-45,in=-110]cycle;
\atoms{vertex,vcb}{m0/p=m}
\end{tikzpicture}
\rightarrow
\begin{tikzpicture}
\atoms{vertex,vca}{0/, 1/p={1.6,0}, 2/p={0,1.6}, 3/p={1.6,1.6}}
\atoms{void}{m/p={0.8,0.8}}
\draw (0)edge[mark=arr](2) (2)edge[mark=arr](3) (1)edge[mark=arr](3) (0)edge[mark=arr](1) (m)edge[mark=arr](3) (0)edge[bend right=40,mark=arr](3) (0)edge[bend left=40,mark=arr](3) (m)edge[bend left,mark={arr,-}](0) (m)edge[bend right,mark={arr,-}](0);
\draw[anyon1gon,mark=arr] (m-c)to[out=-160,in=135]++(-135:0.5)to[out=-45,in=-110]cycle;
\atoms{vertex,vcb}{m0/p=m}
\end{tikzpicture}
\rightarrow
\begin{tikzpicture}
\atoms{void}{0/, 1/p={1.6,0}, 2/p={0,1.6}, 3/p={1.6,1.6}}
\draw (0)edge[mark=arr](2) (2)edge[mark=arr](3) (1)edge[mark=arr](3) (0)edge[mark=arr](1) (0)edge[bend right,mark=arr](3) (0)edge[bend left,mark=arr](3);
\draw[anyon1gon,mark=arr] (3-c)to[out=-160,in=135]++(-135:0.5)to[out=-45,in=-110]cycle;
\atoms{vertex,vca}{0x/p=0, 1x/p=1, 2x/p=2, 3x/p=3}
\end{tikzpicture}
\\
\rightarrow
\begin{tikzpicture}
\atoms{void}{0/, 1/p={1.6,0}, 2/p={0,1.6}, 3/p={1.6,1.6}}
\atoms{void}{m/p={0.8,0.8}}
\draw (0)edge[mark=arr](2) (2)edge[mark=arr](3) (1)edge[mark=arr](3) (0)edge[mark=arr](1) (m)edge[mark={arr,-}](0) (0)edge[bend right,mark=arr](3) (0)edge[bend left,mark=arr](3) (m)edge[bend left,mark={arr}](3) (m)edge[bend right,mark={arr}](3);
\draw[anyon1gon,mark=arr] (3-c)to[out=-160,in=135]++(-135:0.5)to[out=-45,in=-110]cycle;
\atoms{vertex,vcb}{m0/p=m}
\atoms{vertex,vca}{0x/p=0, 1x/p=1, 2x/p=2, 3x/p=3}
\end{tikzpicture}
\rightarrow
\begin{tikzpicture}
\atoms{void}{0/, 1/p={1.6,0}, 2/p={0,1.6}, 3/p={1.6,1.6}}
\atoms{void}{m/p={0.8,0.8}}
\draw (0)edge[mark=arr](2) (2)edge[mark=arr](3) (1)edge[mark=arr](3) (0)edge[mark=arr](1) (m)edge[mark={arr,-}](0) (m)edge[mark={arr,-}](1) (m)edge[mark={arr,-}](2) (m)edge[bend left,mark={arr}](3) (m)edge[bend right,mark={arr}](3);
\draw[anyon1gon,mark=arr] (3-c)to[out=-160,in=135]++(-135:0.5)to[out=-45,in=-110]cycle;
\atoms{vertex,vcb}{m0/p=m}
\atoms{vertex,vca}{0x/p=0, 1x/p=1, 2x/p=2, 3x/p=3}
\end{tikzpicture}\;.
\end{multlined}
\end{equation}
In the first step we glue two tetrahedra, applying twice a 5-qubit operators $U_1$.
The same happens in the last step with an operator $R_1$.
$U_1$ and $R_1$ are the same as $V_1$ shown in Eq.~\eqref{eq:v1_definition} except that the involved edges have different directions.
In the second step, the volume we glue can be cellulated with an anyon tube together with two tetrahedra, defining an operator $S_x$ acting on 6 qubits,
\begin{equation}
\label{eq:s_definition}
\begin{multlined}
S_x
\Ket{
\begin{tikzpicture}
\atoms{vertex,vca}{0/, 3/p={1.6,1.6}}
\atoms{void}{m/p={0.8,0.8}}
\draw (m)edge[mark=arr,mark={slab=$\scriptstyle a$}](3) (0)edge[bend right=40,mark=arr,mark={slab=$\scriptstyle b$,r}](3) (0)edge[bend left=40,mark=arr,mark={slab=$\scriptstyle c$}](3) (m)edge[bend left,mark={arr,-},mark={slab=$\scriptstyle d$,p=0.3,sideoff=-0.1}](0) (m)edge[bend right,mark={arr,-},mark={slab=$\scriptstyle e$,r,p=0.3}](0);
\draw[anyon1gon,mark=arr,mark={slab=$\scriptstyle f$,r}] (m-c)to[out=-160,in=135]++(-135:0.5)to[out=-45,in=-110]cycle;
\atoms{vertex,vcb}{m0/p=m}
\end{tikzpicture}
}
=
\delta_{d+a,b}\delta_{e+a,c}
\\
\rho^x_{f,a}\omega_{e,f,a}\omega_{e,a,f}
\pcoc
\Ket{
\begin{tikzpicture}
\atoms{void}{0/, 3/p={1.6,1.6}}
\draw (0)edge[bend right,mark=arr,mark={slab=$\scriptstyle b$,r}](3) (0)edge[bend left,mark=arr,mark={slab=$\scriptstyle c$}](3);
\draw[anyon1gon,mark=arr,mark={slab=$\scriptstyle f$,r}] (3-c)to[out=-160,in=135]++(-135:0.5)to[out=-45,in=-110]cycle;
\atoms{vertex,vca}{0x/p=0, 3x/p=3}
\end{tikzpicture}
}\;.
\end{multlined}
\end{equation}
In the third step, we glue a tetrahedron at a single face, yielding a 6-qubit operator $W_1$,
\begin{equation}
\label{eq:w_definition}
\begin{multlined}
W_1
\Ket{
\begin{tikzpicture}
\atoms{vertex,vca}{1/p={-135:0.8}, 2/p={-10:0.8}, 3/p={100:0.8}}
\draw (1)edge[mark=arr,mark={slab=$\scriptstyle a$,r}](2) (1)edge[mark=arr,mark={slab=$\scriptstyle b$}](3) (2)edge[mark={arr,-},mark={slab=$\scriptstyle c$,r}](3);
\end{tikzpicture}
}
\\
=
\sum_y
\omega_{y,y+b,c}
\pcoc
\Ket{
\begin{tikzpicture}
\atoms{vertex,vcb}{0/}
\atoms{vertex,vca}{1/p={-135:1.2}, 2/p={-20:1.2}, 3/p={110:1.2}}
\draw (1)edge[mark=arr,mark={slab=$\scriptstyle a$,r}](2) (1)edge[mark=arr,mark={slab=$\scriptstyle b$}](3) (2)edge[mark={arr,-},mark={slab=$\scriptstyle c$,r}](3);
\draw (0)edge[mark={arr,-},mark={slab=$\scriptstyle y$,r}](1) (0)edge[mark={arr,p=0.6},mark={slab=$\scriptstyle y+a$,p=0.3,r,sideoff=-0.1}](2) (0)edge[mark=arr,mark={slab=$\scriptstyle y+b$,p=0.2,sideoff=-0.1}](3);
\end{tikzpicture}
}\;.
\end{multlined}
\end{equation}
As discussed before, we now arbitrarily extend $\mathbf U$, $\mathbf R$, $\mathbf S$, and $\mathbf W$ into isometries $\widetilde{\mathbf U}$, $\widetilde{\mathbf R}$, $\widetilde{\mathbf S}$, and $\widetilde{\mathbf W}$ supported on the full Hilbert space.
Then, we replace the instrument $I[\widetilde{\mathbf T}]$ by a sequence of up-to-6-qubit instruments
\begin{equation}
(I[\widetilde{\mathbf U}], I[\widetilde{\mathbf U}]) \rightarrow I[\widetilde{\mathbf S}] \rightarrow I[\widetilde{\mathbf W}]
\rightarrow (I[\widetilde{\mathbf R}], I[\widetilde{\mathbf R}])\;.
\end{equation}
To extend the operators, we essentially just remove the $\pcoc$ terms from the corresponding definitions.
This way, $V_1$ in Eq.~\eqref{eq:v1_definition} becomes a unitary
\begin{equation}
\widetilde V_1\ket{d,a,b,e} = \omega_{d,a,b} \ket{d,a,b,e+d+b}\;,
\end{equation}
acting trivially on the label $c$.
This unitary can be written as a circuit of controlled-$X$ and controlled-controlled-$Z$ gates,
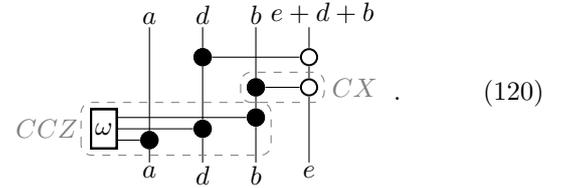
\begin{equation}
\begin{tikzpicture}
\atoms{square,lab={t=$\omega$,p=0:0},yscale=1.5}{om/p={-0.6,0.45}}
\atoms{delta}{d0/p={1.4,1}, d1/p={0.7,1.4}, dx0/p={0,0.3}, dx1/p={0.7,0.45}, dx2/p={1.4,0.6}}
\atoms{z2}{z0/p={2.1,1}, z1/p={2.1,1.4}}
\draw (dx0)edge[ind=$a$](0,0) (dx1)edge[ind=$d$](0.7,0) (dx2)edge[ind=$b$](1.4,0) (z0)edge[ind=$e$](2.1,0) (z1)edge[ind=$\quad e+d+b$](2.1,1.8) (d1)edge[ind=$d$](0.7,1.8) (d0)edge[ind=$b$](1.4,1.8) (dx0)edge[ind=$a$](0,1.8);
\draw (z0)--(d0) (z1)--(d1) (d1)--(dx1) (d0)--(dx2) (z0)--(z1) (dx0)--([sy=-0.15]om-r) (dx1)--(om-r) (dx2)--([sy=0.15]om-r);
\draw[rc,dashed,gray] (-0.9,0.1)rectangle(1.6,0.8);
\node[gray] at (-1.35,0.45){$CCZ$};
\draw[rc,dashed,gray] (1.2,0.8)rectangle(2.3,1.2);
\node[gray] at (2.7,1){$CX$};
\end{tikzpicture}\;.
\end{equation}
$S_x$ in Eq.~\eqref{eq:s_definition} is a map from 6 to 3 qubits.
Since there are 4 anyons and thus 4 measurement results $x$, we need to measure one further qubit to turn $\mathbf S$ into an isometry on the full Hilbert space.
In order to fulfil Definition~\ref{def:path_integral_code}, the measurement outcome for this further qubit must be deterministic inside the cocycle subspace.
This can be done by measuring the cocycle constraint, e.g., on the $(a,c,e)$ triangle in Eq.~\eqref{eq:s_definition}.
Using $\omega_{e,f,a}\omega_{e,a,f}=1$ and $f=d+e$ inside the cocycle subspace, we obtain an isometry
\begin{equation}
\widetilde S_x\ket{c,e,d,a} = \rho^x_{d+e,a}\ket{c,c+e+a}\;,
\end{equation}
acting trivially on $b$ and $f$.
$\widetilde S_x$ be expressed as a circuit,
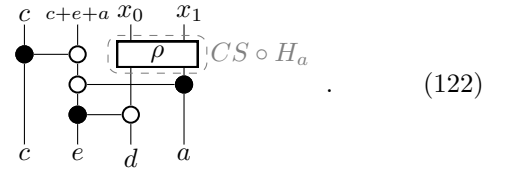
\begin{equation}
\begin{tikzpicture}
\atoms{delta}{d0/p={0.7,0.4}, d1/p={2.1,0.8}, d2/p={0,1.2}}
\atoms{z2}{z0/p={1.4,0.4}, z1/p={0.7,0.8}, z2/p={0.7,1.2}}
\atoms{square,lab={t=$\rho$,p=0:0}}{{rho/p={1.75,1.2},xscale=3}}
\draw[rc] (d0)--(z0) (d1)--(z1) (d2)--(z2) (d0)--(z1) (z1)--(z2) (z0)--([sx=-0.35]rho-b) (d1)--([sx=0.35]rho-b);
\draw (d2)edge[ind=$c$](0,0) (d0)edge[ind=$e$](0.7,0) (z0)edge[ind=$d$](1.4,0) (d1)edge[ind=$a$](2.1,0) (z2)edge[ind=$\scriptstyle c+e+a$](0.7,1.6) (d2)edge[ind=$c$](0,1.6) ([sx=-0.35]rho-t)edge[ind=$x_0$](1.4,1.6) ([sx=0.35]rho-t)edge[ind=$\hspace{1.5mm}x_1$](2.1,1.6);
\draw[dashed,gray,rc] (1.1,0.95)rectangle(2.4,1.45);
\node[gray] at (3.1,1.2){$CS\circ H_a$};
\end{tikzpicture}\;.
\end{equation}
Here we have split $x\rightarrow (x_0,x_1)$ into two qubits using $1\rightarrow (0,0)$, $s\rightarrow (1,0)$, $\bar s\rightarrow (1,1)$, and $s\bar s\rightarrow (0,1)$.
So the qubits labeled $x_0$, $x_1$, and $c+e+a$ are measured after applying the above isometry.
$\rho$ is a 2-qubit gate which in fact equals a Hadamard on the $a$ qubit followed by a controlled-$S$ gate.
The operator $W_1$ in Eq.~\eqref{eq:w_definition} becomes an isometry
\begin{equation}
\widetilde W_1\ket{a,b,c} = \sum_y \omega_{y,y+b,c} \ket{a,b,c,y,y+b,y+a}\;.
\end{equation}
$\widetilde W_1$ can be written as a circuit,
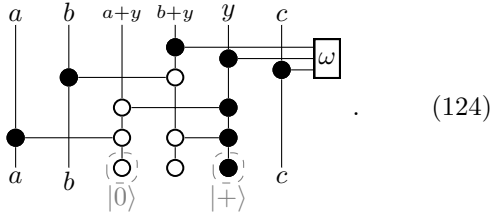
\begin{equation}
\begin{tikzpicture}
\atoms{square,lab={t=$\omega$,p=0:0},yscale=1.5}{om/p={4.8,1.45}}
\atoms{delta}{d0/p={0.7,0.4}, d1/p={1.4,1.2}, d2/p={3.5,0.4}, d3/p={3.5,0.8}, dx0/p={4.2,1.3}, dx1/p={3.5,1.45}, dx2/p={2.8,1.6}, dplus/p={3.5,0}}
\atoms{z2}{z0/p={2.1,0.4}, z1/p={2.8,1.2}, z2/p={2.8,0.4}, z3/p={2.1,0.8}, zp0/p={2.1,0}, zp1/p={2.8,0}}
\draw (dx0)edge[ind=$c$](4.2,1.9) (dx1)edge[ind=$y$](3.5,1.9) (dx2)edge[ind=$\scriptstyle b+y$](2.8,1.9) (d0)edge[ind=$a$](0.7,0) (d1)edge[ind=$b$](1.4,0) (z3)edge[ind=$\scriptstyle a+y$](2.1,1.9) (d0)edge[ind=$a$](0.7,1.9) (d1)edge[ind=$b$](1.4,1.9) (dx0)edge[ind=$c$](4.2,0);
\draw (dx0)--([sy=-0.15]om-l) (dx1)--(om-l) (dx2)--([sy=0.15]om-l) (d3)--(dx1) (z1)--(dx2) (d0)--(z0) (d1)--(z1) (d2)--(z2) (d3)--(z3) (zp0)--(z0) (z0)--(z3) (zp1)--(z2) (z2)--(z1) (dplus)--(d2) (d2)--(d3);
\draw[rc,dashed,gray] (1.9,-0.2)rectangle(2.3,0.2);
\node[gray] at (2.1,-0.4){$\ket0$};
\draw[rc,dashed,gray] (3.3,-0.2)rectangle(3.7,0.2);
\node[gray] at (3.5,-0.4){$\ket+$};
\end{tikzpicture}\;.
\end{equation}
We have thus decomposed our QEC process as a circuit of common 2 or 3-qubit gates.
For a practical implementation it might again be useful to write this circuit in terms of measurements and unitaries acting on qubits on a fixed spatial lattice.
This is straight-forward, but might involve auxiliary qubits and swap operations.

Finally, let us briefly discuss how the decoding from Proposition~\ref{prop:path_integral_decoding} applies to the present code.
The recorded syndrome at a time $T\sim L$ consists of labelings of the $x+y+t$-edges by the anyon types $1$, $s$, $\bar s$, and $s\bar s$.
By interpreting the anyon types as elements of $\zz_2\times \zz_2$, generated by $s$ and $\bar s$, the syndrome defines a 1-chain in the spacetime cellulation, only supported on the $x+y+t$-edges.
In the absence of noise, this 1-chain is a $\zz_2\times \zz_2$-valued 1-cycle, and hence every edge on the same $x+y+t$-line has the same value.
With noise, the syndrome is a 1-chain whose boundary is a $\zz_2\times \zz_2$-valued 0-cycle.
We let the decoder $D$ choose a minimum-weight fix of the 1-cycle, supported on all of the spacetime lattice and not just the $x+y+t$-edges.
Restricted to the spatial lattice at time $T$, the syndrome yields a $\zz_2\times \zz_2$ 0-cycle corresponding to a pattern of anyons in space.
We then apply corrections that match up the anyons through ``horizontal'' (when time goes upwards) worldlines such that we obtain a homologically trivial syndrome in spacetime.
The correction operators are well-known as \emph{ribbon operators} in the double-semion model \cite{Levin2004}.
In our language, these operators can be obtained by embedding an anyon tube segment as in Eq.~\eqref{eq:anyon_segment} horizontally into the spacetime cellulation instead of vertically as in Eq.~\eqref{eq:anyon_projector}, or diagonally as in Eq.~\eqref{eq:diagonal_tube_segment}.

\section{Discussion and outlook}
\label{sec:outlook}
In this paper we have proposed a perspective on topological quantum error correction based on topological fixed-point path integrals.
Our approach provides a unified view on topological stabilizer, subsystem, and Floquet codes, as demonstrated in Section~\ref{sec:examples}.
In particular, we have seen that the stabilizer toric code, subsystem toric code, and CSS honeycomb Floquet code can be considered the same code on different spacetime lattices.
The approach can also describe topological QEC codes that are not based on Pauli/Clifford operations as we have demonstrated in Section~\ref{sec:double_semion_code}.
As summarized in Definition~\ref{def:path_integral_code} and Proposition~\ref{prop:path_integral_decoding}, we have given a simple unified criterion for when a circuit of measurements forms a fault-tolerant topological error-correcting code.
Namely that, for every spacetime history of measurement outcomes, we obtain a topological fixed-point path integral including syndrome defects.

Our framework provides a way to systematically construct new codes.
To this end, we start with some known fixed-point path integral, and possibly apply some tensor-network equations to obtain another path integral in the same fixed-point phase.
Then we interpret this path integral as a circuit of operators by setting a time direction.
Dressing every operator with segments of syndrome defects, we obtain a circuit of instruments with the desired properties.
We have demonstrated this at hand of two examples in Section~\ref{sec:new_examples}.
First, we have presented a Floquet version of the $3+1$-dimensional toric code, by considering the tensor-network path integral on a hypercubic lattice and traversing it in the $t=x+y+z+w$ direction.
The model has qubits living on the right-handed tetrahedra of a triangulation with 4-colored vertices.
The code cycles through 8 rounds, in each of which we perform 2-body measurements among the qubits adjacent to edges of a certain type.
Second, we have constructed a Floquet code based on the double-semion string net.
This code is not designed to be particularly practical for implementation, but is decomposed into a sequence of common 2 or 3-qubit gates.

While this paper was being finalized, Ref.~\cite{Bombin2023} appeared on the arXiv which proposes a similar perspective based on the $ZX$ calculus.
In that reference, it was independently recognized that the tensor-network diagrams for the stabilizer toric code and CSS honeycomb Floquet code are the same, just traversed in a different direction.
In addition to this, our work provides a clear physical interpretation of the tensor networks as topological fixed-point path integrals including topological defects.
We also give a neat geometric interpretation of the phaseless $ZX$ diagrams as cellulations, the $ZX$ rules as topological invariance, and the \emph{Pauli webs} or detection cells as volumes and vertices.
As can be seen from Ref.~\cite{Bombin2023}, also fusion-based topological quantum computation \cite{Bartolucci2021} is described by our formalism.
This holds true for topological measurement-based quantum computing \cite{Raussendorf2007} in general.
A relation between the fusion-based model and the CSS honeycomb Floquet code has also been pointed out in Ref.~\cite{Paesani2022}.
In contrast to all of the above examples, our formalism is not limited to the $ZX$ calculus or stabilizer framework, but works for arbitrary tensor-network path integrals, as demonstrated in Section~\ref{sec:double_semion_code}.
\footnote{Even though any tensor can be written as a $ZX$ diagram, it can be beneficial to work with elementary operations that are not elementary $ZX$ tensors.}

The framework can be generalized in various directions.
First, topological state-sum path integrals do not cover all zero-correlation length path integrals, and similarily not all gapped phases.
Exemptions can be obtained from topological path integrals by inserting a rigid network of topological defects, which we refer to as \emph{foliation defects}.
To this end, we choose some cubic ``superlattice'' with a potentially larger unit cell than the topological path integral.
Then (in 2+1 dimensions) we introduce domain walls at all superlattice faces, which meet at 1-dimensional foliation defects along the edges, which in turn meet at the vertices.
Examples for this in 2+1 dimensions seem to yield topological path integrals again after choosing a larger unit cell, and thus correspond to a ``weak breaking of translation symmetry'', as we have seen in Section~\ref{sec:honeycomb_floquet}.
In 3+1 dimensions however, topological defect networks can describe fracton phases \cite{Aasen2020}, and potentially more if we also insert foliation defects perpendicular to time \cite{Williamson2022}.
Floquet codes based on fracton phases have been presented in Refs.~\cite{Davydova2022,Zhang2022}.
So all in all, topological defects can play three different roles in our formalism, namely computational defects, syndrome defects, and foliation defects.

A second straight-forward generalization is to consider spacetime lattices that change with time.
By changing the topology of the spatial configuration, we obtain circuits that do not only fault-tolerantly store, but also process logical information.
Both storing and processing of logical information becomes much more versatile if we equip the topological path integral with computational defects such as boundaries, domain walls, or other sorts of interfaces and defects.
For example, we can then perform computation via braiding with anyons or via lattice surgery with boundaries.

Another direction is to consider path integrals where the defects that we use for error correction (such as anyons) do not possess abelian fusion rules.
In this case the scheme of Proposition~\ref{prop:path_integral_decoding} outlined in Eq.~\eqref{eq:qec_circuit} cannot work, since there is not necessarily a unique way to perform a correction.
For example, consider a path integral QEC circuit based on the double-Fibonacci phase, and assume we measure the following spacetime syndrome on a torus,
\begin{equation}
\begin{tikzpicture}
\fill[cyan,opacity=0.2] (0.8,1)rectangle++(2,2);
\fill[cyan,opacity=0.2] (0,0)--++(0.8,1)--++(0,2)--++(-0.8,-1)--cycle;
\draw[cyan] (0.8,1)--++(0,2) (2.8,1)--++(0,2);
\fill[red,opacity=0.2] (0,0)--++(2,0)--++(0.8,1)--++(-2,0)--cycle;
\draw[thick] (2.4,1.2)to[out=180,in=-30]($(1.4,1.5)+(-30:0.2)$) (2.4,1.8)to[out=180,in=30]($(1.4,1.5)+(30:0.2)$) (0.4,1.2)to[out=0,in=-150]($(1.4,1.5)+(-150:0.2)$) (0.4,1.8)to[out=0,in=150]($(1.4,1.5)+(150:0.2)$);
\draw[red,dashed] (1.4,1.5)circle(0.2);
\fill[red,opacity=0.2] (0,2)--++(2,0)--++(0.8,1)--++(-2,0)--cycle;
\fill[cyan,opacity=0.2] (2,0)--++(0.8,1)--++(0,2)--++(-0.8,-1)--cycle;
\fill[cyan,opacity=0.2] (0,0)rectangle++(2,2);
\draw[cyan] (0,0)--++(0,2) (2,0)--++(0,2);
\end{tikzpicture}\;,
\end{equation}
with the left and right, as well as front and back identified.
There are two ways of fixing the syndrome inside the red dashed circle, namely
\begin{equation}
\begin{tikzpicture}
\draw[red,dashed] (0,0)circle(0.5);
\draw[thick] (-150:0.5)to[bend left=20](-30:0.5) (150:0.5)to[bend right=20](30:0.5);
\end{tikzpicture}
\;,\qquad
\begin{tikzpicture}
\draw[red,dashed] (0,0)circle(0.5);
\atoms{vertex}{0/p=180:0.2, 1/p=0:0.2}
\draw[thick] (0)--(1) (-150:0.5)--(0) (1)--(-30:0.5) (150:0.5)--(0) (1)--(30:0.5);
\end{tikzpicture}\;,
\end{equation}
which correspond to different logical operations acting on the ground space on a torus.
There is no way to find out which superposition of these logical operations will correctly undo the error that occured.
A decoding strategy that has been tested successfully is based on a hierachical decomposition of the lattice into \emph{colonies} \cite{Dauphinais2016,Schotte2022}.
A different strategy that might work is to ``continuously'' apply small corrections in every timestep instead of one large correction after a large time $T\sim L$.
That is, in every time step, we choose a new low-weight fix of the spacetime syndrome in all of its past.
Then we consider the set of string operators that could be used to close the repaired spacetime syndrome in a cohomologically trivial way inside the current spatial cut.
We pick a low (e.g., minimum) weight representative from this set.
Then, we apply only a single segment of this closing string operator near each of its endpoints.
Independent of the choice of classical decoder, it will be interesting to see whether and how our framework can be used to construct syndrome extraction circuits for arbitrary non-abelian phases.

Another very interesting question concerns chiral phases, that is, topological phases in $2+1$ dimensions whose anyon theory is described by a unitary modular tensor category that is not a Drinfeld center.
It is a common believe that chiral phases do not allow for exactly solvable fixed-point zero-correlation length descriptions, and no such descriptions are known to date.
Concretely, it has been shown that chiral phases do not admit commuting-projector Hamiltonian models \cite{Kapustin2019}.
However, there are indications that going from Hamiltonians to discrete path integrals might resolve this problem \cite{universal_liquid}.
In contrast to condensed matter physics, discrete path integrals (i.e., circuits) are the much more common in topological QEC.
Thus, it is natural to look there for candidates of chiral topological fixed-point path integrals.
Indeed, subsystem codes based on chiral topological phases exist.
Already more than a decade ago, Ref.~\cite{Bombin2009} presented a subsystem code that appears to be related to the 3-fermion phase.
Recently, subsystem codes based on arbitrary (including chiral) abelian anyon theories have been constructed in Ref.~\cite{Ellison2022} using a mechanism of ``gauging out'' anyons.
While it seems to be possible to construct a measurement schedule such that the logical dimension of the instantaneous stabilizer group equals the ground-state dimension of the chiral phase, it is unclear whether the path integral corresponding to these codes genuinely represent the chiral phase.
It might not be a robust topological path integral at all, in which case it seems unlikely that the resulting QEC protocol is fault tolerant.
Or, it might correspond to some larger non-chiral phase (such as the doubled phase), in which case these codes should not be called ``chiral'' when thought of as concrete QEC protocols.
If the resulting path integral is indeed a chiral fixed point, this would solve an important open problem in TQFT.
In any way, it will be highly interesting to analyze subsystem codes related to chiral anyon theories using the path integral formalism, and shed light on their fault-tolerance properties.

\subsubsection*{Acknowledgments}
I would like to thank Julio Magdalena de la Fuente, Alex Townsend-Teague, Alexander Nietner, Ansgar Burchards, Jens Eisert, Margarita Davydova, Shankar Balasubramanian, and David Aasen for helpful conversations and comments on the manuscript, and especially Markus Kesselring for fruitful discussions on 3-dimensional tessellations.
This work was supported by the DFG (CRC 183 project B01), the BMBF (RealistiQ, QSolid), the Munich Quantum Valley (K-8), and the BMWK (PlanQK).

\bibliographystyle{quantum}

\end{document}